\documentclass[prl,twocolumn,twoside,preprintnumbers,superscriptaddress,nofootinbib,letterpaper]{revtex4-1}

\usepackage{amsmath,slashed}
\usepackage{graphicx, graphics, color}
\usepackage{dcolumn}
\usepackage{xspace}
\usepackage{enumerate}
\usepackage{varwidth}
\usepackage{microtype}

\usepackage{physics}
\usepackage{amssymb}
\usepackage{mathtools}
\usepackage{dsfont}
\usepackage{multirow}

\usepackage[hyperfootnotes=true,colorlinks=false,citecolor=blue,linkcolor=blue,urlcolor=blue]{hyperref}

\newcommand{\GeV}{\,\text{GeV}}
\newcommand{\MeV}{\,\text{MeV}}

\newcommand{\mpi}{M_\pi}

\newcommand{\eps}{\epsilon}

\newcommand{\Order}{\mathcal{O}}
\newcommand{\beq}{\begin{equation}}
\newcommand{\eeq}{\end{equation}}
\newcommand{\Mhad}{M_\text{had}}

\allowdisplaybreaks[1]

\begin{document}

\title{Pion \texorpdfstring{$\boldsymbol{\beta}$}{} decay and \texorpdfstring{$\boldsymbol{\tau\to\pi\pi\nu_\tau}$}{} beyond leading logarithms}

\author{Vincenzo Cirigliano}
\affiliation{Institute for Nuclear Theory, University of Washington, Seattle WA 91195-1550, USA}
\author{Martin Hoferichter}
\affiliation{Albert Einstein Center for Fundamental Physics, Institute for Theoretical Physics, University of Bern, Sidlerstrasse 5, 3012 Bern, Switzerland}
\author{Nicola Valori}
\affiliation{Institute for Nuclear Theory, University of Washington, Seattle WA 91195-1550, USA}
\affiliation{Instituto de F\'isica Corpuscular, Universidad de Valencia and CSIC, Edificio Institutos Investigaci\'on,
C/Catedr\'atico Jos\'e Beltr\'an 2, 46980 Paterna, Spain}

\begin{abstract}
   The consistent matching of short-distance contributions and hadronic matrix elements is crucial for precise predictions of weak processes involving hadrons. In this Letter, we address this point for charged-current processes involving two pions---pion $\beta$ decay $\pi^\pm\to\pi^0 e^\pm\nu_e$ and hadronic $\tau$ decays $\tau^\pm\to\pi^\pm\pi^0\nu_\tau$---whose decay rates depend on the so-called $\gamma W$ box correction. Using recent results from lattice QCD, we show how to formulate the matching beyond leading-logarithmic accuracy, in particular, how to cancel the dependence on the scheme choice for evanescent operators. As main results, we
  obtain a prediction for the decay rate of pion $\beta$ decay with theory uncertainties improved by a factor of three, which renders theory uncertainties negligible for future determinations of $V_{ud}$ even beyond the reach of the PIONEER experiment, and
  an evaluation of isospin-breaking corrections to $\tau\to\pi\pi\nu_\tau$ with negligible uncertainty from the short-distance matching, as necessary for a future $\tau$-based determination of the hadronic-vacuum-polarization contribution to the anomalous magnetic moment of the muon.
\end{abstract}

\maketitle

\emph{Introduction}---A precision determination of the Cabibbo--Kobayashi--Maskawa (CKM)~\cite{Cabibbo:1963yz,Kobayashi:1973fv} matrix element $V_{ud}$, either from neutron decay or superallowed nuclear $\beta$ decays, crucially relies on a careful consideration of radiative corrections~\cite{Hardy:2020qwl}, which thus play a key role in the first-row unitarity test of the CKM matrix.
Despite a long history in their evaluation~\cite{Kinoshita:1958ru,Sirlin:1967zz,Sirlin:1967zza,Abers:1968zz,Jaus:1970tah,Sirlin:1977sv,Sirlin:1981ie,Wilkinson:1982hu,Sirlin:1986cc,Towner:1992xm,
Wilkinson:1993hxz,Wilkinson:1993fva,Czarnecki:2004cw,Marciano:2005ec,Towner:2010zz}, it has proven challenging to quantify and ideally remove the remaining model dependence in these corrections, and substantial effort has been invested in recent years to address this issue using dispersion relations, lattice QCD, effective field theory (EFT), and ab-initio nuclear-structure methods~\cite{Seng:2018yzq,Seng:2018qru,Gorchtein:2018fxl,Seng:2022cnq,Ma:2023kfr,Seng:2023cvt,Hill:2023acw,Cirigliano:2023fnz,Cirigliano:2024rfk,Cirigliano:2024msg,Borah:2024ghn,Gennari:2024sbn,VanderGriend:2025mdc,King:2025fph,Cao:2025zxs,Crosas:2025xyv,Gorbahn:2025ssv}. Such robust uncertainty quantification is critical to either corroborate current hints for a unitarity deficit or impose stringent constraints on potential physics beyond the Standard Model (SM)
~\cite{Belfatto:2019swo,Coutinho:2019aiy,Cheung:2020vqm,Belfatto:2021jhf,Branco:2021vhs,Crivellin:2021bkd,Crivellin:2020ebi,Kirk:2020wdk,Crivellin:2021njn,Crivellin:2020lzu,Crivellin:2020klg,Capdevila:2020rrl,Crivellin:2021sff,Crivellin:2020oup,Marzocca:2021azj,Alok:2021ydy, Cirigliano:2022qdm,Cirigliano:2023nol,Dawid:2024wmp}.

Already in the single-nucleon system, this program requires the calculation of perturbative corrections and their resummation~\cite{Cirigliano:2023fnz,VanderGriend:2025mdc,Gorbahn:2025ssv}, but also the incorporation of nonperturbative matrix elements, most notably the so-called $\gamma W$ box corrections~\cite{Marciano:2005ec,Seng:2018yzq,Seng:2018qru,Czarnecki:2019mwq,Seng:2020wjq,Hayen:2020cxh,Shiells:2020fqp,Cirigliano:2022yyo,Ma:2023kfr}. Extending previous analyses in the meson sector~\cite{Descotes-Genon:2005wrq} to the nucleon case, in Ref.~\cite{Cirigliano:2023fnz} it was shown how to formulate the traditional master formula for the neutron decay rate based on an EFT approach, allowing for the consistent resummation of large logarithms. This approach requires the matching of low-energy EFT (LEFT) to chiral perturbation theory (ChPT), to separate low-energy constants (LECs) into short-distance components and nonperturbative matrix elements. As a key feature, the scheme dependence due to the choice of evanescent operators manifestly cancels between LEFT Wilson coefficients and the matrix elements, leading to a result for the decay rate in which the scheme dependence as well as the dependence on the LEFT scale $\mu$ and the chiral scale $\mu_\chi$ cancel up to the order in the fine-structure constant $\alpha$ and the strong coupling $\alpha_s$  to which the perturbative calculation has been performed.

While not competitive at present, an alternative avenue towards the determination of $V_{ud}$ proceeds via pion $\beta$ decay, $\pi^\pm\to\pi^0e^\pm\nu_e$~\cite{Pocanic:2003pf,Cirigliano:2002ng,Czarnecki:2019iwz}, which is sensitive to the same LEFT operator as the $\beta$ decays involving nucleons, in such a way that the same challenges arise in its theoretical description. In this case, input from lattice QCD~\cite{Feng:2020zdc,Ma:2021azh,Yoo:2023gln} is available to determine the nonperturbative parts, and given ongoing efforts to improve the experimental measurement of the branching fraction at PIONEER~\cite{PIONEER:2022yag,PIONEER:2025idw} it is timely to work out the matching beyond leading-logarithmic (LL) accuracy for this process as well, by adapting the methods from Refs.~\cite{Descotes-Genon:2005wrq,Cirigliano:2023fnz} accordingly.

Similarly, by crossing lepton and charged pion, it is evident that the exact same procedure also determines the short-distance contribution to hadronic $\tau$ decays $\tau^\pm\to\pi^\pm\pi^0\nu_\tau$, a crucial ingredient for the evaluation of isospin-breaking (IB) corrections in the context of the hadronic-vacuum-polarization (HVP) contribution to the anomalous magnetic moment of the muon $a_\mu$~\cite{Alemany:1997tn,Cirigliano:2001er,Cirigliano:2002pv,Flores-Baez:2006yiq,Davier:2010fmf,Miranda:2020wdg,Castro:2024prg,Aoyama:2020ynm,Aliberti:2025beg,Hertzog:2025ssc}. While a $\tau$-based estimate is provided in Ref.~\cite{Aliberti:2025beg}, based on input from Refs.~\cite{Davier:2023fpl,Davier:2010fmf,Castro:2024prg,Colangelo:2022prz,Hoferichter:2023sli}, several key short-comings need to be addressed to allow for a full uncertainty quantification of the  IB corrections. One of them concerns the scheme dependence of the short-distance contribution, which boils down to the exact same LECs required for pion $\beta$ decay, as a direct consequence of lepton flavor universality.

Accordingly, we first provide a general discussion of the pion matrix element, matching relations, and renormalization-group (RG) resummation, highlighting the universal form of the resulting correction to the decay rate, and then discuss the phenomenological applications to pion $\beta$ decay and hadronic $\tau$ decays. To obtain a complete result for the decay rates, we also need a calculation of the corresponding long-range radiative corrections, which due to the small phase space for pion $\beta$ decay can proceed in ChPT~\cite{Cirigliano:2002ng}, while for $\tau$ decays a full calculation of the structure-dependent corrections has only recently become available~\cite{Colangelo:2025iad,Colangelo:2025ivq}.

\emph{Renormalization group and matrix elements}---We use EFT methods to efficiently compute  $\beta$ decay rates in the SM
beyond the LL approximation.
Integrating out the weak gauge bosons, the Higgs boson, and the top quark,
we write the pertinent part of the LEFT Lagrangian with
$n_f$ active quark flavors in the convention
\begin{align}
{\mathcal L}_\text{LEFT}&=-2\sqrt{2}G_F\bar e_L\gamma_\rho\mu_L\,\bar\nu_{\mu L}\gamma^\rho \nu_{eL}
\label{eq:LEFT1}
\\
 &-2\sqrt{2}G_F V_{ud} \ C^{(f)}_\beta(\mu,a) \ \bar e_L\gamma_\rho \nu_{eL}\,\bar u_L\gamma^\rho d_L+\text{h.c.},
\notag
\end{align}
i.e., the Fermi constant $G_F$ is defined in muon decay~\cite{MuLan:2012sih}, and a Wilson coefficient $C_\beta^{(f)} (\mu,a)=1+\Order(\alpha)$ describes the semi-leptonic operator in the $n_f$-flavor theory. It depends on the LEFT scale $\mu$ in the $\overline{\text{MS}}$ scheme as well as the scheme chosen for the evanescent operator, defined by
$E(a)$\footnote{We use here
the evanescent scheme parameter $a$ as defined in Ref.~\cite{Gorbahn:2025ssv}, which is related to the conventions of Ref.~\cite{Cirigliano:2023fnz} by $a = 2 a^\text{\cite{Cirigliano:2023fnz}} + 3$.}
\beq
\gamma^\alpha\gamma^\rho\gamma^\beta P_L\otimes \gamma_\alpha \gamma_\rho\gamma_\beta P_L=4\big(4 -a \epsilon \big) \gamma^\rho P_L\otimes \gamma_\rho P_L + E(a),
\eeq
where $d=4-2\eps$ is the dimensional regulator, and naive dimensional regularization for $\gamma_5$ is used. In these conventions one finds for the Wilson coefficient
in the five-flavor theory at $\mu \simeq M_{W, Z}$~\cite{Sirlin:1981ie,Brod:2008ss,Dekens:2019ept,Hill:2019xqk,Cirigliano:2023fnz,Gorbahn:2025ssv}
\begin{align}
C_\beta^{(5)}(\mu,a)&= 1+\frac{\alpha}{\pi}\log\frac{M_Z}{\mu}+\frac{\alpha}{4 \pi}B(a)-\frac{\alpha\alpha_s}{4\pi^2}\log\frac{M_Z}{\mu}\notag\\
&+ \frac{\alpha\alpha_s}{(4\pi)^2} B_s (a)
- \frac{\alpha\alpha_s}{4\pi^2} \frac{1}{s_{W}^{2}}\Bigg( \frac{c_{W}^{2}}{s_W^2} \log\frac{M_W}{M_Z}+1\Bigg)\notag\\
&+\Order\big(\alpha^2,\alpha\alpha_s^{2}\big) ,
\label{eq:Cmatch}
\end{align}
where $B(a)=\frac{a}{3}-4$ and $B_s(a)=\frac{41}{6}-\frac{14}{9}a$
parameterize the scheme dependence, and $s_{W}$, $c_{W}$ are the sine and cosine of the weak mixing angle.
The effective coupling $C_\beta^{(f)} (\mu, a)$ can be evolved down to the hadronic scale
using the RG equations (RGE). The solution is known to  next-to-leading-logarithmic (NLL)
$\big[\Order \big(\alpha^{n+1} \log^n\frac{M_Z}{\mu}\big)\big]$~\cite{Cirigliano:2023fnz}
and $\text{NLL}_s$  $\big[\Order \big(\alpha \, \alpha_s^{n+1} \log^n\frac{M_Z}{\mu}\big)\big]$~\cite{Gorbahn:2025ssv} accuracy~\cite{Supp}.\nocite{Moussallam:1997xx,Buras:1989xd,FlavourLatticeAveragingGroupFLAG:2024oxs,Maltman:2008bx,PACS-CS:2009zxm,McNeile:2010ji,Chakraborty:2014aca,Bruno:2017gxd,Bazavov:2019qoo,Cali:2020hrj,Ayala:2020odx,Petreczky:2020tky,DallaBrida:2022eua,Parker:2018vye,Morel:2020dww,Fan:2022eto,Sturm:2013uka,Davier:2019can,Keshavarzi:2019abf,Ce:2022eix,Erler:2023hyi,Conigli:2025qvh,Chetyrkin:1996cf,Fanchiotti:1992tu}
The final result for the decay amplitude
is written  in the three-flavor theory in terms of a scheme-independent Wilson coefficient
$\bar C_\beta^{(f)}  (\mu)  \equiv C^{(f)}_\beta (\mu, a)/J_f (\mu,a)$
defined in terms of the quantity $J_f(\mu,a)$ that naturally appears in the
solution of the RGE to NLL~\cite{Supp}.

\emph{Matching to ChPT}---Starting from Eq.~\eqref{eq:LEFT1}, to achieve  NLL (and NLL$_s$)  accuracy one needs to compute the matrix element
of $O_\beta \equiv     \bar e_L\gamma_\rho \nu_{eL}\,\bar u_L\gamma^\rho d_L$ between the appropriate external states
to  $\Order(\alpha)$ (and   $\Order(\alpha \alpha_s)$),  including finite, nonlogarithmically enhanced terms, whose scheme dependence will
cancel the one in $C^{(f)}_\beta (\mu,a)$.
In the EFT approach,   one   computes   $\langle f | O_\beta | i \rangle$ in ChPT,
the low-energy EFT in which the degrees of freedom are mesons and baryons,  along with light  particles involved in the electroweak interactions  (photons, charged leptons, and neutrinos).  Through tree-level and loop amplitudes,  ChPT  captures the infrared (IR) physics to a given order in the low-energy expansion, while  encoding the
short-distance  and  nonperturbative  effects of hard virtual photons in appropriate LECs.
We next provide a representation of the LECs contributing to pion $\beta$ decay and $\tau\to\pi\pi\nu_\tau$, capturing the short-distance contributions
to NLL and NLL$_s$ accuracy as well as nonperturbative effects originating at  the hadronic scale.
Next, we present the main results, while more details of the derivation are discussed in the End Matter.

In the spirit of EFT,
we  factorize the dependence on the LEFT Wilson coefficient from the chiral LECs~\cite{Cirigliano:2023fnz}, as a result of which these LECs carry a dependence on both scales $\mu$ and $\mu_\chi$, and on the evanescent scheme parameter $a$. The relevant linear combination of the $X_i$~\cite{Knecht:1999ag} and $K_i$~\cite{Urech:1994hd} reads
\beq
X_\ell(\mu_\chi,\mu, a)
\equiv \bigg(X_6^r -4K_{12}^r+\frac{4}{3}X_1^r\bigg)(\mu_\chi,\mu, a).
\label{eq:LECs}
\eeq
Matching LEFT with three  quark flavors and ChPT,
and using the results from Refs.~\cite{Cirigliano:2023fnz,Gorbahn:2025ssv}
for the $\Order(\alpha)$ and $\Order (\alpha \alpha_s)$ terms, respectively,
we find
\begin{align}
& -2\pi\alpha_\chi(\mu_\chi)  X_\ell (\mu_\chi, \mu,a)  \equiv   \bar\Box^{V(3)}_\pi(\mu_0)
\notag \\
 & \qquad   -\frac{\alpha^{(3)}(\mu)}{4\pi}\bigg(\frac{5}{4}+\frac{3}{2}\log\frac{\mu_\chi^2}{\mu_0^2}+ B(a)
 +2\log\frac{\mu_0^2}{\mu^2}\bigg)
 \notag \\
 & \qquad + \frac{\alpha\alpha_{s}^{(3)}(\mu)}{(4\pi)^2}\bigg(-B_{s}(a)
 +2-2\log\frac{\mu^2}{\mu_{0}^{2}}\bigg),
   \label{eq:final_res0}
\end{align}
where
$\bar \Box_\pi^{V(3)} (\mu_0)$ represents the nonperturbative matrix element,
$\alpha_\chi(\mu_\chi)$ denotes the fine-structure constant in ChPT~\cite{Cirigliano:2023fnz}, while the couplings on the right-hand side refer to $\overline{\text{MS}}$.\footnote{The scale in $\alpha_s^{(3)}$ is chosen following
Ref.~\cite{Gorbahn:2025ssv}.
Replacing $\alpha^{(3)}_s (\mu) \to \alpha^{(3)}_s (\mu_0)$
would ensure exact analytic cancellation of $\mu_0$~\cite{Cirigliano:2023fnz}.}
Combining Wilson coefficient and matrix element, we define the scheme-independent short-distance correction to the vector (Fermi)
transitions\footnote{
The usual  $S_\text{EW}$ correction  corresponds to $1 + 2 \Delta \tilde S_\text{EW}(\mu_\chi)$.
The tilde is introduced to indicate that now we include nonperturbative contributions.}
\begin{align}
g_V^\pi (\mu_\chi)  & \equiv 1+\Delta \tilde S_\text{EW}(\mu_\chi)\equiv
\bar C^{(3)}_\beta  (\mu)
\bigg[1+\bar\Box^{V(3)}_\pi(\mu_0)\notag\\
&+ \frac{\alpha^{(3)} (\mu)}{4 \pi}
  \bigg( \frac{17}{6}
  - \frac{3}{2}  \log \frac{\mu_\chi^2}{\mu_0^2}
 + 2  \log\frac{\mu^2}{\mu_0^2}
  \bigg)\notag\\
&+  \frac{\alpha \alpha_s^{(3)} (\mu)}{(4 \pi)^2}
 \bigg(
 - \frac{130}{81}  -  2  \log\frac{\mu^2}{\mu_0^2}
 \bigg)\bigg].
   \label{eq:final_res}
\end{align}
With appropriate replacement of the external states in the transition-dependent nonperturbative matrix element  $\bar\Box^V(\mu_0)$,
this correction universally applies to the Fermi component of all  $\beta$ decays.
The dependence on  $\mu_0$ cancels between
$\bar\Box^V_\pi(\mu_0)$
and the explicit logarithms, up to terms beyond $\text{NLL}_s$ accuracy; the dependence on $\mu$ cancels between the
scheme-independent Wilson coefficient $\bar C^{(3)}_\beta (\mu)$~\cite{Supp} and the explicit logarithms;
 finally the dependence on  $\mu_\chi$
cancels once one combines $g_V^\pi (\mu_\chi)$ with the chiral loops. Beyond one-loop order, RG corrections need to be included, see End Matter for details.

\begin{figure}[t]
    \centering
    \includegraphics[width=0.4\textwidth]{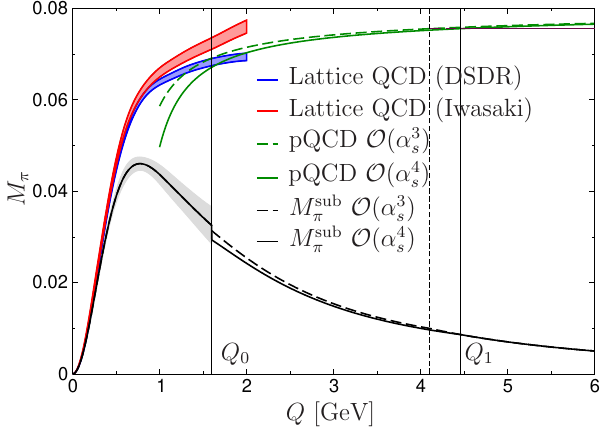}\\
    \includegraphics[width=0.4\textwidth]{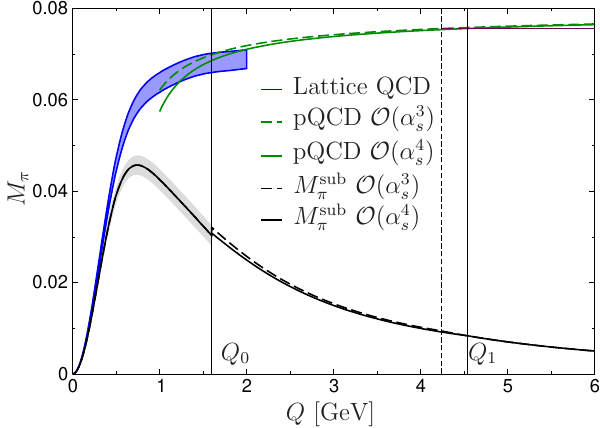}
    \caption{Resulting integrand $M_\pi(Q^2)$ for the lattice-QCD inputs from Ref.~\cite{Feng:2020zdc} (upper, $n_f=3$) and Ref.~\cite{Yoo:2023gln} (lower, $n_f=4$). In both cases, we show the lattice-QCD results up to $2\GeV$, the pQCD curves above $1\GeV$ (green dashed and solid for three- and four-loop order), and the constant one-loop result above the respective values of $Q_1$  (maroon dashed and solid). The resulting subtracted integrand as it enters in Eq.~\eqref{box_EFT} is represented by the black curves (dashed and solid for three- and four-loop pQCD between $Q_0$ and $Q_1$). The dashed vertical line indicates the value of $Q_1$ for the three-loop case.}
    \label{fig:Mpi}
\end{figure}

\emph{Nonperturbative matrix element}---Following Ref.~\cite{Cirigliano:2023fnz}
and taking into account the factorization of short-distance
QCD corrections from Ref.~\cite{Gorbahn:2025ssv},
the nonperturbative input is collected in
\begin{align}
 \bar\Box^V_\pi(\mu_0)
 &=-ie^2\int\frac{d^4q}{(2\pi)^4}\frac{\nu^2+Q^2}{Q^4}\notag\\
 &\times\bigg[\frac{T_3^\pi(\nu,Q^2)}{2M_\pi\nu}-\frac{2}{3}\frac{1-\frac{\alpha_s(\mu_0)}{\pi}}{Q^2+\mu_0^2}\bigg],
\label{eq:box1}
\end{align}
where  $\mu_0$ is a factorization scale that allows one to subtract the asymptotic part of the $VA$ correlator, known from the operator product expansion (OPE).
Its scalar projection is written as in Ref.~\cite{Cirigliano:2023fnz}, translating to the conventions of Ref.~\cite{Feng:2020zdc} via
\beq
\label{T3_iden}
\frac{1}{\sqrt{2}}T_3^\text{\cite{Feng:2020zdc}}(-i\nu,Q^2)=\frac{T_3^\pi(\nu,Q^2)}{4\mpi\nu},
\eeq
where $\nu=q_0$, $Q^2=-q^2$. Accordingly, we can express $\bar\Box^V_\pi(\mu_0)$ in terms of the integrand $M_\pi(Q^2)$ defined in Ref.~\cite{Feng:2020zdc}
\beq
\label{box_EFT}
\bar\Box^V_\pi(\mu_0)
=\frac{3\alpha}{2\pi}\int_0^\infty \frac{dQ^2}{Q^2}\bigg[M_\pi(Q^2)-\frac{1}{12}\frac{Q^2\big(1-\frac{\alpha_s(\mu_0)}{\pi}\big)}{Q^2+\mu_0^2}\bigg].
\eeq
For sufficiently small values of $Q^2\leq Q_0^2$, $M_\pi(Q^2)$ is known from lattice QCD~\cite{Feng:2020zdc,Yoo:2023gln}, while asymptotically its form is determined by perturbative QCD (pQCD)
\beq
\label{pQCD}
M_\pi(Q^2)=\frac{C_d(Q^2)}{12}, \qquad
C_d(Q^2)=1-\frac{\alpha_s}{\pi}+\sum_{n\geq 2} c_n\Big(\frac{\alpha_s}{\pi}\Big)^n,
\eeq
with coefficients $c_{n}$ known up to $n=4$~\cite{Larin:1991tj,Baikov:2010je}. The subtraction in Eq.~\eqref{box_EFT}, however, is only accurate at $\Order(\alpha_s)$, so that, asymptotically, one needs to assume this form in $M_\pi(Q^2)$ starting at some value $Q_1^2$. Accordingly, Eq.~\eqref{box_EFT} takes the form
\beq
\bar\Box^V_\pi(\mu_0)=\frac{3\alpha}{2\pi}\int_0^{Q_1^2} \frac{dQ^2}{Q^2}M_\pi(Q^2)+\frac{\alpha\big(1-\frac{\alpha_s(\mu_0)}{\pi}\big)}{8\pi}\log\frac{\mu_0^2}{Q_1^2}.
\eeq
To improve the matching between lattice QCD and the $\Order(\alpha_s)$ pQCD asymptotics, we take $Q_1$ larger than $Q_0$ and use the full four-loop pQCD result for $Q_0\leq Q\leq Q_1$. The subtraction scale $\mu_0$ is identified with $Q_0$, while $Q_1$ is determined as the point at which $C_d(Q_1^2)=1-\frac{\alpha_s(\mu_0)}{\pi}$ to ensure a smooth transition to the asymptotic form. We observe that the convergence of the pQCD expansion prefers a slightly higher matching point than $Q_0^2=2\GeV^2$~\cite{Feng:2020zdc}, at the expense of slightly larger uncertainties from lattice artifacts. We therefore choose $\mu_0=Q_0=1.6\GeV$, and evolve $\alpha_s$ down from $\alpha_s^{(5)}(M_Z)$~\cite{Chetyrkin:2000yt,Herren:2017osy}, at the loop order that corresponds to the accuracy of $C_d(Q^2)$ used. Given that the calculations in Refs.~\cite{Feng:2020zdc,Yoo:2023gln} are performed with $2+1$ and $2+1+1$ active quark flavors, respectively, we repeat the analysis for $n_f=3$ and $n_f=4$, leading to $Q_1^{(3)}=4.46\GeV$ and $Q_1^{(4)}=4.53\GeV$. Altogether, we obtain
\begin{align}
\label{lattice_box}
10^3\times\bar\Box^{V(3)}_\pi(\mu_0)\big|_\text{\cite{Feng:2020zdc}}&=
0.714(12)(28)-0.023(6)\\
&=0.691(31),\notag\\
10^3\times\bar\Box^{V(4)}_\pi(\mu_0)\big|_\text{\cite{Yoo:2023gln}}&=0.709(27)-0.020(4)=0.690(28),\notag
\end{align}
where the two terms refer to the contribution of the integral below $Q_0$ and the remainder, respectively. For the latter, the uncertainty derives from the difference to the three-loop pQCD result, while for the integral the uncertainties are propagated from the lattice-QCD calculations.
In the case of Ref.~\cite{Feng:2020zdc}, this error is split into statistical and systematic components, where the dominant systematic effect concerns lattice artifacts, estimated as the difference of the two discretizations considered.\footnote{We define the central value for Ref.~\cite{Feng:2020zdc} by the average of the two discretizations therein. The systematic error given in Ref.~\cite{Feng:2020zdc} also accounts for higher-twist effects, which are found to be subleading compared to the discretization uncertainties. For increased $Q_0$, the impact of such power corrections should be reduced further.} The corrections in Eq.~\eqref{lattice_box} that arise from the improved matching between the lattice-QCD results and the $\Order(\alpha_s)$ pQCD asymptotics are of the same size as the lattice-QCD errors, and thus control over the $\Order(\alpha_s^2)$ terms in the OPE subtraction in Eq.~\eqref{box_EFT} would be required to further improve the precision. The resulting integrand $M_\pi(Q^2)$ and its subtracted form according to Eq.~\eqref{box_EFT}  are shown in Fig.~\ref{fig:Mpi}.

\emph{Pion $\beta$ decay}---The master formula for pion $\beta$ decay reads~\cite{Cirigliano:2002ng,Czarnecki:2019iwz}\footnote{$G_F$ can be identified with its value extracted from muon decay $G_F=1.1663785(6)\times 10^{-5}\GeV^{-2}$~\cite{ParticleDataGroup:2026,MuLan:2012sih}, since the only process-dependent corrections not accounted for via the conventions~\eqref{eq:LEFT1} concern mass corrections to the $W$ propagator $2\Delta G_F^W\simeq 3m_\mu^2/(5M_W^2)\simeq 1\times 10^{-6}$~\cite{Ferroglia:2013dga,Fael:2013pja}.}
\begin{align}
\Gamma\big[\pi^+\to\pi^0 e^+\nu_e(\gamma)\big]&=
\frac{G_F^2|V_{ud}|^2\mpi^5|f_+^\pi(0)|^2}{64\pi^3}\notag\\
&\times\big(1+\Delta_\text{RC}^{\pi\ell}\big)I_{\pi\ell},
\end{align}
where $\mpi$ denotes the mass of the charged pion. The form factor $f_+^\pi(0)\simeq 1-7\times 10^{-6}$ deviates from unity by a negligible amount due to the Behrends--Sirlin--Ademollo--Gatto theorem~\cite{Behrends:1960nf,Ademollo:1964sr,Gasser:1984ux}, which ensures that IB corrections can only enter quadratically. The phase-space factor evaluates to
\beq
\label{phase_space}
I_{\pi\ell}=7.3767(41)\times 10^{-8},
\eeq
where the uncertainty derives from the pion mass difference, $\mpi-M_{\pi^0}=4.59364(48)\MeV$~\cite{Crawford:1990jc}.\footnote{This value includes corrections from the slope parameter $\lambda_+=0.036(3)$~\cite{Gasser:1984ux}, with uncertainties derived from higher-order effects in the LEC $L_6^r(M_\rho)=6.6(6)\times 10^{-3}$~\cite{Gasser:1984gg,Bijnens:2002hp,Colangelo:2021moe}.} Radiative corrections are included in $\Delta_\text{RC}^{\pi\ell}$, which decomposes as
\beq
\label{Delta_RC_def}
\frac{1+\Delta_\text{RC}^{\pi\ell}}{\big[1+\Delta \tilde S_\text{EW}(\mu_\chi)\big]^2}
=1+
\frac{3\alpha_\chi(\mu_\chi)}{2\pi}\log\frac{\mu_\chi}{2E_0}+\Delta I_{\pi\ell},
\eeq
where $\Delta I_{\pi\ell}=-0.00378(8)$~\cite{Cirigliano:2002ng}, including an estimate of higher chiral orders $\simeq\frac{\alpha}{\pi}\big(\frac{\mpi}{M_\rho}\big)^2\simeq 7.5\times 10^{-5}$.
In this form, we isolated the only large logarithm that remains in
 the phase-space correction $\Delta I_{\pi\ell}$ by virtue of the Kinoshita--Lee--Nauenberg theorem~\cite{Passera:2011ae,Kinoshita:1962ur,Lee:1964is}, sensitive to the endpoint energy
$E_0=(\mpi^2-M_{\pi^0}^2+m_e^2)/(2\mpi)=4.51894(48)\MeV$.
While we use the full relativistic calculation from Ref.~\cite{Cirigliano:2002ng}, the value of $\Delta I_{\pi\ell}$ can be understood from the integral over the Sirlin function, see End Matter.
Using Eq.~\eqref{eq:final_res}, we find
for the effective $\pi \pi$ vector coupling
\begin{equation}
\label{gVpi_Mrho}
    g_V^\pi (M_\rho)  = 1.01084(2)_{\bar \Box_\pi} (3)_{\mu}[4]_\text{tot},
\end{equation}
where the first uncertainty is propagated from Eq.~\eqref{lattice_box}\footnote{For Ref.~\cite{Yoo:2023gln}, we use the four-flavor analog of Eq.~\eqref{eq:final_res}~\cite{Supp}, yielding a $g_V^\pi(M_\rho)$ that only differs by $3 \times 10^{-6}$ from the three-flavor result derived from Ref.~\cite{Feng:2020zdc}. Accordingly, we quote the average as our final value.} and the second one indicates the residual perturbative uncertainty estimated by varying the  decoupling and matching scales as in Ref.~\cite{Gorbahn:2025ssv}.
For the radiative-correction factor $\Delta_\text{RC}^{\pi\ell}$ we further need to consider the evolution to $\mu_\chi\simeq 2E_0$, which gives
\begin{equation}
\label{Delta_RC}
\Delta_\text{RC}^{\pi\ell} = 0.03403(4)_{\bar \Box_\pi} (5)_{\mu} (5)_{\mu_\chi}(8)_\text{ChPT}[11]_\text{tot},
\end{equation}
where the uncertainty denoted by $\mu_\chi$ comprises a variation of $\mu_\chi$ within $[2E_0/\sqrt{2}, \sqrt{2}\, 2E_0]$ and of the matching scale within $[M_\rho/\sqrt{2},\sqrt{2}M_\rho]$, dominated by the latter.

Our result~\eqref{Delta_RC} differs substantially from $\Delta_\text{RC}^{\pi\ell}|_\text{\cite{Feng:2020zdc}}=0.0332(1)_{\bar \Box_\pi}(3)_\text{QED}$, primarily due to the low-energy RG corrections and the quadratic terms in Eq.~\eqref{Delta_RC_def}. In a fully linearized form,
the direct application of Eq.~\eqref{gVpi_Mrho} would give $\Delta_\text{RC}^{\pi\ell}=0.0335$, which then increases by
\begin{align}
\Delta_\text{RC}^{\pi\ell}\big|_\text{LL+quad.}&\simeq \bigg(\frac{\alpha}{\pi}\bigg)^2\bigg[\frac{17}{16}\log^2\frac{M_\rho}{2E_0}\\
&+\bigg(\log\frac{M_Z}{M_\rho}+\frac{3}{4}\log\frac{M_\rho}{2E_0}\bigg)^2\bigg]
\simeq 4.7\times 10^{-4}\notag
\end{align}
due to LL and quadratic terms.
The final result in Eq.~\eqref{Delta_RC} is obtained from the full RG solution, including another $\simeq 0.6\times 10^{-4}$ from NLL.

Our improved evaluation of $\Delta_\text{RC}^{\pi\ell}$ given in Eq.~\eqref{Delta_RC} implies
\beq
\label{Vud_final}
V_{ud}^\pi=0.97346(281)_\text{Br}(9)_{\tau_\pi}(5)_{\Delta_\text{RC}^{\pi\ell}}(27)_{I_{\pi\ell}}[283]_\text{tot},
\eeq
where the uncertainties refer to the branching fraction $\text{Br}[\pi^\pm\to\pi^0e^\pm\nu_e(\gamma)]=1.038(6)\times 10^{-8}$~\cite{Pocanic:2003pf} (slightly adjusted to the current normalization from $\text{Br}[\pi^\pm\to e^\pm\nu_e(\gamma)]=1.2326(23)\times 10^{-4}$~\cite{ParticleDataGroup:2026,PiENu:2015seu,Czapek:1993kc,Britton:1992pg}), the pion lifetime $\tau_\pi$~\cite{ParticleDataGroup:2026,Koptev:1995je,Numao:1995qf}, $\Delta_\text{RC}^{\pi\ell}$, and the pion mass difference~\eqref{phase_space}, respectively. Accordingly, the theory uncertainty in the extraction of $V_{ud}$ due to $\Delta_\text{RC}^{\pi\ell}$ is reduced by a factor of three, while all other sources of uncertainty are of experimental origin. The PIONEER experiment aims at improving the uncertainty in the branching fraction by an order of magnitude, at which level the uncertainty in the pion mass difference becomes the limiting factor.

\emph{$\tau\to\pi\pi\nu_\tau$}---The photon-inclusive  $\tau\to\pi\pi\nu_\tau$ decay rate is expressed as
\begin{equation}
\frac{d\Gamma\big[\tau^\pm\to\pi^\pm\pi^0\nu_\tau(\gamma)\big]}{K_\Gamma(s)ds}=\big[\beta_{\pi\pi^0}(s)\big]^3|f_+(s)|^2 S_\text{EW}^{\pi\pi}G_\text{EM}(s),
\label{decay_rate}
\end{equation}
where
\beq
K_\Gamma(s)= \frac{\Gamma_e \vert V_{ud}\vert^2}{2 m_\tau^2} \bigg(1-\frac{s}{m_\tau^2}\bigg)^2\bigg(1+\frac{2s}{m_\tau^2}\bigg),
\eeq
$\Gamma_e\equiv \Gamma[\tau\to e\nu_\tau\bar{\nu}_e]$, $\beta_{\pi\pi^0}$ denotes the phase-space factor
\begin{align}
\beta_{\pi\pi^0}(s)&=\lambda^{1/2}\bigg(1,\frac{\mpi^2}{s},\frac{M_{\pi^0}^2}{s}\bigg),\notag\\ \lambda(a,b,c)&=a^2+b^2+c^2-2(ab+ac+bc),
\end{align}
and $S_\text{EW}^{\pi\pi}$~\cite{Sirlin:1981ie,Marciano:1985pd,Marciano:1988vm,Marciano:1993sh,Braaten:1990ef,Erler:2002mv,Davier:2002dy} includes short-distance effects.
The rate is normalized to
the leptonic decay width $\Gamma_e$, which requires that the respective radiative corrections $\Delta S_\text{EW}^e=\frac{\alpha}{2\pi}\big(\frac{25}{4}-\pi^2\big)$ that connect it to $G_F$ be removed
\beq
\label{SEW}
S^{\pi\pi}_\text{EW}= \big[1+\Delta\tilde S_\text{EW}(\mu_\chi)\big]^2-\Delta S_\text{EW}^{e}.
\eeq
Besides providing a convenient normalization channel, this convention also avoids mass corrections $\simeq m_\tau^2/M_W^2\simeq 5\times 10^{-4}$~\cite{Ferroglia:2013dga,Fael:2013pja}, as would arise when using $G_F$ from muon decay instead. The chiral LECs are contained in $G_\text{EM}(s)$, depending linearly on $X_\ell(\mu_\chi)$~\cite{Colangelo:2025ivq}
\beq
\label{GEM_Xell}
   G_\text{EM}(s)\big|_{X_\ell(\mu_\chi)=0}
= G_\text{EM}(s) \big|_{X_\ell(\mu_\chi) = \bar{X}_\ell(\mu_\chi)} + e^2 \bar{X}_\ell(\mu_\chi).
\eeq
The analysis of $G_\text{EM}(s)$ in
Refs.~\cite{Colangelo:2025iad,Colangelo:2025ivq} is performed for the reference value $\bar{X}_\ell(M_\rho)=14\times 10^{-3}$~\cite{Ma:2021azh}, which can be easily adjusted to an arbitrary value by means of Eq.~\eqref{GEM_Xell}. It is evident that only the product $S_\text{EW}^{\pi\pi}G_\text{EM}(s)$ is scale and scheme independent, and in the following we will therefore only consider the combined effect.
Since both $S_\text{EW}^{\pi\pi}$ and the relevant terms in $G_\text{EM}(s)$ do not depend on the $\pi\pi$ invariant mass $s$, the respective IB corrections for the HVP contribution to $a_\mu$ can simply be obtained by scaling $\Delta \bar a_\mu[S_\text{EW}^{\pi\pi}]=-12.166(56)(8)\times 10^{-10}$ accordingly,\footnote{Quantities taken from Refs.~\cite{Colangelo:2025iad,Colangelo:2025ivq} are indicated by a bar.} so that the analysis of Refs.~\cite{Colangelo:2025iad,Colangelo:2025ivq} can be updated via
\begin{align}
  &\Delta a_\mu[\pi\pi,\tau]-\Delta \bar a_\mu[\pi\pi,\tau]\notag\\
  &=
  \Big[\frac{\Delta S_\text{EW}^{\pi\pi}-\Delta \bar G_\text{EM}}{\Delta \bar S_\text{EW}^{\pi\pi}}-1\Big]\Delta \bar a_\mu[S_\text{EW}^{\pi\pi}]\notag\\
  &=-0.07(4)\times 10^{-10},
\end{align}
where $\Delta S_\text{EW}^{\pi\pi}=S_\text{EW}^{\pi\pi}-1$ from Eq.~\eqref{SEW}, $\Delta \bar S_\text{EW}^{\pi\pi}=0.0233$, and
\beq
\Delta \bar G_\text{EM}=-e^2\bigg(\bar X_\ell(M_\rho)-\frac{1}{4\pi^2}\log\frac{m_\tau^2}{M_\rho^2}\bigg).
\eeq
Including phase-space corrections in addition to $S_\text{EW}^{\pi\pi}$ and $G_\text{EM}(s)$, we obtain
\beq
\Delta a_\mu[\pi\pi,\tau]=-24.9(1)_\text{exp}(5)_\text{th}(1)_\text{SD}\times 10^{-10},
\eeq
with central value almost identical to Refs.~\cite{Colangelo:2025iad,Colangelo:2025ivq}, but the uncertainty from the short-distance matching, previously estimated as $1.3\times 10^{-10}$~\cite{Aliberti:2025beg},  has become negligible. Accordingly, the uncertainty of isospin breaking in the matrix elements, $|F_\pi^V/f_+|$, estimated as $4.7\times 10^{-10}$ in Ref.~\cite{Aliberti:2025beg}, now constitutes by far the dominant limitation for the use of hadronic $\tau$ data in the evaluation of the HVP contribution to $a_\mu$.

\emph{Conclusions}---In this Letter, we presented
the matching of short-distance contributions and hadronic matrix elements for pion $\beta$ decay and $\tau\to\pi\pi\nu_\tau$ at NLL accuracy. To this end, we first derived the required matching relations between ChPT and LEFT, both with the spurion method, which yields explicit expressions for the relevant LECs, and based on general properties of the amplitude related to Ward identities, current algebra, and the IR structure, which confirms the spurion analysis and reveals a remarkable universality feature of the matching relation.

For the phenomenological analysis, we combined our matching relations with recent results for the hadronic matrix elements in lattice QCD and NLL corrections in the strong coupling. As key results, we substantially improved the radiative corrections to pion $\beta$ decay and the short-distance contributions to $\tau\to\pi\pi\nu_\tau$. The remaining theory uncertainty in $V_{ud}$ extracted from pion $\beta$ decay now lies a factor of five below the precision goal of the PIONEER experiment, rendering the phase-space factor and thus the pion mass difference the limiting factor besides the measurement of the branching fraction.  For $\tau\to\pi\pi\nu_\tau$, the main application concerns IB corrections for the HVP contribution to $a_\mu$, in which case the remaining uncertainties from the short-distance matching are now negligible.

\emph{Acknowledgments}---We thank W.~Dekens, X.~Feng, M.~Gorbahn, E.~Mereghetti, F.~Moretti, and C.-Y.~Seng for valuable discussions and correspondence as well as comments on the manuscript, and X.~Feng and J.-S.~Yoo for providing the results of Refs.~\cite{Feng:2020zdc,Yoo:2023gln}, respectively.
Financial support by  the SNSF (Project No.\ TMCG-2\_213690),  the U.S.\ DOE under Grant No.\ DE-FG02-00ER41132, the AEI and the EU (Projects No.\ CNS2022-135595 and PID2023-151418NB-I00) is gratefully acknowledged. MH and NV thank the INT at the University of Washington for its hospitality during visits when parts of this work were performed.
We acknowledge support from the DOE Topical Collaboration ``Nuclear Theory for New Physics,'' award No.\ DE-SC0023663.

\section{End Matter}

\emph{Derivation of the matching conditions and universality}---We arrived at  the matching conditions~\eqref{eq:final_res0} and~\eqref{eq:final_res}
between LEFT and ChPT in two ways:
first, following Refs.~\cite{Descotes-Genon:2005wrq,Cirigliano:2023fnz} we  performed the matching at the level of
Green's functions obtained by  differentiation of the generating functional with respect to charge and weak spurions.
The results could not be immediately read off of Ref.~\cite{Descotes-Genon:2005wrq}, because for the LEFT
a Pauli--Villars regulator with on-shell subtraction scheme was used, while
 we work in dimensional regularization and $\overline{\text{MS}}$ subtraction scheme (to resum logarithms to $\text{NLL}_s$ accuracy).
Following the logic of  Ref.~\cite{Descotes-Genon:2005wrq},   we arrived at an expression for  $X_\ell(\mu_\chi,\mu,a)$
in terms of the correlation function   $\int dx\, e^{i k\cdot x}  \langle 0 | T \big\{V_\mu^a (x) V_\nu^b (0)\big\} | \pi^c \rangle$~\cite{Supp},
where $a,b,c$ are
SU(3) flavor indices.
Finally, using a  soft-pion theorem we  related  this matrix element to the object appearing in  the $\gamma W$ box,
namely $\int dx\, e^{i k\cdot x}  \langle  \pi^d | T \big\{V_\mu^a (x) A_\nu^b (0)\big\} | \pi^c \rangle$, up to higher-order chiral corrections.
One of the advantages of the  spurion method  is that it allows one to derive representations for any individual LEC,
and hence the combination  $X_\ell$.

Second, we matched physical amplitudes
for $\pi^+ \to \pi^0 e^+ \nu_e$ in LEFT and ChPT~\cite{Supp}.
The key point is that the $\Order(\alpha)$ term in the matrix element $\langle f | O_\beta | i \rangle$  in LEFT
can be written in terms of convolutions of known kernels
with  nonperturbative matrix elements of  weak and electromagnetic quark currents between hadronic states.
Current algebra relates three-point and two-point correlation functions~\cite{Abers:1968zz,Sirlin:1977sv}.
Inspection of the IR behavior of these convolution integrals~\cite{Abers:1968zz,Sirlin:1967zz}
allows one to identify the parts of the LEFT amplitude that are  contained in the ChPT loops,
leaving the rest as contributions to the LECs.
While this  analysis gives us access only to the linear combination of LECs contributing to the decay rate,
it  reveals   universal features of the matching:
Eq.~\eqref{eq:final_res}  formally applies to the  Fermi component of any $\beta$ decay,
provided one computes the $T_3$ form factor appearing in
$\bar\Box^V (\mu_0)$, as defined in Eq.~\eqref{eq:box1}, in the appropriate initial and final state.
Up to effects suppressed by $E_e/\Mhad$, $m_e/\Mhad$, where $\Mhad$ denotes
the mass of the decaying hadronic  state
and $m_e$, $E_e$ mass and energy of the electron, the long-distance corrections
are also universal (independent of the mass and spin of the hadrons involved in the decay) and provided
by the so-called Sirlin function~\cite{Sirlin:1967zz},
up to a constant  that depends on the choice of ultraviolet (UV) regulator and is fixed by the matching procedure.
Consistently with the explicit calculation of the neutron decay
\cite{Cirigliano:2022hob,Cirigliano:2023fnz} we find for the pion $\beta$ decay
\begin{align}
\frac{d \Gamma}{d \Gamma^{(0)} } & =  [g_V^\pi (\mu_\chi)]^2
\left[ 1 + \frac{\alpha}{2 \pi} \bigg( \frac{5}{4}  + 3 \log \frac{\mu_\chi}{m_e} + \hat g(E_e, E_0)   \bigg) \right]
\nonumber \\
& = 1 + 2 \, \bar\Box^V_\pi(\mu_0)
+ \frac{\alpha}{2 \pi} \bigg(
4 \Big(1-\frac{\alpha_s(\mu_0)}{4\pi}\Big)\log\frac{M_Z}{\mu_0}
\nonumber \\
&+ 3 \log \frac{\mu_0}{m_e} + \hat g(E_e, E_0)  \bigg) +
\Order(\alpha^2, \alpha \alpha_s) ,
\label{eq:rate0}
\end{align}
where in the second line we have provided the fixed-order expression
(up to $\Order (\alpha \alpha_s)$ terms that are not logarithmically enhanced)
and  $\hat g(E_e,E_0)$ is the ``subtracted'' Sirlin function~\cite{Sirlin:1967zz},
from which the large logarithm of $\Mhad/m_e$ has been subtracted:
\begin{align}
\label{Sirlin}
\hat g (E_e, E_0)  & = - \frac{3}{4}
+ \bigg[\frac{1 + \beta^2}{\beta}
+ \frac{1}{12 \beta} \bigg( \frac{\bar E}{E_e}\bigg)^2\bigg]  \log \frac{1+\beta}{1-\beta}
\nonumber \\
&+4 \left[ \frac{1}{2 \beta} \log \frac{1+\beta}{1-\beta} - 1 \right] \left[ \log \frac{2 \bar E}{m_e}  - \frac{3}{2} + \frac{\bar E}{3 E_e} \right]
\nonumber \\
&+
\frac{1}{\beta} \left[ - 4 \,  \text{Li}_2 \left( \frac{2 \beta}{1 + \beta} \right) - \log^2 \left( \frac{1+\beta}{1-\beta} \right)   \right].
\end{align}
Here  $\beta = |\mathbf{p}_e|/E_e$ and $\bar E = E_0 - E_e$, with the electron endpoint energy  $E_0$, which, to this order, is the only source through which the hadron mass  enters the radiative corrections.
The ChPT results~\cite{Cirigliano:2002ng} capture additional $\mpi$-dependent nonuniversal corrections, but the integral over the Sirlin function,
\begin{align}
\Delta I_{\pi\ell}&\simeq \frac{\alpha}{2\pi}\bigg[\frac{5}{4}+3\log\frac{2E_0}{m_e}\notag\\
&\qquad+\frac{64}{\mpi^5I_{\pi\ell}}\int_{m_e}^{E_0}dE_e\,E_e^2\bar E^2 \beta \,\hat g (E_e, E_0)\bigg]\notag\\
&\simeq\frac{\alpha}{\pi}\bigg(\frac{187}{40}-\frac{2\pi^2}{3}\bigg)\simeq -0.0044,
\end{align}
explains the main features of the phase-space correction $\Delta I_{\pi\ell}$. In particular, the last line, applicable in the limit $m_e\to 0$ and neglecting recoil corrections, demonstrates that with the choice $\log\frac{\mu_\chi}{2E_0}$ in Eq.~\eqref{Delta_RC_def} indeed all logarithms disappear in $\Delta I_{\pi\ell}$~\cite{Passera:2011ae}.

\emph{Low-energy RG}---The low-energy RG for $X_\ell(\mu_\chi,\mu,a)$ needs to be formulated at the level of $g_V^\pi(\mu_\chi)$ to account for the evolution of $\alpha_\chi(\mu_\chi)$, leading to
\begin{align}
\label{gvpi_RG}
\mu_\chi\frac{dg_V^\pi(\mu_\chi)}{d\mu_\chi}&=\gamma(\alpha_\chi)g_\pi^V(\mu_\chi),\notag\\
\gamma(\alpha_\chi)&=\tilde\gamma_0\frac{\alpha_\chi}{4\pi}+\tilde\gamma_1\frac{\alpha^2_\chi}{(4\pi)^2}+\Order\big(\alpha_\chi^3\big),\notag\\
\tilde\gamma_0&=-3,\qquad\tilde\gamma_1=\frac{10\tilde n}{3}+\frac{5}{2}-\frac{8\pi^2}{3},
\end{align}
where
\beq
\tilde n(\mu_\chi)=\sum_{\ell=e,\mu} Q_\ell^2 n_\ell\theta\big(\mu_\chi-m_\ell\big)+\frac{1}{4}\sum_\pi Q_\pi^2 n_\pi\theta\big(\mu_\chi-M_\pi)
\eeq
counts the number of active spin-$1/2$ ($\ell$) and spin-$0$ ($\pi$) degrees of freedom weighted by their charges $Q_{\ell,\pi}$, and $\alpha_\chi(\mu_\chi)$ is evaluated as in Ref.~\cite{Cirigliano:2023fnz}. To obtain the $\Order(\alpha_\chi^2)$ term in Eq.~\eqref{gvpi_RG} we used universality of the corresponding anomalous dimension~\cite{Borah:2024ghn,Ji:1991pr}, which allows us to also resum NLL in the low-energy theory. At one-loop order the dependence of $\alpha_\chi$ on $\mu_\chi$ can be ignored, and $\tilde\gamma_0$ reproduces the RG relation
obtained using ChPT beta functions~\cite{Knecht:1999ag,Urech:1994hd}
\beq
\mu_\chi\frac{\partial}{\partial\mu_\chi}X_\ell(\mu_\chi,\mu,a)=\frac{3}{8\pi^2}.
\eeq

\section{Supplemental material}

\appendix

\section{Matching  with  the spurion method}
\label{app:matching}

The LECs entering Eq.~\eqref{eq:LECs} were obtained by matching LEFT correlators onto ChPT ones at the one-loop level, following the spurion method outlined in Refs.~\cite{Descotes-Genon:2005wrq,Cirigliano:2023fnz}. This method consists of promoting the parameters responsible for the chiral symmetry breaking, i.e.,  quark masses and electroweak charges,
to dynamical fields transforming under SU(3)$_{L}\times$ SU(3)$_{R}$ in such a way that the LEFT Lagrangian preserves the chiral symmetry. Then, following Refs.~\cite{Urech:1994hd,Knecht:1999ag}, it is possible to build a complete basis of effective operators involving leptons, photons, light mesons, and spurion sources up to order $p^{4}$ in the chiral expansion and respecting  chiral symmetry.
By computing correlators at the one-loop level in both theories, it is then possible to extract the renormalized LECs, which encode the short-distance physics and the nonperturbative matrix elements at a given order.

These correlators are defined by taking the functional derivatives of the generating functional of the connected diagrams ($W$) with respect to the spurion sources and computing matrix elements between lepton and pion external states as
\begin{equation}
    \int d^{4}x\,\bigg\langle l(p)\bar{\nu}(q) \bigg| \frac{\delta^{2} W(\mathbf{q}_{A},\mathbf{q}_{V},\mathbf{q}_{W})}{\delta\mathbf{q}^{b}_{V(A)}(x)\delta\mathbf{q}_{W}^{c}(0)}\bigg|\pi^{c}(r)\bigg\rangle,
\end{equation}
where $\mathbf{q}_{V(A)}$ are the spurion sources associated with the vector and axial-vector electric charge, while $\mathbf{q}_{W}$ is the spurion field associated with the weak current, both expressed in the adjoint representation of SU(3). We adopt the LECs basis proposed in Ref.~\cite{Descotes-Genon:2005wrq}, which allows one to separate the electroweak contribution ($\hat{X}_{6}$) from the strong interaction one ($\hat{X}_{1}$, $\hat{X}_{2}$) in the charged lepton kinetic term ($X_{6}$) as
\begin{equation}
    X_{6} = \hat{X}_{6} + \frac{4}{3} \hat{X}_{1}+4\hat{X}_{2}.
\end{equation}
By adapting the results of Refs.~\cite{Descotes-Genon:2005wrq,Moussallam:1997xx} to the $\overline{\text{MS}}$ (LEFT) and  $\overline{\text{MS}}_\chi$ (ChPT) schemes, respectively, we find:
\begin{align}
\label{eq:LECs_separate}
    \hat{X}_{6} &= \frac{1}{16 \pi^2}\bigg(1- \log\frac{\mu^{2}_{\chi}}{\mu^{2}}\bigg),\\
    \hat{X}_{1}+X_{1} &= i\,\frac{3}{4}\int \frac{d^{d}k}{(2 \pi)^{d}} \bigg[ \frac{\Gamma_{VV}(k,r)}{k^{2}}-\frac{1}{k^{2}(k^{2}-\mu_{0}^{2})}\bigg] \nonumber \\
    &\qquad+ \frac{1}{16\pi^{2}}\bigg[-\frac{3}{4}\log\frac{\mu^{2}}{\mu_{0}^{2}}+\frac{3}{2}B(a)+\frac{33}{8} \notag\\
    &\qquad+ \frac{\alpha_{s}}{4\pi}\bigg( 3\log\frac{\mu^{2}}{\mu_{0}^{2}}+ \frac{3}{2} B_{s}(a) -3 \bigg)\bigg],\notag\\
    \hat{X}_{2} &= i\, \frac{3}{8} \int \frac{d^{d}k}{(2 \pi)^{d}}\bigg[\Xi(k^2)-\frac{1}{k^2(k^2-\mu_{0}^{2})} \bigg]\nonumber\\
    &\qquad+\frac{1}{16\pi^{2}}\bigg(-\frac{5}{4}\log\frac{\mu^{2}}{\mu_{0}^{2}}-\frac{7}{8}\log\frac{\mu_{0}^{2}}{\mu_{\chi}^{2}}-\frac{19}{16}\bigg),\notag\\
    K_{12} &= i\, \frac{3}{8} \int \frac{d^{d}k}{(2 \pi)^{d}}\bigg[\Xi(k^2)-\frac{1}{k^2(k^2-\mu_{0}^{2})}\bigg] \nonumber\\
    &\qquad+\frac{1}{16\pi^{2}}\bigg(-\frac{1}{4}\log\frac{\mu^{2}}{\mu_{0}^{2}}+\frac{1}{8}\log\frac{\mu_{0}^{2}}{\mu_{\chi}^{2}}-\frac{3}{16}\bigg),\notag
\end{align}
where $\Xi(k^2) = k^{2}\big[g_{2}(k^2)-g_{1}(k^2)\big] - \big[f_{2}(k^{2})-f_{1}(k^2)\big]$, $f_{i}(k^2)$ and $g_{i}(k^2)$ are hadronic functions defined in Ref.~\cite{Descotes-Genon:2005wrq}, and the correlator $\Gamma_{VV}(k,r)$ is given by
\begin{align}
\label{TVV_corr}
    &\int d^{4}x\, e^{ik\cdot x} \big\langle 0 \big|T\big\{V^{b}_{\mu}(x) V^{c}_{\nu}(0)\big\} \big|\pi^{a}(r)\big\rangle\notag\\
    &\qquad=i  F_{0} \, \epsilon_{\mu\nu\rho\gamma}\,k^{\rho}r^{\gamma} d^{abc}\, \Gamma_{VV}(k,r),
\end{align}
in terms of the pion decay constant in the chiral limit, $F_0$, and the structure constant $d^{abc}$ of the SU(3) generators.
 We find that the $g_{i}$ and $f_{i}$ terms cancel out in Eq.~\eqref{eq:LECs}, leaving $\Gamma_{VV}$ as the only nonperturbative term.
In order to use lattice data from Refs.~\cite{Feng:2020zdc,Yoo:2023gln} for the box-diagram term presented in Eq.~\eqref{eq:box1}, we need to relate the pion-to-vacuum amplitude with the pion-to-pion one.
This can be achieved by
generalizing
the \emph{soft-pion theorem}
\begin{equation}
    \lim_{p^{\mu}\,\rightarrow \,0} \langle\pi^{a}(p)\beta|O|\alpha\rangle = -\frac{i}{F_{0}}\langle\beta |[Q_{5}^{a},O]|\alpha\rangle,
\end{equation}
where $\beta$, $\alpha$ are arbitrary external states,   $Q_{5}^{a}= \int d^{3}x\, A_{0}^{a}(x)$ is the axial charge,
and $O$ is a local operator,  to the case in which $O$ is replaced by the time-ordered product of
two operators.
In this way we
obtain the relation
$\Gamma_{VV}=-3/\sqrt{2}\,T_3^\text{\cite{Feng:2020zdc}}$.
Using this in Eq.~\eqref{eq:LECs_separate} and inserting the relations~\eqref{eq:LECs_separate}  into Eq.~\eqref{eq:LECs}
we arrive at Eq.~\eqref{eq:final_res0}, with $\bar \Box_\pi^{V}$ given in Eq.~\eqref{eq:box1}.

\section{Matching at the amplitude level}
\label{app:matching1}

\begin{figure}[t]
    \centering
    \includegraphics[width=0.45\textwidth]{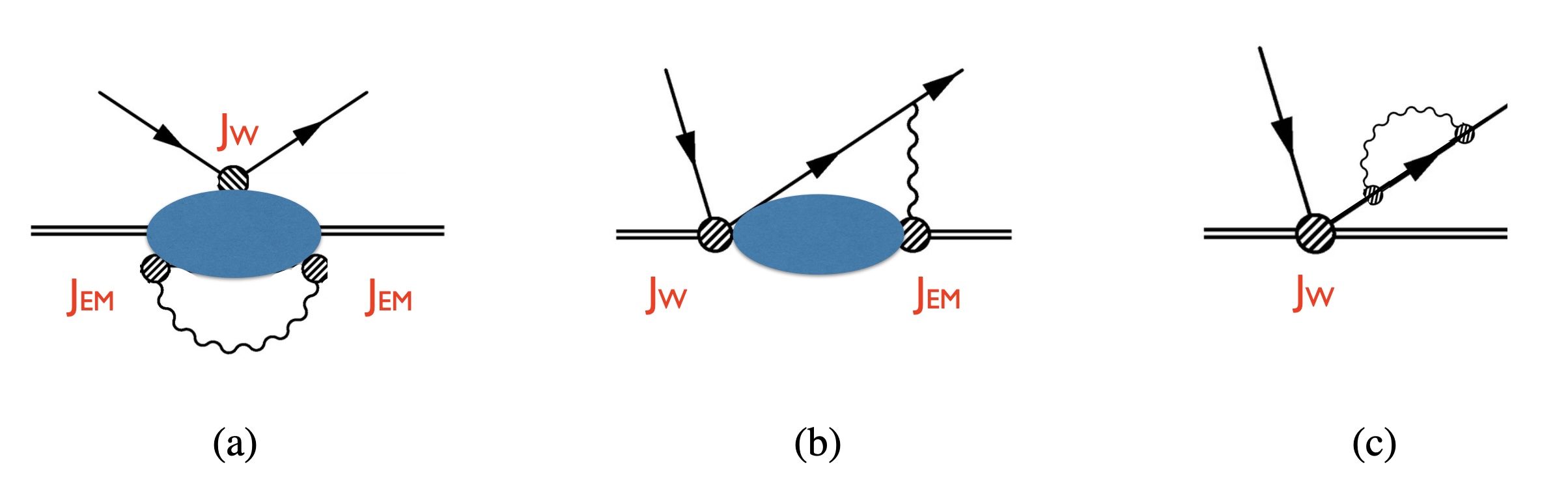}
    \caption{Diagrams contributing to the $\Order(\alpha)$ corrections to $\beta$ decays in LEFT.
    Double lines represent hadronic states,  single lines leptons, and wavy lines photons.
    The weak and electromagnetic current insertions are represented by hatched circles. The blue blobs denote hadronic matrix elements.}
    \label{fig:diagrams}
\end{figure}

Introducing the weak vector and axial-vector currents as $V_\mu = \bar u \gamma_\mu d$ and $A_\mu = \bar u \gamma_\mu \gamma_5 d$,
the Fermi  amplitude  for the process
$h_i (p_i ) \to h_f (p_f) + e (p_e) + \bar \nu_e (p_\nu)$
at tree level reads
\beq
A_0 = - \sqrt{2} G_F V_{ud}
C_\beta^{(3)}
J_\text{lept}^\mu  \langle h_f (p_f) | V_\mu (0) | h_i (p_i) \rangle,
\label{eq:app1}
\eeq
where $J_\text{lept}^\mu  =  \bar u(p_e) \gamma^\mu P_L v (p_\nu)$.
The following analysis assumes that the hadronic states $h_{i,f}$ are
 members of the same isospin $I$ multiplet (characterized by the mass scale $\Mhad$):
 this holds for neutron decay ($I=1/2$),  pion decay ($I=1$), and $0^+ \to 0^+$  nuclear decays ($I=1$).
 Therefore,  the mass splitting, the  momentum transfer $q = p_i - p_f$, and
 the electron endpoint energy $E_0$ are all quantities of first order in IB,
 $\Order\big(\alpha, (m_u-m_d)/\Lambda_\text{QCD}\big)$.
In terms of the correlators   of the electromagnetic and weak vector or axial-vector currents ($W=V,A$),\footnote{To ensure consistency with the original references~\cite{Descotes-Genon:2005wrq,Abers:1968zz}, the conventions for the correlator in Eqs.~\eqref{TVV_corr} and~\eqref{eq:app2} are kept different, related by $k\to -k$ and an exponential $e^{i(p_f-p_i)\cdot x}$ due to translational invariance.}
\begin{align}
W^{\mu \nu} (k) &= i \int d^d x  \,  e^{-i k \cdot x} \notag\\  &\times\big\langle h_f (p_f) \big|  \, T  \big\{W^\mu (0) J_\text{em}^\nu (x) \big\}  \big| h_i (p_i) \big\rangle,
\label{eq:app2}
\end{align}
including the $\Order (\alpha)$ electromagnetic corrections depicted in
Fig.~\ref{fig:diagrams},
the LEFT  $\beta$ decay amplitude
$A = - 2  \sqrt{2} G_F V_{ud}
C_\beta^{(3)} \,   \langle \bar \nu_e e  h_f  | O_\beta  | h_i \rangle$
 can be written in Feynman gauge as
 \begin{align}
 A & = A_0 \Bigg[ \sqrt{Z_e}  - 2 e^2  i \int \frac{d^dk}{(2 \pi)^d} \frac{1}{(k^2 + i \epsilon) (\tilde k^2-m_e^2 + i \epsilon)} \Bigg]\notag\\
&+
   \int \frac{d^dk}{(2 \pi)^d} \frac{2e^2i \sqrt{2} G_F V_{ud} C^{(3)}_\beta   J^\mu_\text{lept} p_e^\nu \bar V_{\mu\nu}(k)
 }{ (k^2 + i \epsilon) (\tilde k^2-m_e^2 + i \epsilon)}
 \nonumber \\
 & + \sqrt{2} G_F V_{ud} C^{(3)}_\beta  \bar u(p_e) \frac{ \gamma^\nu \gamma^\alpha \gamma^\mu - \gamma^\mu \gamma^\alpha \gamma^\nu}{2}  P_L  v (p_\nu)\notag\\
&\times
  e^2  \int \frac{d^dk}{(2 \pi)^d} \frac{ k_\alpha  A_{\mu \nu}  (k) }{ (k^2 + i \epsilon)^2 (\tilde k^2-m_e^2 + i \epsilon)},
  \label{eq:app3}
\end{align}
where $\tilde k=k+p_e$ and $\bar V_{\mu\nu}(k)=V_{\mu \nu} (k)
+ V^\alpha_\alpha (k)  \frac{k_\mu k_\nu}{k^2}$.
Throughout, we use dimensional regularization with anticommuting $\gamma_5$,
and use modified minimal subtraction   $\overline{\text{MS}}$  to remove the UV divergences.
Apart from this,   the derivation of Eq.~\eqref{eq:app3} completely parallels the analysis of Ref.~\cite{Abers:1968zz},
which was performed in the $V-A$ theory of weak interactions, i.e., LEFT at operator dimension six.
Reference~\cite{Abers:1968zz} heavily relied on Ward identities deriving from current algebra~\cite{Abers:1968zz,Sirlin:1977sv},
which also hold in  dimensionally regulated LEFT.
$Z_e$ denotes the electron wavefunction renormalization depicted by diagram $(c)$ of Fig.~\ref{fig:diagrams}.
The remaining terms represent contributions from diagrams $(a)$ and $(b)$ in Fig.~\ref{fig:diagrams}:
diagram $(b)$ leads to  the second term in the first line of Eq.~\eqref{eq:app3} (after using the
Ward identity $k_\mu V^{\nu \mu} = k_\mu V^{\mu\nu} = \langle h_f | V^\nu | h_i \rangle + \Order (\alpha, q/\Mhad)$),
the term involving $V_{\mu \nu}(k)$  in the second line of  Eq.~\eqref{eq:app3},
and all the remaining terms involving $A_{\mu \nu}(k)$;
upon using the Ward identity relating three- and two-current correlation functions~\cite{Abers:1968zz},
diagram $(a)$ combines with one term in diagram $(b)$, resulting in
the term involving $V^\alpha_\alpha (k)$ in the second line of  Eq.~\eqref{eq:app3}.
Finally,  following  Ref.~\cite{Abers:1968zz}  we explicitly
identified the contributions from  the vector and axial-vector current,
and note that only the isoscalar electromagnetic current contributes in the integration over $A_{\mu \nu} (k)$.

We next  discuss the  matrix element  $\langle \bar \nu_e e  h_f  | O_\beta  | h_i \rangle$
in the  $\overline{\text{MS}}$  scheme.
The first line of Eq.~\eqref{eq:app3} presents no difficulty.
For the contributions from the vector correlation function $V_{\mu \nu}$,
as observed in Refs.~\cite{Abers:1968zz,Sirlin:1977sv},  the second line of Eq.~\eqref{eq:app3} will only contribute terms unsuppressed by
$E_e/\Mhad$  or $|\mathbf{p}_e|/\Mhad$ if  the correlator  has an IR singularity of the type $V_{\mu \nu} (k)  \propto 1/k$, which in turn can only arise from ``elastic'' intermediate
states.   Hence, the form of $V_{\mu \nu}$ that provides the desired correction is uniquely fixed  in terms of known matrix elements.
References~\cite{Abers:1968zz, Sirlin:1977sv}  and the ChPT analyses of pion and neutron $\beta$ decay~\cite{Cirigliano:2002ng,Cirigliano:2022hob}
use slightly different forms for $V_{\mu \nu}(k)$, which, however, all share
the universal  IR singularity, irrespective of hadron spin:
\beq
V^\text{IR}_{\mu \nu} (k)
 =    \langle h_f (p_f) | V_\mu (0) | h_i (p_i) \rangle \ \frac{v_\nu}{ v \cdot k}  +  O(k^0),
\eeq
where $v_\nu$ is the four-velocity of the charged hadron involved in the $\beta$ decay.
We have written this in a form that makes it clear that the hadron mass  (or spin) does not play any role in these  leading corrections.
The loops evaluated with the universal term in $V^\text{IR}_{\mu \nu} (k)$ are UV divergent and the finite parts
defined in  $\overline{\text{MS}}$,
 when combined with the leading Low contribution to real photon
emission, lead to the usual Sirlin function~\cite{Sirlin:1967zza} up to a constant term.
The terms  of  $\Order(k^0)$  lead to corrections proportional to $|\mathbf{p}_e|/\Mhad$, usually neglected.\footnote{They are kept in the full  relativistic loop
function in the ChPT analysis of pion $\beta$ decay~\cite{Cirigliano:2002ng}.}

The treatment of the contributions  from  the axial-vector  weak current and  isoscalar electromagnetic current
($A_{\mu \nu}$) is more subtle, as this term  involves evanescent structures (as seen from Eq.~\eqref{eq:app3})
and receives $\Order(\alpha_s)$ corrections~\cite{Sirlin:1981ie,Gorbahn:2025ssv}.
In order to make contact with nonperturbative input from lattice QCD or  dispersive calculations,
in the context of neutron decay Refs.~\cite{Cirigliano:2023fnz,Gorbahn:2025ssv} introduced an intermediate renormalization
scheme that  subtracts  the large-momentum part of the photon loop integral
involving  $A_{\mu \nu}$,  controlled by the OPE.
This amounts to replacing the last two lines in  Eq.~\eqref{eq:app3}
with the  UV convergent quantity
\begin{align}
 &\sqrt{2} G_F V_{ud} C^{(3)}_\beta
  \epsilon^{\nu \alpha \mu \rho}         J_\rho^\text{lept}\notag\\
  &\times
  e^2 \int \frac{d^4 k}{(2 \pi)^4} \frac{ k_\alpha  \big[ A_{\mu \nu}  (k) - A_{\mu \nu}^\text{OPE} (k, \mu_0) \big] }{ (k^2 + i \epsilon)^2 (\tilde k^2-m_e^2 + i \epsilon)},
\end{align}
in which the Dirac algebra has  been performed in $d=4$ dimensions.
This scheme also  introduces the arbitrary scale $\mu_0$  in the OPE subtraction~\cite{Cirigliano:2023fnz},
which serves as a factorization scale and regulates IR divergences.
The crucial final step is to connect this finite operator to the  $\overline{\text{MS}}$ operator,
which  requires an additional finite and evanescent-scheme dependent renormalization.
This scheme-matching conversion was worked out to $\Order (\alpha)$ in Ref.~\cite{Cirigliano:2023fnz}
and to  $\Order (\alpha \alpha_s)$ in Ref.~\cite{Gorbahn:2025ssv}, to which we refer for further details.

Having studied the amplitude in LEFT, we are now ready to match to ChPT.
Based on the above discussion,  one can see  that  the ChPT amplitude (for any process, such as   pion or neutron decay)   coincides  with the
first two lines of Eq.~\eqref{eq:app3}, up to corrections of order $|\mathbf{p}_e|/\Mhad$, irrelevant at the order at which we work.
In fact,  the machinery of  ChPT builds up $V_{\mu \nu}$ in terms of low-lying states, thus capturing the leading IR behavior,
and incorporates the Ward identities, so that the ChPT  loops must respect the form of  Eq.~\eqref{eq:app3}.\footnote{In the case of neutron decay,
the second term in the first line of Eq.~\eqref{eq:app3} and part of the second line correspond to the  vertex correction (triangle diagram).
The proton wavefunction renormalization is included in the $V^\alpha_\alpha$ term in the second line.
Similar arguments apply to pion $\beta$ decay. In that case, more diagrams  appear in ChPT due to a $\gamma W$ tadpole needed to satisfy the
Ward identities. When everything is expressed in terms of $V_{\mu \nu}$, the expressions match.}
If one  used the same regularization and subtraction  scheme  in LEFT and ChPT amplitudes, the  first two lines of Eq.~\eqref{eq:app3}
would be identical in the two theories and contribute nothing to the matching.
Since indeed one uses different dimensional-regularization   scales in the two theories ($\mu$ and $\mu_\chi$) and the
two minimal-subtraction schemes differ by a constant ($\overline{\text{MS}}$ versus  $\overline{\text{MS}}_\chi$),
there is a (trivial) contribution to the matching from the IR loops.
In the case of neutron decay this contribution to the matching leads to
$-X_6/2 + 2 (V_1 + V_2) - g_9 = (3/2)/(4\pi)^2 (1 - \log (\mu_\chi^2/\mu^2))$~\cite{Cirigliano:2023fnz}.
Similarly, in the case of pion decay we obtain  $-\hat X_6/2 -2 (\hat X_2 - K_{12}) = (3/2)/(4\pi)^2 (1 - \log (\mu_\chi^2/\mu^2))$.
On the other hand, the integrals involving the axial-vector current
(usually referred to as the ``$\gamma W$ box'' contribution, even though at low energy
after integrating out the $W$ boson  the box reduces to a triangle diagram)
are not dominated by IR scales and in fact there is no corresponding term  in the ChPT loops---they correspond to local operators in ChPT.
Therefore, the $A_{\mu \nu}$ terms in Eq.~\eqref{eq:app3} contribute to the matching both a finite nonperturbative term and
a scheme-dependent perturbative term.
In the case of neutron decay the $\gamma W$ box contribution is encoded
in the combination of LECs $2(V_3 + V_4)$ (see  Eq.~(73) of Ref.~\cite{Cirigliano:2023fnz}),
while for pion decay it is encoded by the combination $-(2/3) (X_1 +  \hat X_1)$
(see Eq.~\eqref{eq:LECs_separate}).

The arguments sketched above lead to the  universal form for the matching condition in Eq.~\eqref{eq:final_res}
and  the universal shift to the decay rate given in  Eq.~\eqref{eq:rate0},
which agree with the previous results obtained in the specific case of neutron decay~\cite{Cirigliano:2022hob,Cirigliano:2023fnz,Gorbahn:2025ssv}.

\section{Renormalization group to NLL}
\label{app:RGE}

The Wilson coefficient  $C_\beta^{(f)}$ and  the QED and QCD couplings satisfy the following
RGE:
\begin{align}
\mu \frac{d C^{(f)}_\beta}{d \mu} & = \gamma (\alpha, \alpha_s) \, C^{(f)}_\beta,
\nonumber \\
\mu \frac{d \alpha^{(f)}}{d \mu} & = - 2 \alpha^{(f)} \bigg[ \beta^{(f)}_0 \frac{\alpha^{(f)}}{4 \pi}  + \beta^{(f)}_1 \bigg( \frac{\alpha^{(f)}}{4 \pi} \bigg)^2
\bigg],
\nonumber \\
\mu \frac{d \alpha^{(f)}_s  }{d \mu} & = - 2 \alpha^{(f)}_s \bigg[\beta^{(f)}_{0s} \frac{\alpha^{(f)}_s}{4 \pi}  + \beta^{(f)}_{1s} \bigg( \frac{\alpha^{(f)}_s}{4 \pi} \bigg)^2
\bigg],
\end{align}
with~\cite{Sirlin:1981ie,Buras:1989xd,Cirigliano:2023fnz,Gorbahn:2025ssv}
\begin{align}
\gamma (\alpha, \alpha_s)   &=  \gamma_0  \frac{\alpha (\mu) }{4 \pi}  + \gamma_{1}^{(f)}  \, \frac{\alpha^2 (\mu)}{(4 \pi)^2}\notag\\
&+\tilde \gamma_0
\frac{\alpha \alpha_s (\mu)}{(4 \pi)^2}  + \tilde \gamma_{1}^{(f)}  \, \frac{\alpha \alpha_s^2 (\mu)}{(4 \pi)^3},
\end{align}
and
\begin{align}
\gamma_0 &=  - 4, &
\gamma_1^{(f)} &= \frac{8}{9} \, \overline{Q^2} \, (a - 2),
\nonumber \\
\tilde \gamma_0 &=  4, &
\tilde \gamma_1^{(f)} &=  a \, \frac{28}{27} \, (33 - 2 n_f)  + \frac{10}{9} n_f - 33,
\nonumber \\
\beta_0^{(f)} &= - \frac{4}{3} \, \overline{Q^2},&
\beta_1^{(f)} &= - 4 \,  \overline{Q^4},
\nonumber \\
\beta_{0s}^{(f)} &= \frac{33 - 2 n_f}{3},&
\beta_{1s}^{(f)} &= 102 - \frac{38}{3} n_f,
\end{align}
where $n_f = n_u + n_d$ is the total number of active quark flavors,
(with $n_u$ and $n_d$ the number of up and down quark flavors)
and   $\overline{Q^n} = 3 (n_u Q_u^n + n_d Q_d^n) + n_\ell Q_\ell^n$
is the appropriate average of fermionic charges
($Q_u = 2/3$, $Q_d = -1/3$,  $Q_\ell = -1$).

The solution to the RGE to $\text{NLL}_s$  accuracy can be written as~\cite{Gorbahn:2025ssv}
\begin{align}
C_\beta^{(f)} (\mu, a) &=  J_f (\mu, a) \, u_f (\mu) \, u_f^{-1} (\mu_0) \, J_f^{-1} (\mu_0,a) \notag\\
&\qquad\times C_\beta^{(f)} (\mu_0, a),\notag\\
u_f (\mu) &=
\left(\frac{\alpha^{(f)}(\mu)}{\alpha (M_Z)} \right)^{- \frac{\gamma_0}{2 \beta_0^{(f)}}}  \
\left(\frac{\alpha_s^{(f)}(\mu)}{\alpha_s (M_Z)} \right)^{- \frac{\alpha}{4 \pi} \frac{\tilde \gamma_{0}}{2 \beta_{0s}^{(f)}   }},
\nonumber \\
J_f (\mu,a) &=
1 -\frac{\alpha^{(f)} (\mu)}{4 \pi}
\left(  \frac{\gamma_1^{(f)}}{2 \beta_0^{(f)}}
- \frac{\gamma_0 \beta_1^{(f)}
}{2 \big[\beta_0^{(f)}\big]^2}
 \right)
 \nonumber \\
 &- \frac{\alpha \alpha_s^{(f)} (\mu)}{(4 \pi)^2}
\left(  \frac{\tilde \gamma_{1s}^{(f)}}{2 \beta_{0s}^{(f)}}
- \frac{ \tilde \gamma_0 \beta_{1s}^{(f)}
}{2 \big[\beta_{0s}^{(f)}\big]^2}
 \right).
\end{align}

Solving the RGE with initial condition given in Eq.~\eqref{eq:Cmatch} for $C_\beta^{(5)} (\mu,a)$ in the $n_f=5$ flavor theory
at the weak scale  and taking into account the various heavy-quark thresholds one obtains for the three-quark theory
(defined by integrating out simultaneously the charm quark and $\tau$ lepton)~\cite{Gorbahn:2025ssv}
\beq
C_\beta^{(3)} (\mu, a) =  J_3 (\mu, a) \, u_3 (\mu)  \, \hat M_4 \, \hat M_5 \, \hat C,
\eeq
with
\begin{align}
\hat C &= u_5^{-1} (\mu) \, J_5^{-1} (\mu, a)  \, C^{(5)} (\mu),
\nonumber \\
\hat M_f &= u_{f-1}^{-1} (\mu) \, J_{f-1}^{-1} (\mu, a)   \, J_{f}  (\mu, a)  \, u_f (\mu).
\end{align}
Neither $\hat C$ nor $\hat M_f$  depend on $\mu$ and $a$ to $\text{NLL}_s$ accuracy~\cite{Gorbahn:2025ssv}.

Matching to ChPT as discussed in \ref{app:matching} and
 \ref{app:matching1}, combining the results from Refs.~\cite{Cirigliano:2023fnz,Gorbahn:2025ssv},
we can write
\begin{align}
\label{eq:matchG25}
g_V^\pi (\mu_\chi) &=
\Big[ 1+\bar\Box^{V(3)}_\pi(\mu_0) \Big]
\times
\left( Z_{O,O}^{s.c.} (\mu, \mu_0, a) \right)^{-1}
 \\
& \times
\Bigg[ 1 + \frac{3 \alpha^{(3)} (\mu)}{8 \pi}    \bigg( 1 - \log \frac{\mu_\chi^2}{\mu^2} \bigg)  \Bigg]
\times C^{(3)}_\beta (\mu, a).
\notag
\end{align}
Here  the entire expression multiplying $C^{(3)}_\beta (\mu, a)$
can be considered as the scheme- and scale-dependent  matrix elements of the Fermi operator $O_\beta$
in the three-flavor theory
(up to soft-photon contributions that go into the Sirlin function).
The scheme-dependent finite correction $ Z_{O,O}^{s.c.} (\mu, \mu_0, a)$
was worked out to $\Order (\alpha)$ in
Ref.~\cite{Cirigliano:2023fnz}
and to $\Order (\alpha \alpha_s)$    in Ref.~\cite{Gorbahn:2025ssv}:
\begin{align}
 Z_{O,O}^{s.c.} (\mu, \mu_0, a) &=  1 + \frac{\alpha}{4 \pi} \Bigg[
 \frac{a}{3} - \frac{5}{4} - \log \frac{\mu}{\mu_0}
  \nonumber \\
&+ \frac{\alpha_s}{4 \pi} \bigg( - \frac{14 a}{9}  + \frac{29}{6} + 4  \log \frac{\mu}{\mu_0} \bigg)
\Bigg].
\end{align}

As discussed in \ref{app:matching1}, the first line in Eq.~\eqref{eq:matchG25} corresponds to the combination
of LECs $1 -(2 e^2/3) (X_1 +  \hat X_1)$ (arising from insertion of the isoscalar electromagnetic current), while the term in the second line multiplying $C^{(3)}_\beta (\mu, a)$
corresponds to $1 - e^2 \hat X_6/2 -2 e^2 (\hat X_2 - K_{12})$ (arising from insertions of the isovector electromagnetic current).

Finally, one can write
$g_V^\pi (\mu_\chi) =  \bar  C^{(f)}_\beta (\mu)   \ \bar {O}^{(f)}_\beta (\mu, \mu_\chi)$
as the product of the separately  scheme-independent  (but scale-dependent)  factors
\begin{align}
 \bar C^{(3)}_\beta  (\mu) & =  u_3 (\mu) \, \hat M_4 \, \hat M_5  \,  \hat C  =   J_3 (\mu,a)^{-1} \, C^{(3)}_\beta (\mu, a),
 \nonumber \\
\bar {O}^{(3)}_\beta  (\mu, \mu_\chi)  &=
\Big[ 1+\bar\Box^{V(3)}_\pi(\mu_0) \Big] \notag\\
&\times
\Bigg[ 1 + \frac{3 \alpha^{(3)} (\mu)}{8 \pi}   \left( 1 - \log \frac{\mu_\chi^2}{\mu^2} \right)  \Bigg]
\nonumber \\
& \times  \left(  Z_{O,O}^{s.c.}  (\mu, \mu_0, a) \right)^{-1} \times  J_3 (\mu, a).
\end{align}
Using the explicit form of $  Z_{O,O}^{s.c.}  (\mu, \mu_0, a)$ and   $ J_3 (\mu, a) $
we arrive at Eq.~\eqref{eq:final_res}.
All the terms of $\Order (\alpha)$ and the logarithmic terms proportional to $\alpha \alpha_s$
agree with Ref.~\cite{Cirigliano:2023fnz}, once one takes into account that
the definitions of scheme-independent Wilson coefficients adopted here and
in that reference differ, which
affects the nonlogarithmic coefficient of $\Order (\alpha)$.
The nonlogarithmic terms of $\Order(\alpha \alpha_s)$ arise from the $\text{NLL}_s$ analysis of
Ref.~\cite{Gorbahn:2025ssv}.

In order to use the lattice-QCD results from Ref.~\cite{Yoo:2023gln} we also  match ChPT to LEFT with four active quark flavors.
The intermediate steps of the analysis are formally identical to what has been described above, except that at $\mu \simeq 1.8\GeV$
we integrate out only the $\tau$ and not the charm quark.
Renaming the Wilson coefficient and coupling constants below the $\tau$ threshold
as $C_\beta^{(\tilde{4})}$, $\alpha^{(\tilde{4})}$, $\alpha_s^{(\tilde{4})}$,
we  find that Eq.~\eqref{eq:final_res} is replaced by
\begin{align}
g_V^\pi (\mu_\chi)  & =
\bar C^{(\tilde{4})}_\beta  (\mu)
\bigg[1+\bar\Box^{V(4)}_\pi(\mu_0)\notag\\
&+ \frac{\alpha^{(\tilde{4})} (\mu)}{4 \pi}
  \bigg( \frac{499}{192}
  - \frac{3}{2}  \log \frac{\mu_\chi^2}{\mu_0^2}
 + 2  \log\frac{\mu^2}{\mu_0^2}
  \bigg)\notag\\
&+  \frac{\alpha \alpha_s^{(\tilde{4})} (\mu)}{(4 \pi)^2}
 \bigg(
 - \frac{1026}{625}  -  2  \log\frac{\mu^2}{\mu_0^2}
 \bigg)\bigg].
   \label{eq:final_res_4quarks}
\end{align}

The input for the solution of the RGE is determined as follows. For $\alpha_s$ we use~\cite{FlavourLatticeAveragingGroupFLAG:2024oxs,ParticleDataGroup:2026,Maltman:2008bx,PACS-CS:2009zxm,McNeile:2010ji,Chakraborty:2014aca,Bruno:2017gxd,Bazavov:2019qoo,Cali:2020hrj,Ayala:2020odx,Petreczky:2020tky,DallaBrida:2022eua}
\beq
\alpha_s^{(5)}(M_Z)=0.1181(7),
\eeq
which we evolve~\cite{Chetyrkin:2000yt,Herren:2017osy} using decoupling scales $\mu_b=m_b$ and $\mu_c=1.8\GeV$. For $\alpha$, we start from the Thomson limit $\alpha^{-1}=137.0359991(1)$~\cite{Parker:2018vye,Morel:2020dww,Fan:2022eto} and evolve it to $M_Z$ in the on-shell scheme, using the four-loop leptonic running $\Delta\alpha_\text{lept}=314.979(2)\times 10^{-4}$~\cite{Sturm:2013uka} and a range of hadronic contributions $\Delta\alpha_\text{had}=277.5(1.5)\times 10^{-4}$ that covers recent evaluations both from lattice QCD and phenomenology~\cite{Davier:2019can,Keshavarzi:2019abf,Ce:2022eix,Erler:2023hyi,Conigli:2025qvh}. Converting to $\overline{\text{MS}}$ using $\Delta \alpha^{(5)}\big|_{\overline{\text{MS}}}-\Delta \alpha^{(5)}\big|_\text{on-shell}=71.22(5)\times 10^{-4}$~\cite{ParticleDataGroup:2026,Chetyrkin:1996cf}, we obtain
\beq
1/\alpha^{(5)}(M_Z)=127.94(2).
\eeq
By convention, this $\overline{\text{MS}}$ value at $M_Z$ includes $W^\pm$ loops, which thus need to be removed to obtain the initial condition for the RG in QED+QCD~\cite{Fanchiotti:1992tu,Cirigliano:2023fnz}
\beq
1/\alpha^{(5)}(M_W)=127.925(20).
\eeq
Throughout, we use masses and $s_W^2=0.23122(6)$ from Ref.~\cite{ParticleDataGroup:2026}.

\bibliography{ref}

\begin{thebibliography}{144}%
\makeatletter
\providecommand \@ifxundefined [1]{%
 \@ifx{#1\undefined}
}%
\providecommand \@ifnum [1]{%
 \ifnum #1\expandafter \@firstoftwo
 \else \expandafter \@secondoftwo
 \fi
}%
\providecommand \@ifx [1]{%
 \ifx #1\expandafter \@firstoftwo
 \else \expandafter \@secondoftwo
 \fi
}%
\providecommand \natexlab [1]{#1}%
\providecommand \enquote  [1]{``#1''}%
\providecommand \bibnamefont  [1]{#1}%
\providecommand \bibfnamefont [1]{#1}%
\providecommand \citenamefont [1]{#1}%
\providecommand \href@noop [0]{\@secondoftwo}%
\providecommand \href [0]{\begingroup \@sanitize@url \@href}%
\providecommand \@href[1]{\@@startlink{#1}\@@href}%
\providecommand \@@href[1]{\endgroup#1\@@endlink}%
\providecommand \@sanitize@url [0]{\catcode `\\12\catcode `\$12\catcode
  `\&12\catcode `\#12\catcode `\^12\catcode `\_12\catcode `\%12\relax}%
\providecommand \@@startlink[1]{}%
\providecommand \@@endlink[0]{}%
\providecommand \url  [0]{\begingroup\@sanitize@url \@url }%
\providecommand \@url [1]{\endgroup\@href {#1}{\urlprefix }}%
\providecommand \urlprefix  [0]{URL }%
\providecommand \Eprint [0]{\href }%
\providecommand \doibase [0]{http://dx.doi.org/}%
\providecommand \selectlanguage [0]{\@gobble}%
\providecommand \bibinfo  [0]{\@secondoftwo}%
\providecommand \bibfield  [0]{\@secondoftwo}%
\providecommand \translation [1]{[#1]}%
\providecommand \BibitemOpen [0]{}%
\providecommand \bibitemStop [0]{}%
\providecommand \bibitemNoStop [0]{.\EOS\space}%
\providecommand \EOS [0]{\spacefactor3000\relax}%
\providecommand \BibitemShut  [1]{\csname bibitem#1\endcsname}%
\let\auto@bib@innerbib\@empty
%</preamble>
\bibitem [{\citenamefont {Cabibbo}(1963)}]{Cabibbo:1963yz}%
  \BibitemOpen
  \bibfield  {author} {\bibinfo {author} {\bibfnamefont {N.}~\bibnamefont
  {Cabibbo}},\ }\href {\doibase 10.1103/PhysRevLett.10.531} {\bibfield
  {journal} {\bibinfo  {journal} {Phys. Rev. Lett.}\ }\textbf {\bibinfo
  {volume} {10}},\ \bibinfo {pages} {531} (\bibinfo {year} {1963})}\BibitemShut
  {NoStop}%
\bibitem [{\citenamefont {Kobayashi}\ and\ \citenamefont
  {Maskawa}(1973)}]{Kobayashi:1973fv}%
  \BibitemOpen
  \bibfield  {author} {\bibinfo {author} {\bibfnamefont {M.}~\bibnamefont
  {Kobayashi}}\ and\ \bibinfo {author} {\bibfnamefont {T.}~\bibnamefont
  {Maskawa}},\ }\href {\doibase 10.1143/PTP.49.652} {\bibfield  {journal}
  {\bibinfo  {journal} {Prog. Theor. Phys.}\ }\textbf {\bibinfo {volume}
  {49}},\ \bibinfo {pages} {652} (\bibinfo {year} {1973})}\BibitemShut
  {NoStop}%
\bibitem [{\citenamefont {Hardy}\ and\ \citenamefont
  {Towner}(2020)}]{Hardy:2020qwl}%
  \BibitemOpen
  \bibfield  {author} {\bibinfo {author} {\bibfnamefont {J.~C.}\ \bibnamefont
  {Hardy}}\ and\ \bibinfo {author} {\bibfnamefont {I.~S.}\ \bibnamefont
  {Towner}},\ }\href {\doibase 10.1103/PhysRevC.102.045501} {\bibfield
  {journal} {\bibinfo  {journal} {Phys. Rev. C}\ }\textbf {\bibinfo {volume}
  {102}},\ \bibinfo {pages} {045501} (\bibinfo {year} {2020})}\BibitemShut
  {NoStop}%
\bibitem [{\citenamefont {Kinoshita}\ and\ \citenamefont
  {Sirlin}(1959)}]{Kinoshita:1958ru}%
  \BibitemOpen
  \bibfield  {author} {\bibinfo {author} {\bibfnamefont {T.}~\bibnamefont
  {Kinoshita}}\ and\ \bibinfo {author} {\bibfnamefont {A.}~\bibnamefont
  {Sirlin}},\ }\href {\doibase 10.1103/PhysRev.113.1652} {\bibfield  {journal}
  {\bibinfo  {journal} {Phys. Rev.}\ }\textbf {\bibinfo {volume} {113}},\
  \bibinfo {pages} {1652} (\bibinfo {year} {1959})}\BibitemShut {NoStop}%
\bibitem [{\citenamefont {Sirlin}(1967{\natexlab{a}})}]{Sirlin:1967zz}%
  \BibitemOpen
  \bibfield  {author} {\bibinfo {author} {\bibfnamefont {A.}~\bibnamefont
  {Sirlin}},\ }\href {\doibase 10.1103/PhysRevLett.19.877} {\bibfield
  {journal} {\bibinfo  {journal} {Phys. Rev. Lett.}\ }\textbf {\bibinfo
  {volume} {19}},\ \bibinfo {pages} {877} (\bibinfo {year}
  {1967}{\natexlab{a}})}\BibitemShut {NoStop}%
\bibitem [{\citenamefont {Sirlin}(1967{\natexlab{b}})}]{Sirlin:1967zza}%
  \BibitemOpen
  \bibfield  {author} {\bibinfo {author} {\bibfnamefont {A.}~\bibnamefont
  {Sirlin}},\ }\href {\doibase 10.1103/PhysRev.164.1767} {\bibfield  {journal}
  {\bibinfo  {journal} {Phys. Rev.}\ }\textbf {\bibinfo {volume} {164}},\
  \bibinfo {pages} {1767} (\bibinfo {year} {1967}{\natexlab{b}})}\BibitemShut
  {NoStop}%
\bibitem [{\citenamefont {Abers}\ \emph {et~al.}(1968)\citenamefont {Abers},
  \citenamefont {Dicus}, \citenamefont {Norton},\ and\ \citenamefont
  {Quinn}}]{Abers:1968zz}%
  \BibitemOpen
  \bibfield  {author} {\bibinfo {author} {\bibfnamefont {E.~S.}\ \bibnamefont
  {Abers}}, \bibinfo {author} {\bibfnamefont {D.~A.}\ \bibnamefont {Dicus}},
  \bibinfo {author} {\bibfnamefont {R.~E.}\ \bibnamefont {Norton}}, \ and\
  \bibinfo {author} {\bibfnamefont {H.~R.}\ \bibnamefont {Quinn}},\ }\href
  {\doibase 10.1103/PhysRev.167.1461} {\bibfield  {journal} {\bibinfo
  {journal} {Phys. Rev.}\ }\textbf {\bibinfo {volume} {167}},\ \bibinfo {pages}
  {1461} (\bibinfo {year} {1968})}\BibitemShut {NoStop}%
\bibitem [{\citenamefont {Jaus}\ and\ \citenamefont
  {Rasche}(1970)}]{Jaus:1970tah}%
  \BibitemOpen
  \bibfield  {author} {\bibinfo {author} {\bibfnamefont {W.}~\bibnamefont
  {Jaus}}\ and\ \bibinfo {author} {\bibfnamefont {G.}~\bibnamefont {Rasche}},\
  }\href {\doibase 10.1016/0375-9474(70)90690-1} {\bibfield  {journal}
  {\bibinfo  {journal} {Nucl. Phys. A}\ }\textbf {\bibinfo {volume} {143}},\
  \bibinfo {pages} {202} (\bibinfo {year} {1970})}\BibitemShut {NoStop}%
\bibitem [{\citenamefont {Sirlin}(1978)}]{Sirlin:1977sv}%
  \BibitemOpen
  \bibfield  {author} {\bibinfo {author} {\bibfnamefont {A.}~\bibnamefont
  {Sirlin}},\ }\href {\doibase 10.1103/RevModPhys.50.573} {\bibfield  {journal}
  {\bibinfo  {journal} {Rev. Mod. Phys.}\ }\textbf {\bibinfo {volume} {50}},\
  \bibinfo {pages} {573} (\bibinfo {year} {1978})},\ \bibinfo {note} {[Erratum:
  Rev. Mod. Phys. {\bf 50}, 905 (1978)]}\BibitemShut {NoStop}%
\bibitem [{\citenamefont {Sirlin}(1982)}]{Sirlin:1981ie}%
  \BibitemOpen
  \bibfield  {author} {\bibinfo {author} {\bibfnamefont {A.}~\bibnamefont
  {Sirlin}},\ }\href {\doibase 10.1016/0550-3213(82)90303-0} {\bibfield
  {journal} {\bibinfo  {journal} {Nucl. Phys. B}\ }\textbf {\bibinfo {volume}
  {196}},\ \bibinfo {pages} {83} (\bibinfo {year} {1982})}\BibitemShut
  {NoStop}%
\bibitem [{\citenamefont {Wilkinson}(1982)}]{Wilkinson:1982hu}%
  \BibitemOpen
  \bibfield  {author} {\bibinfo {author} {\bibfnamefont {D.~H.}\ \bibnamefont
  {Wilkinson}},\ }\href {\doibase 10.1016/0375-9474(82)90051-3} {\bibfield
  {journal} {\bibinfo  {journal} {Nucl. Phys. A}\ }\textbf {\bibinfo {volume}
  {377}},\ \bibinfo {pages} {474} (\bibinfo {year} {1982})}\BibitemShut
  {NoStop}%
\bibitem [{\citenamefont {Sirlin}\ and\ \citenamefont
  {Zucchini}(1986)}]{Sirlin:1986cc}%
  \BibitemOpen
  \bibfield  {author} {\bibinfo {author} {\bibfnamefont {A.}~\bibnamefont
  {Sirlin}}\ and\ \bibinfo {author} {\bibfnamefont {R.}~\bibnamefont
  {Zucchini}},\ }\href {\doibase 10.1103/PhysRevLett.57.1994} {\bibfield
  {journal} {\bibinfo  {journal} {Phys. Rev. Lett.}\ }\textbf {\bibinfo
  {volume} {57}},\ \bibinfo {pages} {1994} (\bibinfo {year}
  {1986})}\BibitemShut {NoStop}%
\bibitem [{\citenamefont {Towner}(1992)}]{Towner:1992xm}%
  \BibitemOpen
  \bibfield  {author} {\bibinfo {author} {\bibfnamefont {I.~S.}\ \bibnamefont
  {Towner}},\ }\href {\doibase 10.1016/0375-9474(92)90170-O} {\bibfield
  {journal} {\bibinfo  {journal} {Nucl. Phys. A}\ }\textbf {\bibinfo {volume}
  {540}},\ \bibinfo {pages} {478} (\bibinfo {year} {1992})}\BibitemShut
  {NoStop}%
\bibitem [{\citenamefont {Wilkinson}(1993{\natexlab{a}})}]{Wilkinson:1993hxz}%
  \BibitemOpen
  \bibfield  {author} {\bibinfo {author} {\bibfnamefont {D.~H.}\ \bibnamefont
  {Wilkinson}},\ }\href {\doibase 10.1016/0168-9002(93)90270-R} {\bibfield
  {journal} {\bibinfo  {journal} {Nucl. Instrum. Meth. A}\ }\textbf {\bibinfo
  {volume} {335}},\ \bibinfo {pages} {172} (\bibinfo {year}
  {1993}{\natexlab{a}})}\BibitemShut {NoStop}%
\bibitem [{\citenamefont {Wilkinson}(1993{\natexlab{b}})}]{Wilkinson:1993fva}%
  \BibitemOpen
  \bibfield  {author} {\bibinfo {author} {\bibfnamefont {D.~H.}\ \bibnamefont
  {Wilkinson}},\ }\href {\doibase 10.1016/0168-9002(93)90272-J} {\bibfield
  {journal} {\bibinfo  {journal} {Nucl. Instrum. Meth. A}\ }\textbf {\bibinfo
  {volume} {335}},\ \bibinfo {pages} {201} (\bibinfo {year}
  {1993}{\natexlab{b}})}\BibitemShut {NoStop}%
\bibitem [{\citenamefont {Czarnecki}\ \emph {et~al.}(2004)\citenamefont
  {Czarnecki}, \citenamefont {Marciano},\ and\ \citenamefont
  {Sirlin}}]{Czarnecki:2004cw}%
  \BibitemOpen
  \bibfield  {author} {\bibinfo {author} {\bibfnamefont {A.}~\bibnamefont
  {Czarnecki}}, \bibinfo {author} {\bibfnamefont {W.~J.}\ \bibnamefont
  {Marciano}}, \ and\ \bibinfo {author} {\bibfnamefont {A.}~\bibnamefont
  {Sirlin}},\ }\href {\doibase 10.1103/PhysRevD.70.093006} {\bibfield
  {journal} {\bibinfo  {journal} {Phys. Rev. D}\ }\textbf {\bibinfo {volume}
  {70}},\ \bibinfo {pages} {093006} (\bibinfo {year} {2004})},\ \Eprint
  {http://arxiv.org/abs/hep-ph/0406324} {arXiv:hep-ph/0406324} \BibitemShut
  {NoStop}%
\bibitem [{\citenamefont {Marciano}\ and\ \citenamefont
  {Sirlin}(2006)}]{Marciano:2005ec}%
  \BibitemOpen
  \bibfield  {author} {\bibinfo {author} {\bibfnamefont {W.~J.}\ \bibnamefont
  {Marciano}}\ and\ \bibinfo {author} {\bibfnamefont {A.}~\bibnamefont
  {Sirlin}},\ }\href {\doibase 10.1103/PhysRevLett.96.032002} {\bibfield
  {journal} {\bibinfo  {journal} {Phys. Rev. Lett.}\ }\textbf {\bibinfo
  {volume} {96}},\ \bibinfo {pages} {032002} (\bibinfo {year} {2006})},\
  \Eprint {http://arxiv.org/abs/hep-ph/0510099} {arXiv:hep-ph/0510099}
  \BibitemShut {NoStop}%
\bibitem [{\citenamefont {Towner}\ and\ \citenamefont
  {Hardy}(2010)}]{Towner:2010zz}%
  \BibitemOpen
  \bibfield  {author} {\bibinfo {author} {\bibfnamefont {I.~S.}\ \bibnamefont
  {Towner}}\ and\ \bibinfo {author} {\bibfnamefont {J.~C.}\ \bibnamefont
  {Hardy}},\ }\href {\doibase 10.1088/0034-4885/73/4/046301} {\bibfield
  {journal} {\bibinfo  {journal} {Rept. Prog. Phys.}\ }\textbf {\bibinfo
  {volume} {73}},\ \bibinfo {pages} {046301} (\bibinfo {year}
  {2010})}\BibitemShut {NoStop}%
\bibitem [{\citenamefont {Seng}\ \emph {et~al.}(2018)\citenamefont {Seng},
  \citenamefont {Gorchtein}, \citenamefont {Patel},\ and\ \citenamefont
  {Ramsey-Musolf}}]{Seng:2018yzq}%
  \BibitemOpen
  \bibfield  {author} {\bibinfo {author} {\bibfnamefont {C.-Y.}\ \bibnamefont
  {Seng}}, \bibinfo {author} {\bibfnamefont {M.}~\bibnamefont {Gorchtein}},
  \bibinfo {author} {\bibfnamefont {H.~H.}\ \bibnamefont {Patel}}, \ and\
  \bibinfo {author} {\bibfnamefont {M.~J.}\ \bibnamefont {Ramsey-Musolf}},\
  }\href {\doibase 10.1103/PhysRevLett.121.241804} {\bibfield  {journal}
  {\bibinfo  {journal} {Phys. Rev. Lett.}\ }\textbf {\bibinfo {volume} {121}},\
  \bibinfo {pages} {241804} (\bibinfo {year} {2018})},\ \Eprint
  {http://arxiv.org/abs/1807.10197} {arXiv:1807.10197 [hep-ph]} \BibitemShut
  {NoStop}%
\bibitem [{\citenamefont {Seng}\ \emph {et~al.}(2019)\citenamefont {Seng},
  \citenamefont {Gorchtein},\ and\ \citenamefont
  {Ramsey-Musolf}}]{Seng:2018qru}%
  \BibitemOpen
  \bibfield  {author} {\bibinfo {author} {\bibfnamefont {C.~Y.}\ \bibnamefont
  {Seng}}, \bibinfo {author} {\bibfnamefont {M.}~\bibnamefont {Gorchtein}}, \
  and\ \bibinfo {author} {\bibfnamefont {M.~J.}\ \bibnamefont
  {Ramsey-Musolf}},\ }\href {\doibase 10.1103/PhysRevD.100.013001} {\bibfield
  {journal} {\bibinfo  {journal} {Phys. Rev. D}\ }\textbf {\bibinfo {volume}
  {100}},\ \bibinfo {pages} {013001} (\bibinfo {year} {2019})},\ \Eprint
  {http://arxiv.org/abs/1812.03352} {arXiv:1812.03352 [nucl-th]} \BibitemShut
  {NoStop}%
\bibitem [{\citenamefont {Gorchtein}(2019)}]{Gorchtein:2018fxl}%
  \BibitemOpen
  \bibfield  {author} {\bibinfo {author} {\bibfnamefont {M.}~\bibnamefont
  {Gorchtein}},\ }\href {\doibase 10.1103/PhysRevLett.123.042503} {\bibfield
  {journal} {\bibinfo  {journal} {Phys. Rev. Lett.}\ }\textbf {\bibinfo
  {volume} {123}},\ \bibinfo {pages} {042503} (\bibinfo {year} {2019})},\
  \Eprint {http://arxiv.org/abs/1812.04229} {arXiv:1812.04229 [nucl-th]}
  \BibitemShut {NoStop}%
\bibitem [{\citenamefont {Seng}\ and\ \citenamefont
  {Gorchtein}(2023)}]{Seng:2022cnq}%
  \BibitemOpen
  \bibfield  {author} {\bibinfo {author} {\bibfnamefont {C.-Y.}\ \bibnamefont
  {Seng}}\ and\ \bibinfo {author} {\bibfnamefont {M.}~\bibnamefont
  {Gorchtein}},\ }\href {\doibase 10.1103/PhysRevC.107.035503} {\bibfield
  {journal} {\bibinfo  {journal} {Phys. Rev. C}\ }\textbf {\bibinfo {volume}
  {107}},\ \bibinfo {pages} {035503} (\bibinfo {year} {2023})},\ \Eprint
  {http://arxiv.org/abs/2211.10214} {arXiv:2211.10214 [nucl-th]} \BibitemShut
  {NoStop}%
\bibitem [{\citenamefont {Ma}\ \emph {et~al.}(2024)\citenamefont {Ma},
  \citenamefont {Feng}, \citenamefont {Gorchtein}, \citenamefont {Jin},
  \citenamefont {Liu}, \citenamefont {Seng}, \citenamefont {Wang},\ and\
  \citenamefont {Zhang}}]{Ma:2023kfr}%
  \BibitemOpen
  \bibfield  {author} {\bibinfo {author} {\bibfnamefont {P.-X.}\ \bibnamefont
  {Ma}}, \bibinfo {author} {\bibfnamefont {X.}~\bibnamefont {Feng}}, \bibinfo
  {author} {\bibfnamefont {M.}~\bibnamefont {Gorchtein}}, \bibinfo {author}
  {\bibfnamefont {L.-C.}\ \bibnamefont {Jin}}, \bibinfo {author} {\bibfnamefont
  {K.-F.}\ \bibnamefont {Liu}}, \bibinfo {author} {\bibfnamefont {C.-Y.}\
  \bibnamefont {Seng}}, \bibinfo {author} {\bibfnamefont {B.-G.}\ \bibnamefont
  {Wang}}, \ and\ \bibinfo {author} {\bibfnamefont {Z.-L.}\ \bibnamefont
  {Zhang}},\ }\href {\doibase 10.1103/PhysRevLett.132.191901} {\bibfield
  {journal} {\bibinfo  {journal} {Phys. Rev. Lett.}\ }\textbf {\bibinfo
  {volume} {132}},\ \bibinfo {pages} {191901} (\bibinfo {year} {2024})},\
  \Eprint {http://arxiv.org/abs/2308.16755} {arXiv:2308.16755 [hep-lat]}
  \BibitemShut {NoStop}%
\bibitem [{\citenamefont {Seng}\ and\ \citenamefont
  {Gorchtein}(2024)}]{Seng:2023cvt}%
  \BibitemOpen
  \bibfield  {author} {\bibinfo {author} {\bibfnamefont {C.-Y.}\ \bibnamefont
  {Seng}}\ and\ \bibinfo {author} {\bibfnamefont {M.}~\bibnamefont
  {Gorchtein}},\ }\href {\doibase 10.1103/PhysRevC.109.044302} {\bibfield
  {journal} {\bibinfo  {journal} {Phys. Rev. C}\ }\textbf {\bibinfo {volume}
  {109}},\ \bibinfo {pages} {044302} (\bibinfo {year} {2024})},\ \Eprint
  {http://arxiv.org/abs/2304.03800} {arXiv:2304.03800 [nucl-th]} \BibitemShut
  {NoStop}%
\bibitem [{\citenamefont {Hill}\ and\ \citenamefont
  {Plestid}(2024)}]{Hill:2023acw}%
  \BibitemOpen
  \bibfield  {author} {\bibinfo {author} {\bibfnamefont {R.~J.}\ \bibnamefont
  {Hill}}\ and\ \bibinfo {author} {\bibfnamefont {R.}~\bibnamefont {Plestid}},\
  }\href {\doibase 10.1103/PhysRevLett.133.021803} {\bibfield  {journal}
  {\bibinfo  {journal} {Phys. Rev. Lett.}\ }\textbf {\bibinfo {volume} {133}},\
  \bibinfo {pages} {021803} (\bibinfo {year} {2024})},\ \Eprint
  {http://arxiv.org/abs/2309.07343} {arXiv:2309.07343 [hep-ph]} \BibitemShut
  {NoStop}%
\bibitem [{\citenamefont {Cirigliano}\ \emph
  {et~al.}(2023{\natexlab{a}})\citenamefont {Cirigliano}, \citenamefont
  {Dekens}, \citenamefont {Mereghetti},\ and\ \citenamefont
  {Tomalak}}]{Cirigliano:2023fnz}%
  \BibitemOpen
  \bibfield  {author} {\bibinfo {author} {\bibfnamefont {V.}~\bibnamefont
  {Cirigliano}}, \bibinfo {author} {\bibfnamefont {W.}~\bibnamefont {Dekens}},
  \bibinfo {author} {\bibfnamefont {E.}~\bibnamefont {Mereghetti}}, \ and\
  \bibinfo {author} {\bibfnamefont {O.}~\bibnamefont {Tomalak}},\ }\href
  {\doibase 10.1103/PhysRevD.108.053003} {\bibfield  {journal} {\bibinfo
  {journal} {Phys. Rev. D}\ }\textbf {\bibinfo {volume} {108}},\ \bibinfo
  {pages} {053003} (\bibinfo {year} {2023}{\natexlab{a}})},\ \Eprint
  {http://arxiv.org/abs/2306.03138} {arXiv:2306.03138 [hep-ph]} \BibitemShut
  {NoStop}%
\bibitem [{\citenamefont {Cirigliano}\ \emph
  {et~al.}(2024{\natexlab{a}})\citenamefont {Cirigliano}, \citenamefont
  {Dekens}, \citenamefont {de~Vries}, \citenamefont {Gandolfi}, \citenamefont
  {Hoferichter},\ and\ \citenamefont {Mereghetti}}]{Cirigliano:2024rfk}%
  \BibitemOpen
  \bibfield  {author} {\bibinfo {author} {\bibfnamefont {V.}~\bibnamefont
  {Cirigliano}}, \bibinfo {author} {\bibfnamefont {W.}~\bibnamefont {Dekens}},
  \bibinfo {author} {\bibfnamefont {J.}~\bibnamefont {de~Vries}}, \bibinfo
  {author} {\bibfnamefont {S.}~\bibnamefont {Gandolfi}}, \bibinfo {author}
  {\bibfnamefont {M.}~\bibnamefont {Hoferichter}}, \ and\ \bibinfo {author}
  {\bibfnamefont {E.}~\bibnamefont {Mereghetti}},\ }\href {\doibase
  10.1103/PhysRevLett.133.211801} {\bibfield  {journal} {\bibinfo  {journal}
  {Phys. Rev. Lett.}\ }\textbf {\bibinfo {volume} {133}},\ \bibinfo {pages}
  {211801} (\bibinfo {year} {2024}{\natexlab{a}})},\ \Eprint
  {http://arxiv.org/abs/2405.18469} {arXiv:2405.18469 [hep-ph]} \BibitemShut
  {NoStop}%
\bibitem [{\citenamefont {Cirigliano}\ \emph
  {et~al.}(2024{\natexlab{b}})\citenamefont {Cirigliano}, \citenamefont
  {Dekens}, \citenamefont {de~Vries}, \citenamefont {Gandolfi}, \citenamefont
  {Hoferichter},\ and\ \citenamefont {Mereghetti}}]{Cirigliano:2024msg}%
  \BibitemOpen
  \bibfield  {author} {\bibinfo {author} {\bibfnamefont {V.}~\bibnamefont
  {Cirigliano}}, \bibinfo {author} {\bibfnamefont {W.}~\bibnamefont {Dekens}},
  \bibinfo {author} {\bibfnamefont {J.}~\bibnamefont {de~Vries}}, \bibinfo
  {author} {\bibfnamefont {S.}~\bibnamefont {Gandolfi}}, \bibinfo {author}
  {\bibfnamefont {M.}~\bibnamefont {Hoferichter}}, \ and\ \bibinfo {author}
  {\bibfnamefont {E.}~\bibnamefont {Mereghetti}},\ }\href {\doibase
  10.1103/PhysRevC.110.055502} {\bibfield  {journal} {\bibinfo  {journal}
  {Phys. Rev. C}\ }\textbf {\bibinfo {volume} {110}},\ \bibinfo {pages}
  {055502} (\bibinfo {year} {2024}{\natexlab{b}})},\ \Eprint
  {http://arxiv.org/abs/2405.18464} {arXiv:2405.18464 [nucl-th]} \BibitemShut
  {NoStop}%
\bibitem [{\citenamefont {Borah}\ \emph {et~al.}(2024)\citenamefont {Borah},
  \citenamefont {Hill},\ and\ \citenamefont {Plestid}}]{Borah:2024ghn}%
  \BibitemOpen
  \bibfield  {author} {\bibinfo {author} {\bibfnamefont {K.}~\bibnamefont
  {Borah}}, \bibinfo {author} {\bibfnamefont {R.~J.}\ \bibnamefont {Hill}}, \
  and\ \bibinfo {author} {\bibfnamefont {R.}~\bibnamefont {Plestid}},\ }\href
  {\doibase 10.1103/PhysRevD.109.113007} {\bibfield  {journal} {\bibinfo
  {journal} {Phys. Rev. D}\ }\textbf {\bibinfo {volume} {109}},\ \bibinfo
  {pages} {113007} (\bibinfo {year} {2024})},\ \Eprint
  {http://arxiv.org/abs/2402.13307} {arXiv:2402.13307 [hep-ph]} \BibitemShut
  {NoStop}%
\bibitem [{\citenamefont {Gennari}\ \emph {et~al.}(2025)\citenamefont
  {Gennari}, \citenamefont {Drissi}, \citenamefont {Gorchtein}, \citenamefont
  {Navratil},\ and\ \citenamefont {Seng}}]{Gennari:2024sbn}%
  \BibitemOpen
  \bibfield  {author} {\bibinfo {author} {\bibfnamefont {M.}~\bibnamefont
  {Gennari}}, \bibinfo {author} {\bibfnamefont {M.}~\bibnamefont {Drissi}},
  \bibinfo {author} {\bibfnamefont {M.}~\bibnamefont {Gorchtein}}, \bibinfo
  {author} {\bibfnamefont {P.}~\bibnamefont {Navratil}}, \ and\ \bibinfo
  {author} {\bibfnamefont {C.-Y.}\ \bibnamefont {Seng}},\ }\href {\doibase
  10.1103/PhysRevLett.134.012501} {\bibfield  {journal} {\bibinfo  {journal}
  {Phys. Rev. Lett.}\ }\textbf {\bibinfo {volume} {134}},\ \bibinfo {pages}
  {012501} (\bibinfo {year} {2025})},\ \Eprint
  {http://arxiv.org/abs/2405.19281} {arXiv:2405.19281 [nucl-th]} \BibitemShut
  {NoStop}%
\bibitem [{\citenamefont {Vander~Griend}\ \emph {et~al.}(2025)\citenamefont
  {Vander~Griend}, \citenamefont {Cao}, \citenamefont {Hill},\ and\
  \citenamefont {Plestid}}]{VanderGriend:2025mdc}%
  \BibitemOpen
  \bibfield  {author} {\bibinfo {author} {\bibfnamefont {P.}~\bibnamefont
  {Vander~Griend}}, \bibinfo {author} {\bibfnamefont {Z.}~\bibnamefont {Cao}},
  \bibinfo {author} {\bibfnamefont {R.~J.}\ \bibnamefont {Hill}}, \ and\
  \bibinfo {author} {\bibfnamefont {R.}~\bibnamefont {Plestid}},\ }\href
  {\doibase 10.1016/j.physletb.2025.139678} {\bibfield  {journal} {\bibinfo
  {journal} {Phys. Lett. B}\ }\textbf {\bibinfo {volume} {868}},\ \bibinfo
  {pages} {139678} (\bibinfo {year} {2025})},\ \Eprint
  {http://arxiv.org/abs/2501.17916} {arXiv:2501.17916 [hep-ph]} \BibitemShut
  {NoStop}%
\bibitem [{\citenamefont {King}\ \emph {et~al.}(2026)\citenamefont {King},
  \citenamefont {Carlson}, \citenamefont {Flores}, \citenamefont {Gandolfi},
  \citenamefont {Mereghetti}, \citenamefont {Pastore}, \citenamefont
  {Piarulli},\ and\ \citenamefont {Wiringa}}]{King:2025fph}%
  \BibitemOpen
  \bibfield  {author} {\bibinfo {author} {\bibfnamefont {G.~B.}\ \bibnamefont
  {King}}, \bibinfo {author} {\bibfnamefont {J.}~\bibnamefont {Carlson}},
  \bibinfo {author} {\bibfnamefont {A.~R.}\ \bibnamefont {Flores}}, \bibinfo
  {author} {\bibfnamefont {S.}~\bibnamefont {Gandolfi}}, \bibinfo {author}
  {\bibfnamefont {E.}~\bibnamefont {Mereghetti}}, \bibinfo {author}
  {\bibfnamefont {S.}~\bibnamefont {Pastore}}, \bibinfo {author} {\bibfnamefont
  {M.}~\bibnamefont {Piarulli}}, \ and\ \bibinfo {author} {\bibfnamefont
  {R.~B.}\ \bibnamefont {Wiringa}},\ }\href {\doibase 10.1103/l863-z5rj}
  {\bibfield  {journal} {\bibinfo  {journal} {Phys. Rev. C}\ }\textbf {\bibinfo
  {volume} {114}},\ \bibinfo {pages} {015501} (\bibinfo {year} {2026})},\
  \Eprint {http://arxiv.org/abs/2509.07310} {arXiv:2509.07310 [nucl-th]}
  \BibitemShut {NoStop}%
\bibitem [{\citenamefont {Cao}\ \emph {et~al.}(2025)\citenamefont {Cao},
  \citenamefont {Hill}, \citenamefont {Plestid},\ and\ \citenamefont
  {Vander~Griend}}]{Cao:2025zxs}%
  \BibitemOpen
  \bibfield  {author} {\bibinfo {author} {\bibfnamefont {Z.}~\bibnamefont
  {Cao}}, \bibinfo {author} {\bibfnamefont {R.~J.}\ \bibnamefont {Hill}},
  \bibinfo {author} {\bibfnamefont {R.}~\bibnamefont {Plestid}}, \ and\
  \bibinfo {author} {\bibfnamefont {P.}~\bibnamefont {Vander~Griend}},\
  }\href@noop {} {\  (\bibinfo {year} {2025})},\ \Eprint
  {http://arxiv.org/abs/2511.05446} {arXiv:2511.05446 [hep-ph]} \BibitemShut
  {NoStop}%
\bibitem [{\citenamefont {Crosas}\ and\ \citenamefont
  {Mereghetti}(2026)}]{Crosas:2025xyv}%
  \BibitemOpen
  \bibfield  {author} {\bibinfo {author} {\bibfnamefont {{\`O}.~L.}\
  \bibnamefont {Crosas}}\ and\ \bibinfo {author} {\bibfnamefont
  {E.}~\bibnamefont {Mereghetti}},\ }\href {\doibase 10.1007/JHEP02(2026)114}
  {\bibfield  {journal} {\bibinfo  {journal} {JHEP}\ }\textbf {\bibinfo
  {volume} {02}},\ \bibinfo {pages} {114} (\bibinfo {year} {2026})},\ \Eprint
  {http://arxiv.org/abs/2511.05481} {arXiv:2511.05481 [hep-ph]} \BibitemShut
  {NoStop}%
\bibitem [{\citenamefont {Gorbahn}\ \emph {et~al.}(2025)\citenamefont
  {Gorbahn}, \citenamefont {Moretti},\ and\ \citenamefont
  {J{\"a}ger}}]{Gorbahn:2025ssv}%
  \BibitemOpen
  \bibfield  {author} {\bibinfo {author} {\bibfnamefont {M.}~\bibnamefont
  {Gorbahn}}, \bibinfo {author} {\bibfnamefont {F.}~\bibnamefont {Moretti}}, \
  and\ \bibinfo {author} {\bibfnamefont {S.}~\bibnamefont {J{\"a}ger}},\
  }\href@noop {} {\  (\bibinfo {year} {2025})},\ \Eprint
  {http://arxiv.org/abs/2510.27648} {arXiv:2510.27648 [hep-ph]} \BibitemShut
  {NoStop}%
\bibitem [{\citenamefont {Belfatto}\ \emph {et~al.}(2020)\citenamefont
  {Belfatto}, \citenamefont {Beradze},\ and\ \citenamefont
  {Berezhiani}}]{Belfatto:2019swo}%
  \BibitemOpen
  \bibfield  {author} {\bibinfo {author} {\bibfnamefont {B.}~\bibnamefont
  {Belfatto}}, \bibinfo {author} {\bibfnamefont {R.}~\bibnamefont {Beradze}}, \
  and\ \bibinfo {author} {\bibfnamefont {Z.}~\bibnamefont {Berezhiani}},\
  }\href {\doibase 10.1140/epjc/s10052-020-7691-6} {\bibfield  {journal}
  {\bibinfo  {journal} {Eur. Phys. J. C}\ }\textbf {\bibinfo {volume} {80}},\
  \bibinfo {pages} {149} (\bibinfo {year} {2020})},\ \Eprint
  {http://arxiv.org/abs/1906.02714} {arXiv:1906.02714 [hep-ph]} \BibitemShut
  {NoStop}%
\bibitem [{\citenamefont {Coutinho}\ \emph {et~al.}(2020)\citenamefont
  {Coutinho}, \citenamefont {Crivellin},\ and\ \citenamefont
  {Manzari}}]{Coutinho:2019aiy}%
  \BibitemOpen
  \bibfield  {author} {\bibinfo {author} {\bibfnamefont {A.~M.}\ \bibnamefont
  {Coutinho}}, \bibinfo {author} {\bibfnamefont {A.}~\bibnamefont {Crivellin}},
  \ and\ \bibinfo {author} {\bibfnamefont {C.~A.}\ \bibnamefont {Manzari}},\
  }\href {\doibase 10.1103/PhysRevLett.125.071802} {\bibfield  {journal}
  {\bibinfo  {journal} {Phys. Rev. Lett.}\ }\textbf {\bibinfo {volume} {125}},\
  \bibinfo {pages} {071802} (\bibinfo {year} {2020})},\ \Eprint
  {http://arxiv.org/abs/1912.08823} {arXiv:1912.08823 [hep-ph]} \BibitemShut
  {NoStop}%
\bibitem [{\citenamefont {Cheung}\ \emph {et~al.}(2020)\citenamefont {Cheung},
  \citenamefont {Keung}, \citenamefont {Lu},\ and\ \citenamefont
  {Tseng}}]{Cheung:2020vqm}%
  \BibitemOpen
  \bibfield  {author} {\bibinfo {author} {\bibfnamefont {K.}~\bibnamefont
  {Cheung}}, \bibinfo {author} {\bibfnamefont {W.-Y.}\ \bibnamefont {Keung}},
  \bibinfo {author} {\bibfnamefont {C.-T.}\ \bibnamefont {Lu}}, \ and\ \bibinfo
  {author} {\bibfnamefont {P.-Y.}\ \bibnamefont {Tseng}},\ }\href {\doibase
  10.1007/JHEP05(2020)117} {\bibfield  {journal} {\bibinfo  {journal} {JHEP}\
  }\textbf {\bibinfo {volume} {05}},\ \bibinfo {pages} {117} (\bibinfo {year}
  {2020})},\ \Eprint {http://arxiv.org/abs/2001.02853} {arXiv:2001.02853
  [hep-ph]} \BibitemShut {NoStop}%
\bibitem [{\citenamefont {Belfatto}\ and\ \citenamefont
  {Berezhiani}(2021)}]{Belfatto:2021jhf}%
  \BibitemOpen
  \bibfield  {author} {\bibinfo {author} {\bibfnamefont {B.}~\bibnamefont
  {Belfatto}}\ and\ \bibinfo {author} {\bibfnamefont {Z.}~\bibnamefont
  {Berezhiani}},\ }\href {\doibase 10.1007/JHEP10(2021)079} {\bibfield
  {journal} {\bibinfo  {journal} {JHEP}\ }\textbf {\bibinfo {volume} {10}},\
  \bibinfo {pages} {079} (\bibinfo {year} {2021})},\ \Eprint
  {http://arxiv.org/abs/2103.05549} {arXiv:2103.05549 [hep-ph]} \BibitemShut
  {NoStop}%
\bibitem [{\citenamefont {Branco}\ \emph {et~al.}(2021)\citenamefont {Branco},
  \citenamefont {Penedo}, \citenamefont {Pereira}, \citenamefont {Rebelo},\
  and\ \citenamefont {Silva-Marcos}}]{Branco:2021vhs}%
  \BibitemOpen
  \bibfield  {author} {\bibinfo {author} {\bibfnamefont {G.~C.}\ \bibnamefont
  {Branco}}, \bibinfo {author} {\bibfnamefont {J.~T.}\ \bibnamefont {Penedo}},
  \bibinfo {author} {\bibfnamefont {P.~M.~F.}\ \bibnamefont {Pereira}},
  \bibinfo {author} {\bibfnamefont {M.~N.}\ \bibnamefont {Rebelo}}, \ and\
  \bibinfo {author} {\bibfnamefont {J.~I.}\ \bibnamefont {Silva-Marcos}},\
  }\href {\doibase 10.1007/JHEP07(2021)099} {\bibfield  {journal} {\bibinfo
  {journal} {JHEP}\ }\textbf {\bibinfo {volume} {07}},\ \bibinfo {pages} {099}
  (\bibinfo {year} {2021})},\ \Eprint {http://arxiv.org/abs/2103.13409}
  {arXiv:2103.13409 [hep-ph]} \BibitemShut {NoStop}%
\bibitem [{\citenamefont {Crivellin}\ \emph
  {et~al.}(2021{\natexlab{a}})\citenamefont {Crivellin}, \citenamefont
  {Hoferichter}, \citenamefont {Kirk}, \citenamefont {Manzari},\ and\
  \citenamefont {Schnell}}]{Crivellin:2021bkd}%
  \BibitemOpen
  \bibfield  {author} {\bibinfo {author} {\bibfnamefont {A.}~\bibnamefont
  {Crivellin}}, \bibinfo {author} {\bibfnamefont {M.}~\bibnamefont
  {Hoferichter}}, \bibinfo {author} {\bibfnamefont {M.}~\bibnamefont {Kirk}},
  \bibinfo {author} {\bibfnamefont {C.~A.}\ \bibnamefont {Manzari}}, \ and\
  \bibinfo {author} {\bibfnamefont {L.}~\bibnamefont {Schnell}},\ }\href
  {\doibase 10.1007/JHEP10(2021)221} {\bibfield  {journal} {\bibinfo  {journal}
  {JHEP}\ }\textbf {\bibinfo {volume} {10}},\ \bibinfo {pages} {221} (\bibinfo
  {year} {2021}{\natexlab{a}})},\ \Eprint {http://arxiv.org/abs/2107.13569}
  {arXiv:2107.13569 [hep-ph]} \BibitemShut {NoStop}%
\bibitem [{\citenamefont {Crivellin}\ \emph {et~al.}(2020)\citenamefont
  {Crivellin}, \citenamefont {Kirk}, \citenamefont {Manzari},\ and\
  \citenamefont {Montull}}]{Crivellin:2020ebi}%
  \BibitemOpen
  \bibfield  {author} {\bibinfo {author} {\bibfnamefont {A.}~\bibnamefont
  {Crivellin}}, \bibinfo {author} {\bibfnamefont {F.}~\bibnamefont {Kirk}},
  \bibinfo {author} {\bibfnamefont {C.~A.}\ \bibnamefont {Manzari}}, \ and\
  \bibinfo {author} {\bibfnamefont {M.}~\bibnamefont {Montull}},\ }\href
  {\doibase 10.1007/JHEP12(2020)166} {\bibfield  {journal} {\bibinfo  {journal}
  {JHEP}\ }\textbf {\bibinfo {volume} {12}},\ \bibinfo {pages} {166} (\bibinfo
  {year} {2020})},\ \Eprint {http://arxiv.org/abs/2008.01113} {arXiv:2008.01113
  [hep-ph]} \BibitemShut {NoStop}%
\bibitem [{\citenamefont {Kirk}(2021)}]{Kirk:2020wdk}%
  \BibitemOpen
  \bibfield  {author} {\bibinfo {author} {\bibfnamefont {M.}~\bibnamefont
  {Kirk}},\ }\href {\doibase 10.1103/PhysRevD.103.035004} {\bibfield  {journal}
  {\bibinfo  {journal} {Phys. Rev. D}\ }\textbf {\bibinfo {volume} {103}},\
  \bibinfo {pages} {035004} (\bibinfo {year} {2021})},\ \Eprint
  {http://arxiv.org/abs/2008.03261} {arXiv:2008.03261 [hep-ph]} \BibitemShut
  {NoStop}%
\bibitem [{\citenamefont {Crivellin}\ \emph
  {et~al.}(2021{\natexlab{b}})\citenamefont {Crivellin}, \citenamefont
  {Hoferichter},\ and\ \citenamefont {Manzari}}]{Crivellin:2021njn}%
  \BibitemOpen
  \bibfield  {author} {\bibinfo {author} {\bibfnamefont {A.}~\bibnamefont
  {Crivellin}}, \bibinfo {author} {\bibfnamefont {M.}~\bibnamefont
  {Hoferichter}}, \ and\ \bibinfo {author} {\bibfnamefont {C.~A.}\ \bibnamefont
  {Manzari}},\ }\href {\doibase 10.1103/PhysRevLett.127.071801} {\bibfield
  {journal} {\bibinfo  {journal} {Phys. Rev. Lett.}\ }\textbf {\bibinfo
  {volume} {127}},\ \bibinfo {pages} {071801} (\bibinfo {year}
  {2021}{\natexlab{b}})},\ \Eprint {http://arxiv.org/abs/2102.02825}
  {arXiv:2102.02825 [hep-ph]} \BibitemShut {NoStop}%
\bibitem [{\citenamefont {Crivellin}\ and\ \citenamefont
  {Hoferichter}(2020)}]{Crivellin:2020lzu}%
  \BibitemOpen
  \bibfield  {author} {\bibinfo {author} {\bibfnamefont {A.}~\bibnamefont
  {Crivellin}}\ and\ \bibinfo {author} {\bibfnamefont {M.}~\bibnamefont
  {Hoferichter}},\ }\href {\doibase 10.1103/PhysRevLett.125.111801} {\bibfield
  {journal} {\bibinfo  {journal} {Phys. Rev. Lett.}\ }\textbf {\bibinfo
  {volume} {125}},\ \bibinfo {pages} {111801} (\bibinfo {year} {2020})},\
  \Eprint {http://arxiv.org/abs/2002.07184} {arXiv:2002.07184 [hep-ph]}
  \BibitemShut {NoStop}%
\bibitem [{\citenamefont {Crivellin}\ \emph
  {et~al.}(2021{\natexlab{c}})\citenamefont {Crivellin}, \citenamefont {Kirk},
  \citenamefont {Manzari},\ and\ \citenamefont {Panizzi}}]{Crivellin:2020klg}%
  \BibitemOpen
  \bibfield  {author} {\bibinfo {author} {\bibfnamefont {A.}~\bibnamefont
  {Crivellin}}, \bibinfo {author} {\bibfnamefont {F.}~\bibnamefont {Kirk}},
  \bibinfo {author} {\bibfnamefont {C.~A.}\ \bibnamefont {Manzari}}, \ and\
  \bibinfo {author} {\bibfnamefont {L.}~\bibnamefont {Panizzi}},\ }\href
  {\doibase 10.1103/PhysRevD.103.073002} {\bibfield  {journal} {\bibinfo
  {journal} {Phys. Rev. D}\ }\textbf {\bibinfo {volume} {103}},\ \bibinfo
  {pages} {073002} (\bibinfo {year} {2021}{\natexlab{c}})},\ \Eprint
  {http://arxiv.org/abs/2012.09845} {arXiv:2012.09845 [hep-ph]} \BibitemShut
  {NoStop}%
\bibitem [{\citenamefont {Capdevila}\ \emph {et~al.}(2021)\citenamefont
  {Capdevila}, \citenamefont {Crivellin}, \citenamefont {Manzari},\ and\
  \citenamefont {Montull}}]{Capdevila:2020rrl}%
  \BibitemOpen
  \bibfield  {author} {\bibinfo {author} {\bibfnamefont {B.}~\bibnamefont
  {Capdevila}}, \bibinfo {author} {\bibfnamefont {A.}~\bibnamefont
  {Crivellin}}, \bibinfo {author} {\bibfnamefont {C.~A.}\ \bibnamefont
  {Manzari}}, \ and\ \bibinfo {author} {\bibfnamefont {M.}~\bibnamefont
  {Montull}},\ }\href {\doibase 10.1103/PhysRevD.103.015032} {\bibfield
  {journal} {\bibinfo  {journal} {Phys. Rev. D}\ }\textbf {\bibinfo {volume}
  {103}},\ \bibinfo {pages} {015032} (\bibinfo {year} {2021})},\ \Eprint
  {http://arxiv.org/abs/2005.13542} {arXiv:2005.13542 [hep-ph]} \BibitemShut
  {NoStop}%
\bibitem [{\citenamefont {Crivellin}\ and\ \citenamefont
  {Hoferichter}(2021)}]{Crivellin:2021sff}%
  \BibitemOpen
  \bibfield  {author} {\bibinfo {author} {\bibfnamefont {A.}~\bibnamefont
  {Crivellin}}\ and\ \bibinfo {author} {\bibfnamefont {M.}~\bibnamefont
  {Hoferichter}},\ }\href {\doibase 10.1126/science.abk2450} {\bibfield
  {journal} {\bibinfo  {journal} {Science}\ }\textbf {\bibinfo {volume}
  {374}},\ \bibinfo {pages} {1051} (\bibinfo {year} {2021})},\ \Eprint
  {http://arxiv.org/abs/2111.12739} {arXiv:2111.12739 [hep-ph]} \BibitemShut
  {NoStop}%
\bibitem [{\citenamefont {Crivellin}\ \emph
  {et~al.}(2021{\natexlab{d}})\citenamefont {Crivellin}, \citenamefont
  {Manzari}, \citenamefont {Alguer{\'o}},\ and\ \citenamefont
  {Matias}}]{Crivellin:2020oup}%
  \BibitemOpen
  \bibfield  {author} {\bibinfo {author} {\bibfnamefont {A.}~\bibnamefont
  {Crivellin}}, \bibinfo {author} {\bibfnamefont {C.~A.}\ \bibnamefont
  {Manzari}}, \bibinfo {author} {\bibfnamefont {M.}~\bibnamefont
  {Alguer{\'o}}}, \ and\ \bibinfo {author} {\bibfnamefont {J.}~\bibnamefont
  {Matias}},\ }\href {\doibase 10.1103/PhysRevLett.127.011801} {\bibfield
  {journal} {\bibinfo  {journal} {Phys. Rev. Lett.}\ }\textbf {\bibinfo
  {volume} {127}},\ \bibinfo {pages} {011801} (\bibinfo {year}
  {2021}{\natexlab{d}})},\ \Eprint {http://arxiv.org/abs/2010.14504}
  {arXiv:2010.14504 [hep-ph]} \BibitemShut {NoStop}%
\bibitem [{\citenamefont {Marzocca}\ and\ \citenamefont
  {Trifinopoulos}(2021)}]{Marzocca:2021azj}%
  \BibitemOpen
  \bibfield  {author} {\bibinfo {author} {\bibfnamefont {D.}~\bibnamefont
  {Marzocca}}\ and\ \bibinfo {author} {\bibfnamefont {S.}~\bibnamefont
  {Trifinopoulos}},\ }\href {\doibase 10.1103/PhysRevLett.127.061803}
  {\bibfield  {journal} {\bibinfo  {journal} {Phys. Rev. Lett.}\ }\textbf
  {\bibinfo {volume} {127}},\ \bibinfo {pages} {061803} (\bibinfo {year}
  {2021})},\ \Eprint {http://arxiv.org/abs/2104.05730} {arXiv:2104.05730
  [hep-ph]} \BibitemShut {NoStop}%
\bibitem [{\citenamefont {Alok}\ \emph {et~al.}(2023)\citenamefont {Alok},
  \citenamefont {Dighe}, \citenamefont {Gangal},\ and\ \citenamefont
  {Kumar}}]{Alok:2021ydy}%
  \BibitemOpen
  \bibfield  {author} {\bibinfo {author} {\bibfnamefont {A.~K.}\ \bibnamefont
  {Alok}}, \bibinfo {author} {\bibfnamefont {A.}~\bibnamefont {Dighe}},
  \bibinfo {author} {\bibfnamefont {S.}~\bibnamefont {Gangal}}, \ and\ \bibinfo
  {author} {\bibfnamefont {J.}~\bibnamefont {Kumar}},\ }\href {\doibase
  10.1103/PhysRevD.108.113005} {\bibfield  {journal} {\bibinfo  {journal}
  {Phys. Rev. D}\ }\textbf {\bibinfo {volume} {108}},\ \bibinfo {pages}
  {113005} (\bibinfo {year} {2023})},\ \Eprint
  {http://arxiv.org/abs/2108.05614} {arXiv:2108.05614 [hep-ph]} \BibitemShut
  {NoStop}%
\bibitem [{\citenamefont {Cirigliano}\ \emph
  {et~al.}(2022{\natexlab{a}})\citenamefont {Cirigliano}, \citenamefont
  {Dekens}, \citenamefont {de~Vries}, \citenamefont {Mereghetti},\ and\
  \citenamefont {Tong}}]{Cirigliano:2022qdm}%
  \BibitemOpen
  \bibfield  {author} {\bibinfo {author} {\bibfnamefont {V.}~\bibnamefont
  {Cirigliano}}, \bibinfo {author} {\bibfnamefont {W.}~\bibnamefont {Dekens}},
  \bibinfo {author} {\bibfnamefont {J.}~\bibnamefont {de~Vries}}, \bibinfo
  {author} {\bibfnamefont {E.}~\bibnamefont {Mereghetti}}, \ and\ \bibinfo
  {author} {\bibfnamefont {T.}~\bibnamefont {Tong}},\ }\href {\doibase
  10.1103/PhysRevD.106.075001} {\bibfield  {journal} {\bibinfo  {journal}
  {Phys. Rev. D}\ }\textbf {\bibinfo {volume} {106}},\ \bibinfo {pages}
  {075001} (\bibinfo {year} {2022}{\natexlab{a}})},\ \Eprint
  {http://arxiv.org/abs/2204.08440} {arXiv:2204.08440 [hep-ph]} \BibitemShut
  {NoStop}%
\bibitem [{\citenamefont {Cirigliano}\ \emph
  {et~al.}(2024{\natexlab{c}})\citenamefont {Cirigliano}, \citenamefont
  {Dekens}, \citenamefont {de~Vries}, \citenamefont {Mereghetti},\ and\
  \citenamefont {Tong}}]{Cirigliano:2023nol}%
  \BibitemOpen
  \bibfield  {author} {\bibinfo {author} {\bibfnamefont {V.}~\bibnamefont
  {Cirigliano}}, \bibinfo {author} {\bibfnamefont {W.}~\bibnamefont {Dekens}},
  \bibinfo {author} {\bibfnamefont {J.}~\bibnamefont {de~Vries}}, \bibinfo
  {author} {\bibfnamefont {E.}~\bibnamefont {Mereghetti}}, \ and\ \bibinfo
  {author} {\bibfnamefont {T.}~\bibnamefont {Tong}},\ }\href {\doibase
  10.1007/JHEP03(2024)033} {\bibfield  {journal} {\bibinfo  {journal} {JHEP}\
  }\textbf {\bibinfo {volume} {03}},\ \bibinfo {pages} {033} (\bibinfo {year}
  {2024}{\natexlab{c}})},\ \Eprint {http://arxiv.org/abs/2311.00021}
  {arXiv:2311.00021 [hep-ph]} \BibitemShut {NoStop}%
\bibitem [{\citenamefont {Dawid}\ \emph {et~al.}(2024)\citenamefont {Dawid},
  \citenamefont {Cirigliano},\ and\ \citenamefont {Dekens}}]{Dawid:2024wmp}%
  \BibitemOpen
  \bibfield  {author} {\bibinfo {author} {\bibfnamefont {M.}~\bibnamefont
  {Dawid}}, \bibinfo {author} {\bibfnamefont {V.}~\bibnamefont {Cirigliano}}, \
  and\ \bibinfo {author} {\bibfnamefont {W.}~\bibnamefont {Dekens}},\ }\href
  {\doibase 10.1007/JHEP08(2024)175} {\bibfield  {journal} {\bibinfo  {journal}
  {JHEP}\ }\textbf {\bibinfo {volume} {08}},\ \bibinfo {pages} {175} (\bibinfo
  {year} {2024})},\ \Eprint {http://arxiv.org/abs/2402.06723} {arXiv:2402.06723
  [hep-ph]} \BibitemShut {NoStop}%
\bibitem [{\citenamefont {Czarnecki}\ \emph {et~al.}(2019)\citenamefont
  {Czarnecki}, \citenamefont {Marciano},\ and\ \citenamefont
  {Sirlin}}]{Czarnecki:2019mwq}%
  \BibitemOpen
  \bibfield  {author} {\bibinfo {author} {\bibfnamefont {A.}~\bibnamefont
  {Czarnecki}}, \bibinfo {author} {\bibfnamefont {W.~J.}\ \bibnamefont
  {Marciano}}, \ and\ \bibinfo {author} {\bibfnamefont {A.}~\bibnamefont
  {Sirlin}},\ }\href {\doibase 10.1103/PhysRevD.100.073008} {\bibfield
  {journal} {\bibinfo  {journal} {Phys. Rev. D}\ }\textbf {\bibinfo {volume}
  {100}},\ \bibinfo {pages} {073008} (\bibinfo {year} {2019})},\ \Eprint
  {http://arxiv.org/abs/1907.06737} {arXiv:1907.06737 [hep-ph]} \BibitemShut
  {NoStop}%
\bibitem [{\citenamefont {Seng}\ \emph {et~al.}(2020)\citenamefont {Seng},
  \citenamefont {Feng}, \citenamefont {Gorchtein},\ and\ \citenamefont
  {Jin}}]{Seng:2020wjq}%
  \BibitemOpen
  \bibfield  {author} {\bibinfo {author} {\bibfnamefont {C.-Y.}\ \bibnamefont
  {Seng}}, \bibinfo {author} {\bibfnamefont {X.}~\bibnamefont {Feng}}, \bibinfo
  {author} {\bibfnamefont {M.}~\bibnamefont {Gorchtein}}, \ and\ \bibinfo
  {author} {\bibfnamefont {L.-C.}\ \bibnamefont {Jin}},\ }\href {\doibase
  10.1103/PhysRevD.101.111301} {\bibfield  {journal} {\bibinfo  {journal}
  {Phys. Rev. D}\ }\textbf {\bibinfo {volume} {101}},\ \bibinfo {pages}
  {111301} (\bibinfo {year} {2020})},\ \Eprint
  {http://arxiv.org/abs/2003.11264} {arXiv:2003.11264 [hep-ph]} \BibitemShut
  {NoStop}%
\bibitem [{\citenamefont {Hayen}(2021)}]{Hayen:2020cxh}%
  \BibitemOpen
  \bibfield  {author} {\bibinfo {author} {\bibfnamefont {L.}~\bibnamefont
  {Hayen}},\ }\href {\doibase 10.1103/PhysRevD.103.113001} {\bibfield
  {journal} {\bibinfo  {journal} {Phys. Rev. D}\ }\textbf {\bibinfo {volume}
  {103}},\ \bibinfo {pages} {113001} (\bibinfo {year} {2021})},\ \Eprint
  {http://arxiv.org/abs/2010.07262} {arXiv:2010.07262 [hep-ph]} \BibitemShut
  {NoStop}%
\bibitem [{\citenamefont {Shiells}\ \emph {et~al.}(2021)\citenamefont
  {Shiells}, \citenamefont {Blunden},\ and\ \citenamefont
  {Melnitchouk}}]{Shiells:2020fqp}%
  \BibitemOpen
  \bibfield  {author} {\bibinfo {author} {\bibfnamefont {K.}~\bibnamefont
  {Shiells}}, \bibinfo {author} {\bibfnamefont {P.~G.}\ \bibnamefont
  {Blunden}}, \ and\ \bibinfo {author} {\bibfnamefont {W.}~\bibnamefont
  {Melnitchouk}},\ }\href {\doibase 10.1103/PhysRevD.104.033003} {\bibfield
  {journal} {\bibinfo  {journal} {Phys. Rev. D}\ }\textbf {\bibinfo {volume}
  {104}},\ \bibinfo {pages} {033003} (\bibinfo {year} {2021})},\ \Eprint
  {http://arxiv.org/abs/2012.01580} {arXiv:2012.01580 [hep-ph]} \BibitemShut
  {NoStop}%
\bibitem [{\citenamefont {Cirigliano}\ \emph
  {et~al.}(2023{\natexlab{b}})\citenamefont {Cirigliano}, \citenamefont
  {Crivellin}, \citenamefont {Hoferichter},\ and\ \citenamefont
  {Moulson}}]{Cirigliano:2022yyo}%
  \BibitemOpen
  \bibfield  {author} {\bibinfo {author} {\bibfnamefont {V.}~\bibnamefont
  {Cirigliano}}, \bibinfo {author} {\bibfnamefont {A.}~\bibnamefont
  {Crivellin}}, \bibinfo {author} {\bibfnamefont {M.}~\bibnamefont
  {Hoferichter}}, \ and\ \bibinfo {author} {\bibfnamefont {M.}~\bibnamefont
  {Moulson}},\ }\href {\doibase 10.1016/j.physletb.2023.137748} {\bibfield
  {journal} {\bibinfo  {journal} {Phys. Lett. B}\ }\textbf {\bibinfo {volume}
  {838}},\ \bibinfo {pages} {137748} (\bibinfo {year} {2023}{\natexlab{b}})},\
  \Eprint {http://arxiv.org/abs/2208.11707} {arXiv:2208.11707 [hep-ph]}
  \BibitemShut {NoStop}%
\bibitem [{\citenamefont {Descotes-Genon}\ and\ \citenamefont
  {Moussallam}(2005)}]{Descotes-Genon:2005wrq}%
  \BibitemOpen
  \bibfield  {author} {\bibinfo {author} {\bibfnamefont {S.}~\bibnamefont
  {Descotes-Genon}}\ and\ \bibinfo {author} {\bibfnamefont {B.}~\bibnamefont
  {Moussallam}},\ }\href {\doibase 10.1140/epjc/s2005-02316-8} {\bibfield
  {journal} {\bibinfo  {journal} {Eur. Phys. J. C}\ }\textbf {\bibinfo {volume}
  {42}},\ \bibinfo {pages} {403} (\bibinfo {year} {2005})},\ \Eprint
  {http://arxiv.org/abs/hep-ph/0505077} {arXiv:hep-ph/0505077} \BibitemShut
  {NoStop}%
\bibitem [{\citenamefont {Po{\v c}ani{\'c}}\ \emph {et~al.}(2004)\citenamefont
  {Po{\v c}ani{\'c}} \emph {et~al.}}]{Pocanic:2003pf}%
  \BibitemOpen
  \bibfield  {author} {\bibinfo {author} {\bibfnamefont {D.}~\bibnamefont
  {Po{\v c}ani{\'c}}} \emph {et~al.},\ }\href {\doibase
  10.1103/PhysRevLett.93.181803} {\bibfield  {journal} {\bibinfo  {journal}
  {Phys. Rev. Lett.}\ }\textbf {\bibinfo {volume} {93}},\ \bibinfo {pages}
  {181803} (\bibinfo {year} {2004})},\ \Eprint
  {http://arxiv.org/abs/hep-ex/0312030} {arXiv:hep-ex/0312030} \BibitemShut
  {NoStop}%
\bibitem [{\citenamefont {Cirigliano}\ \emph {et~al.}(2003)\citenamefont
  {Cirigliano}, \citenamefont {Knecht}, \citenamefont {Neufeld},\ and\
  \citenamefont {Pichl}}]{Cirigliano:2002ng}%
  \BibitemOpen
  \bibfield  {author} {\bibinfo {author} {\bibfnamefont {V.}~\bibnamefont
  {Cirigliano}}, \bibinfo {author} {\bibfnamefont {M.}~\bibnamefont {Knecht}},
  \bibinfo {author} {\bibfnamefont {H.}~\bibnamefont {Neufeld}}, \ and\
  \bibinfo {author} {\bibfnamefont {H.}~\bibnamefont {Pichl}},\ }\href
  {\doibase 10.1140/epjc/s2002-01093-2} {\bibfield  {journal} {\bibinfo
  {journal} {Eur. Phys. J. C}\ }\textbf {\bibinfo {volume} {27}},\ \bibinfo
  {pages} {255} (\bibinfo {year} {2003})},\ \Eprint
  {http://arxiv.org/abs/hep-ph/0209226} {arXiv:hep-ph/0209226} \BibitemShut
  {NoStop}%
\bibitem [{\citenamefont {Czarnecki}\ \emph {et~al.}(2020)\citenamefont
  {Czarnecki}, \citenamefont {Marciano},\ and\ \citenamefont
  {Sirlin}}]{Czarnecki:2019iwz}%
  \BibitemOpen
  \bibfield  {author} {\bibinfo {author} {\bibfnamefont {A.}~\bibnamefont
  {Czarnecki}}, \bibinfo {author} {\bibfnamefont {W.~J.}\ \bibnamefont
  {Marciano}}, \ and\ \bibinfo {author} {\bibfnamefont {A.}~\bibnamefont
  {Sirlin}},\ }\href {\doibase 10.1103/PhysRevD.101.091301} {\bibfield
  {journal} {\bibinfo  {journal} {Phys. Rev. D}\ }\textbf {\bibinfo {volume}
  {101}},\ \bibinfo {pages} {091301} (\bibinfo {year} {2020})},\ \Eprint
  {http://arxiv.org/abs/1911.04685} {arXiv:1911.04685 [hep-ph]} \BibitemShut
  {NoStop}%
\bibitem [{\citenamefont {Feng}\ \emph {et~al.}(2020)\citenamefont {Feng},
  \citenamefont {Gorchtein}, \citenamefont {Jin}, \citenamefont {Ma},\ and\
  \citenamefont {Seng}}]{Feng:2020zdc}%
  \BibitemOpen
  \bibfield  {author} {\bibinfo {author} {\bibfnamefont {X.}~\bibnamefont
  {Feng}}, \bibinfo {author} {\bibfnamefont {M.}~\bibnamefont {Gorchtein}},
  \bibinfo {author} {\bibfnamefont {L.-C.}\ \bibnamefont {Jin}}, \bibinfo
  {author} {\bibfnamefont {P.-X.}\ \bibnamefont {Ma}}, \ and\ \bibinfo {author}
  {\bibfnamefont {C.-Y.}\ \bibnamefont {Seng}},\ }\href {\doibase
  10.1103/PhysRevLett.124.192002} {\bibfield  {journal} {\bibinfo  {journal}
  {Phys. Rev. Lett.}\ }\textbf {\bibinfo {volume} {124}},\ \bibinfo {pages}
  {192002} (\bibinfo {year} {2020})},\ \Eprint
  {http://arxiv.org/abs/2003.09798} {arXiv:2003.09798 [hep-lat]} \BibitemShut
  {NoStop}%
\bibitem [{\citenamefont {Ma}\ \emph {et~al.}(2021)\citenamefont {Ma},
  \citenamefont {Feng}, \citenamefont {Gorchtein}, \citenamefont {Jin},\ and\
  \citenamefont {Seng}}]{Ma:2021azh}%
  \BibitemOpen
  \bibfield  {author} {\bibinfo {author} {\bibfnamefont {P.-X.}\ \bibnamefont
  {Ma}}, \bibinfo {author} {\bibfnamefont {X.}~\bibnamefont {Feng}}, \bibinfo
  {author} {\bibfnamefont {M.}~\bibnamefont {Gorchtein}}, \bibinfo {author}
  {\bibfnamefont {L.-C.}\ \bibnamefont {Jin}}, \ and\ \bibinfo {author}
  {\bibfnamefont {C.-Y.}\ \bibnamefont {Seng}},\ }\href {\doibase
  10.1103/PhysRevD.103.114503} {\bibfield  {journal} {\bibinfo  {journal}
  {Phys. Rev. D}\ }\textbf {\bibinfo {volume} {103}},\ \bibinfo {pages}
  {114503} (\bibinfo {year} {2021})},\ \Eprint
  {http://arxiv.org/abs/2102.12048} {arXiv:2102.12048 [hep-lat]} \BibitemShut
  {NoStop}%
\bibitem [{\citenamefont {Yoo}\ \emph {et~al.}(2023)\citenamefont {Yoo},
  \citenamefont {Bhattacharya}, \citenamefont {Gupta}, \citenamefont {Mondal},\
  and\ \citenamefont {Yoon}}]{Yoo:2023gln}%
  \BibitemOpen
  \bibfield  {author} {\bibinfo {author} {\bibfnamefont {J.-S.}\ \bibnamefont
  {Yoo}}, \bibinfo {author} {\bibfnamefont {T.}~\bibnamefont {Bhattacharya}},
  \bibinfo {author} {\bibfnamefont {R.}~\bibnamefont {Gupta}}, \bibinfo
  {author} {\bibfnamefont {S.}~\bibnamefont {Mondal}}, \ and\ \bibinfo {author}
  {\bibfnamefont {B.}~\bibnamefont {Yoon}},\ }\href {\doibase
  10.1103/PhysRevD.108.034508} {\bibfield  {journal} {\bibinfo  {journal}
  {Phys. Rev. D}\ }\textbf {\bibinfo {volume} {108}},\ \bibinfo {pages}
  {034508} (\bibinfo {year} {2023})},\ \Eprint
  {http://arxiv.org/abs/2305.03198} {arXiv:2305.03198 [hep-lat]} \BibitemShut
  {NoStop}%
\bibitem [{\citenamefont {Altmannshofer}\ \emph {et~al.}(2022)\citenamefont
  {Altmannshofer} \emph {et~al.}}]{PIONEER:2022yag}%
  \BibitemOpen
  \bibfield  {author} {\bibinfo {author} {\bibfnamefont {W.}~\bibnamefont
  {Altmannshofer}} \emph {et~al.} (\bibinfo {collaboration} {PIONEER}),\
  }\href@noop {} {\  (\bibinfo {year} {2022})},\ \Eprint
  {http://arxiv.org/abs/2203.01981} {arXiv:2203.01981 [hep-ex]} \BibitemShut
  {NoStop}%
\bibitem [{\citenamefont {Adelmann}\ \emph {et~al.}(2025)\citenamefont
  {Adelmann} \emph {et~al.}}]{PIONEER:2025idw}%
  \BibitemOpen
  \bibfield  {author} {\bibinfo {author} {\bibfnamefont {A.}~\bibnamefont
  {Adelmann}} \emph {et~al.} (\bibinfo {collaboration} {PIONEER}),\ }\href@noop
  {} {\  (\bibinfo {year} {2025})},\ \Eprint {http://arxiv.org/abs/2504.06375}
  {arXiv:2504.06375 [hep-ex]} \BibitemShut {NoStop}%
\bibitem [{\citenamefont {Alemany}\ \emph {et~al.}(1998)\citenamefont
  {Alemany}, \citenamefont {Davier},\ and\ \citenamefont
  {Hoecker}}]{Alemany:1997tn}%
  \BibitemOpen
  \bibfield  {author} {\bibinfo {author} {\bibfnamefont {R.}~\bibnamefont
  {Alemany}}, \bibinfo {author} {\bibfnamefont {M.}~\bibnamefont {Davier}}, \
  and\ \bibinfo {author} {\bibfnamefont {A.}~\bibnamefont {Hoecker}},\ }\href
  {\doibase 10.1007/s100520050127} {\bibfield  {journal} {\bibinfo  {journal}
  {Eur. Phys. J. C}\ }\textbf {\bibinfo {volume} {2}},\ \bibinfo {pages} {123}
  (\bibinfo {year} {1998})},\ \Eprint {http://arxiv.org/abs/hep-ph/9703220}
  {arXiv:hep-ph/9703220} \BibitemShut {NoStop}%
\bibitem [{\citenamefont {Cirigliano}\ \emph {et~al.}(2001)\citenamefont
  {Cirigliano}, \citenamefont {Ecker},\ and\ \citenamefont
  {Neufeld}}]{Cirigliano:2001er}%
  \BibitemOpen
  \bibfield  {author} {\bibinfo {author} {\bibfnamefont {V.}~\bibnamefont
  {Cirigliano}}, \bibinfo {author} {\bibfnamefont {G.}~\bibnamefont {Ecker}}, \
  and\ \bibinfo {author} {\bibfnamefont {H.}~\bibnamefont {Neufeld}},\ }\href
  {\doibase 10.1016/S0370-2693(01)00764-X} {\bibfield  {journal} {\bibinfo
  {journal} {Phys. Lett. B}\ }\textbf {\bibinfo {volume} {513}},\ \bibinfo
  {pages} {361} (\bibinfo {year} {2001})},\ \Eprint
  {http://arxiv.org/abs/hep-ph/0104267} {arXiv:hep-ph/0104267} \BibitemShut
  {NoStop}%
\bibitem [{\citenamefont {Cirigliano}\ \emph {et~al.}(2002)\citenamefont
  {Cirigliano}, \citenamefont {Ecker},\ and\ \citenamefont
  {Neufeld}}]{Cirigliano:2002pv}%
  \BibitemOpen
  \bibfield  {author} {\bibinfo {author} {\bibfnamefont {V.}~\bibnamefont
  {Cirigliano}}, \bibinfo {author} {\bibfnamefont {G.}~\bibnamefont {Ecker}}, \
  and\ \bibinfo {author} {\bibfnamefont {H.}~\bibnamefont {Neufeld}},\ }\href
  {\doibase 10.1088/1126-6708/2002/08/002} {\bibfield  {journal} {\bibinfo
  {journal} {JHEP}\ }\textbf {\bibinfo {volume} {08}},\ \bibinfo {pages} {002}
  (\bibinfo {year} {2002})},\ \Eprint {http://arxiv.org/abs/hep-ph/0207310}
  {arXiv:hep-ph/0207310} \BibitemShut {NoStop}%
\bibitem [{\citenamefont {Flores-Ba{\'e}z}\ \emph {et~al.}(2006)\citenamefont
  {Flores-Ba{\'e}z}, \citenamefont {Flores-Tlalpa}, \citenamefont
  {L{\'o}pez~Castro},\ and\ \citenamefont
  {Toledo~S{\'a}nchez}}]{Flores-Baez:2006yiq}%
  \BibitemOpen
  \bibfield  {author} {\bibinfo {author} {\bibfnamefont {F.}~\bibnamefont
  {Flores-Ba{\'e}z}}, \bibinfo {author} {\bibfnamefont {A.}~\bibnamefont
  {Flores-Tlalpa}}, \bibinfo {author} {\bibfnamefont {G.}~\bibnamefont
  {L{\'o}pez~Castro}}, \ and\ \bibinfo {author} {\bibfnamefont
  {G.}~\bibnamefont {Toledo~S{\'a}nchez}},\ }\href {\doibase
  10.1103/PhysRevD.74.071301} {\bibfield  {journal} {\bibinfo  {journal} {Phys.
  Rev. D}\ }\textbf {\bibinfo {volume} {74}},\ \bibinfo {pages} {071301}
  (\bibinfo {year} {2006})},\ \Eprint {http://arxiv.org/abs/hep-ph/0608084}
  {arXiv:hep-ph/0608084} \BibitemShut {NoStop}%
\bibitem [{\citenamefont {Davier}\ \emph {et~al.}(2010)\citenamefont {Davier},
  \citenamefont {Hoecker}, \citenamefont {L{\'o}pez~Castro}, \citenamefont
  {Malaescu}, \citenamefont {Mo}, \citenamefont {Toledo~S{\'a}nchez},
  \citenamefont {Wang}, \citenamefont {Yuan},\ and\ \citenamefont
  {Zhang}}]{Davier:2010fmf}%
  \BibitemOpen
  \bibfield  {author} {\bibinfo {author} {\bibfnamefont {M.}~\bibnamefont
  {Davier}}, \bibinfo {author} {\bibfnamefont {A.}~\bibnamefont {Hoecker}},
  \bibinfo {author} {\bibfnamefont {G.}~\bibnamefont {L{\'o}pez~Castro}},
  \bibinfo {author} {\bibfnamefont {B.}~\bibnamefont {Malaescu}}, \bibinfo
  {author} {\bibfnamefont {X.~H.}\ \bibnamefont {Mo}}, \bibinfo {author}
  {\bibfnamefont {G.}~\bibnamefont {Toledo~S{\'a}nchez}}, \bibinfo {author}
  {\bibfnamefont {P.}~\bibnamefont {Wang}}, \bibinfo {author} {\bibfnamefont
  {C.~Z.}\ \bibnamefont {Yuan}}, \ and\ \bibinfo {author} {\bibfnamefont
  {Z.}~\bibnamefont {Zhang}},\ }\href {\doibase 10.1140/epjc/s10052-009-1219-4}
  {\bibfield  {journal} {\bibinfo  {journal} {Eur. Phys. J. C}\ }\textbf
  {\bibinfo {volume} {66}},\ \bibinfo {pages} {127} (\bibinfo {year} {2010})},\
  \Eprint {http://arxiv.org/abs/0906.5443} {arXiv:0906.5443 [hep-ph]}
  \BibitemShut {NoStop}%
\bibitem [{\citenamefont {Miranda}\ and\ \citenamefont
  {Roig}(2020)}]{Miranda:2020wdg}%
  \BibitemOpen
  \bibfield  {author} {\bibinfo {author} {\bibfnamefont {J.~A.}\ \bibnamefont
  {Miranda}}\ and\ \bibinfo {author} {\bibfnamefont {P.}~\bibnamefont {Roig}},\
  }\href {\doibase 10.1103/PhysRevD.102.114017} {\bibfield  {journal} {\bibinfo
   {journal} {Phys. Rev. D}\ }\textbf {\bibinfo {volume} {102}},\ \bibinfo
  {pages} {114017} (\bibinfo {year} {2020})},\ \Eprint
  {http://arxiv.org/abs/2007.11019} {arXiv:2007.11019 [hep-ph]} \BibitemShut
  {NoStop}%
\bibitem [{\citenamefont {Castro}\ \emph {et~al.}(2025)\citenamefont {Castro},
  \citenamefont {Miranda},\ and\ \citenamefont {Roig}}]{Castro:2024prg}%
  \BibitemOpen
  \bibfield  {author} {\bibinfo {author} {\bibfnamefont {G.~L.}\ \bibnamefont
  {Castro}}, \bibinfo {author} {\bibfnamefont {A.}~\bibnamefont {Miranda}}, \
  and\ \bibinfo {author} {\bibfnamefont {P.}~\bibnamefont {Roig}},\ }\href
  {\doibase 10.1103/PhysRevD.111.073004} {\bibfield  {journal} {\bibinfo
  {journal} {Phys. Rev. D}\ }\textbf {\bibinfo {volume} {111}},\ \bibinfo
  {pages} {073004} (\bibinfo {year} {2025})},\ \Eprint
  {http://arxiv.org/abs/2411.07696} {arXiv:2411.07696 [hep-ph]} \BibitemShut
  {NoStop}%
\bibitem [{\citenamefont {Aoyama}\ \emph {et~al.}(2020)\citenamefont {Aoyama}
  \emph {et~al.}}]{Aoyama:2020ynm}%
  \BibitemOpen
  \bibfield  {author} {\bibinfo {author} {\bibfnamefont {T.}~\bibnamefont
  {Aoyama}} \emph {et~al.},\ }\href {\doibase 10.1016/j.physrep.2020.07.006}
  {\bibfield  {journal} {\bibinfo  {journal} {Phys. Rept.}\ }\textbf {\bibinfo
  {volume} {887}},\ \bibinfo {pages} {1} (\bibinfo {year} {2020})},\ \Eprint
  {http://arxiv.org/abs/2006.04822} {arXiv:2006.04822 [hep-ph]} \BibitemShut
  {NoStop}%
\bibitem [{\citenamefont {Aliberti}\ \emph {et~al.}(2025)\citenamefont
  {Aliberti} \emph {et~al.}}]{Aliberti:2025beg}%
  \BibitemOpen
  \bibfield  {author} {\bibinfo {author} {\bibfnamefont {R.}~\bibnamefont
  {Aliberti}} \emph {et~al.},\ }\href {\doibase 10.1016/j.physrep.2025.08.002}
  {\bibfield  {journal} {\bibinfo  {journal} {Phys. Rept.}\ }\textbf {\bibinfo
  {volume} {1143}},\ \bibinfo {pages} {1} (\bibinfo {year} {2025})},\ \Eprint
  {http://arxiv.org/abs/2505.21476} {arXiv:2505.21476 [hep-ph]} \BibitemShut
  {NoStop}%
\bibitem [{\citenamefont {Hertzog}\ and\ \citenamefont
  {Hoferichter}(2025)}]{Hertzog:2025ssc}%
  \BibitemOpen
  \bibfield  {author} {\bibinfo {author} {\bibfnamefont {D.~W.}\ \bibnamefont
  {Hertzog}}\ and\ \bibinfo {author} {\bibfnamefont {M.}~\bibnamefont
  {Hoferichter}},\ }\href@noop {} {\  (\bibinfo {year} {2025})},\ \Eprint
  {http://arxiv.org/abs/2512.16980} {arXiv:2512.16980 [hep-ph]} \BibitemShut
  {NoStop}%
\bibitem [{\citenamefont {Davier}\ \emph {et~al.}(2024)\citenamefont {Davier},
  \citenamefont {Hoecker}, \citenamefont {Lutz}, \citenamefont {Malaescu},\
  and\ \citenamefont {Zhang}}]{Davier:2023fpl}%
  \BibitemOpen
  \bibfield  {author} {\bibinfo {author} {\bibfnamefont {M.}~\bibnamefont
  {Davier}}, \bibinfo {author} {\bibfnamefont {A.}~\bibnamefont {Hoecker}},
  \bibinfo {author} {\bibfnamefont {A.-M.}\ \bibnamefont {Lutz}}, \bibinfo
  {author} {\bibfnamefont {B.}~\bibnamefont {Malaescu}}, \ and\ \bibinfo
  {author} {\bibfnamefont {Z.}~\bibnamefont {Zhang}},\ }\href {\doibase
  10.1140/epjc/s10052-024-12964-7} {\bibfield  {journal} {\bibinfo  {journal}
  {Eur. Phys. J. C}\ }\textbf {\bibinfo {volume} {84}},\ \bibinfo {pages} {721}
  (\bibinfo {year} {2024})},\ \Eprint {http://arxiv.org/abs/2312.02053}
  {arXiv:2312.02053 [hep-ph]} \BibitemShut {NoStop}%
\bibitem [{\citenamefont {Colangelo}\ \emph
  {et~al.}(2022{\natexlab{a}})\citenamefont {Colangelo}, \citenamefont
  {Hoferichter}, \citenamefont {Kubis},\ and\ \citenamefont
  {Stoffer}}]{Colangelo:2022prz}%
  \BibitemOpen
  \bibfield  {author} {\bibinfo {author} {\bibfnamefont {G.}~\bibnamefont
  {Colangelo}}, \bibinfo {author} {\bibfnamefont {M.}~\bibnamefont
  {Hoferichter}}, \bibinfo {author} {\bibfnamefont {B.}~\bibnamefont {Kubis}},
  \ and\ \bibinfo {author} {\bibfnamefont {P.}~\bibnamefont {Stoffer}},\ }\href
  {\doibase 10.1007/JHEP10(2022)032} {\bibfield  {journal} {\bibinfo  {journal}
  {JHEP}\ }\textbf {\bibinfo {volume} {10}},\ \bibinfo {pages} {032} (\bibinfo
  {year} {2022}{\natexlab{a}})},\ \Eprint {http://arxiv.org/abs/2208.08993}
  {arXiv:2208.08993 [hep-ph]} \BibitemShut {NoStop}%
\bibitem [{\citenamefont {Hoferichter}\ \emph {et~al.}(2023)\citenamefont
  {Hoferichter}, \citenamefont {Colangelo}, \citenamefont {Hoid}, \citenamefont
  {Kubis}, \citenamefont {Ruiz~de Elvira}, \citenamefont {Schuh}, \citenamefont
  {Stamen},\ and\ \citenamefont {Stoffer}}]{Hoferichter:2023sli}%
  \BibitemOpen
  \bibfield  {author} {\bibinfo {author} {\bibfnamefont {M.}~\bibnamefont
  {Hoferichter}}, \bibinfo {author} {\bibfnamefont {G.}~\bibnamefont
  {Colangelo}}, \bibinfo {author} {\bibfnamefont {B.-L.}\ \bibnamefont {Hoid}},
  \bibinfo {author} {\bibfnamefont {B.}~\bibnamefont {Kubis}}, \bibinfo
  {author} {\bibfnamefont {J.}~\bibnamefont {Ruiz~de Elvira}}, \bibinfo
  {author} {\bibfnamefont {D.}~\bibnamefont {Schuh}}, \bibinfo {author}
  {\bibfnamefont {D.}~\bibnamefont {Stamen}}, \ and\ \bibinfo {author}
  {\bibfnamefont {P.}~\bibnamefont {Stoffer}},\ }\href {\doibase
  10.1103/PhysRevLett.131.161905} {\bibfield  {journal} {\bibinfo  {journal}
  {Phys. Rev. Lett.}\ }\textbf {\bibinfo {volume} {131}},\ \bibinfo {pages}
  {161905} (\bibinfo {year} {2023})},\ \Eprint
  {http://arxiv.org/abs/2307.02532} {arXiv:2307.02532 [hep-ph]} \BibitemShut
  {NoStop}%
\bibitem [{\citenamefont {Colangelo}\ \emph
  {et~al.}(2026{\natexlab{a}})\citenamefont {Colangelo}, \citenamefont
  {Cottini}, \citenamefont {Hoferichter},\ and\ \citenamefont
  {Holz}}]{Colangelo:2025iad}%
  \BibitemOpen
  \bibfield  {author} {\bibinfo {author} {\bibfnamefont {G.}~\bibnamefont
  {Colangelo}}, \bibinfo {author} {\bibfnamefont {M.}~\bibnamefont {Cottini}},
  \bibinfo {author} {\bibfnamefont {M.}~\bibnamefont {Hoferichter}}, \ and\
  \bibinfo {author} {\bibfnamefont {S.}~\bibnamefont {Holz}},\ }\href {\doibase
  10.1103/ryk1-x6v1} {\bibfield  {journal} {\bibinfo  {journal} {Phys. Rev.
  Lett.}\ }\textbf {\bibinfo {volume} {136}},\ \bibinfo {pages} {101903}
  (\bibinfo {year} {2026}{\natexlab{a}})},\ \Eprint
  {http://arxiv.org/abs/2510.26871} {arXiv:2510.26871 [hep-ph]} \BibitemShut
  {NoStop}%
\bibitem [{\citenamefont {Colangelo}\ \emph
  {et~al.}(2026{\natexlab{b}})\citenamefont {Colangelo}, \citenamefont
  {Cottini}, \citenamefont {Hoferichter},\ and\ \citenamefont
  {Holz}}]{Colangelo:2025ivq}%
  \BibitemOpen
  \bibfield  {author} {\bibinfo {author} {\bibfnamefont {G.}~\bibnamefont
  {Colangelo}}, \bibinfo {author} {\bibfnamefont {M.}~\bibnamefont {Cottini}},
  \bibinfo {author} {\bibfnamefont {M.}~\bibnamefont {Hoferichter}}, \ and\
  \bibinfo {author} {\bibfnamefont {S.}~\bibnamefont {Holz}},\ }\href {\doibase
  10.1007/JHEP02(2026)181} {\bibfield  {journal} {\bibinfo  {journal} {JHEP}\
  }\textbf {\bibinfo {volume} {02}},\ \bibinfo {pages} {181} (\bibinfo {year}
  {2026}{\natexlab{b}})},\ \Eprint {http://arxiv.org/abs/2511.07507}
  {arXiv:2511.07507 [hep-ph]} \BibitemShut {NoStop}%
\bibitem [{\citenamefont {Tishchenko}\ \emph {et~al.}(2013)\citenamefont
  {Tishchenko} \emph {et~al.}}]{MuLan:2012sih}%
  \BibitemOpen
  \bibfield  {author} {\bibinfo {author} {\bibfnamefont {V.}~\bibnamefont
  {Tishchenko}} \emph {et~al.} (\bibinfo {collaboration} {MuLan}),\ }\href
  {\doibase 10.1103/PhysRevD.87.052003} {\bibfield  {journal} {\bibinfo
  {journal} {Phys. Rev. D}\ }\textbf {\bibinfo {volume} {87}},\ \bibinfo
  {pages} {052003} (\bibinfo {year} {2013})},\ \Eprint
  {http://arxiv.org/abs/1211.0960} {arXiv:1211.0960 [hep-ex]} \BibitemShut
  {NoStop}%
\bibitem [{\citenamefont {Brod}\ and\ \citenamefont
  {Gorbahn}(2008)}]{Brod:2008ss}%
  \BibitemOpen
  \bibfield  {author} {\bibinfo {author} {\bibfnamefont {J.}~\bibnamefont
  {Brod}}\ and\ \bibinfo {author} {\bibfnamefont {M.}~\bibnamefont {Gorbahn}},\
  }\href {\doibase 10.1103/PhysRevD.78.034006} {\bibfield  {journal} {\bibinfo
  {journal} {Phys. Rev. D}\ }\textbf {\bibinfo {volume} {78}},\ \bibinfo
  {pages} {034006} (\bibinfo {year} {2008})},\ \Eprint
  {http://arxiv.org/abs/0805.4119} {arXiv:0805.4119 [hep-ph]} \BibitemShut
  {NoStop}%
\bibitem [{\citenamefont {Dekens}\ and\ \citenamefont
  {Stoffer}(2019)}]{Dekens:2019ept}%
  \BibitemOpen
  \bibfield  {author} {\bibinfo {author} {\bibfnamefont {W.}~\bibnamefont
  {Dekens}}\ and\ \bibinfo {author} {\bibfnamefont {P.}~\bibnamefont
  {Stoffer}},\ }\href {\doibase 10.1007/JHEP10(2019)197} {\bibfield  {journal}
  {\bibinfo  {journal} {JHEP}\ }\textbf {\bibinfo {volume} {10}},\ \bibinfo
  {pages} {197} (\bibinfo {year} {2019})},\ \bibinfo {note} {[Erratum: JHEP
  {\bf 11}, 148 (2022)]},\ \Eprint {http://arxiv.org/abs/1908.05295}
  {arXiv:1908.05295 [hep-ph]} \BibitemShut {NoStop}%
\bibitem [{\citenamefont {Hill}\ and\ \citenamefont
  {Tomalak}(2020)}]{Hill:2019xqk}%
  \BibitemOpen
  \bibfield  {author} {\bibinfo {author} {\bibfnamefont {R.~J.}\ \bibnamefont
  {Hill}}\ and\ \bibinfo {author} {\bibfnamefont {O.}~\bibnamefont {Tomalak}},\
  }\href {\doibase 10.1016/j.physletb.2020.135466} {\bibfield  {journal}
  {\bibinfo  {journal} {Phys. Lett. B}\ }\textbf {\bibinfo {volume} {805}},\
  \bibinfo {pages} {135466} (\bibinfo {year} {2020})},\ \Eprint
  {http://arxiv.org/abs/1911.01493} {arXiv:1911.01493 [hep-ph]} \BibitemShut
  {NoStop}%
\bibitem [{Sup()}]{Supp}%
  \BibitemOpen
  \href@noop {} {}\bibinfo {note} {See Supplemental Material for details of the
  matching with the spurion method, at the amplitude level, and the RG to NLL,
  including
  Refs.~\cite{Moussallam:1997xx,Buras:1989xd,FlavourLatticeAveragingGroupFLAG:2024oxs,Maltman:2008bx,PACS-CS:2009zxm,McNeile:2010ji,Chakraborty:2014aca,Bruno:2017gxd,Bazavov:2019qoo,Cali:2020hrj,Ayala:2020odx,Petreczky:2020tky,DallaBrida:2022eua,Parker:2018vye,Morel:2020dww,Fan:2022eto,Sturm:2013uka,Davier:2019can,Keshavarzi:2019abf,Ce:2022eix,Erler:2023hyi,Conigli:2025qvh,Chetyrkin:1996cf,Fanchiotti:1992tu}.}\BibitemShut
  {Stop}%
\bibitem [{\citenamefont {Moussallam}(1997)}]{Moussallam:1997xx}%
  \BibitemOpen
  \bibfield  {author} {\bibinfo {author} {\bibfnamefont {B.}~\bibnamefont
  {Moussallam}},\ }\href {\doibase 10.1016/S0550-3213(97)00464-1} {\bibfield
  {journal} {\bibinfo  {journal} {Nucl. Phys. B}\ }\textbf {\bibinfo {volume}
  {504}},\ \bibinfo {pages} {381} (\bibinfo {year} {1997})},\ \Eprint
  {http://arxiv.org/abs/hep-ph/9701400} {arXiv:hep-ph/9701400} \BibitemShut
  {NoStop}%
\bibitem [{\citenamefont {Buras}\ and\ \citenamefont
  {Weisz}(1990)}]{Buras:1989xd}%
  \BibitemOpen
  \bibfield  {author} {\bibinfo {author} {\bibfnamefont {A.~J.}\ \bibnamefont
  {Buras}}\ and\ \bibinfo {author} {\bibfnamefont {P.~H.}\ \bibnamefont
  {Weisz}},\ }\href {\doibase 10.1016/0550-3213(90)90223-Z} {\bibfield
  {journal} {\bibinfo  {journal} {Nucl. Phys. B}\ }\textbf {\bibinfo {volume}
  {333}},\ \bibinfo {pages} {66} (\bibinfo {year} {1990})}\BibitemShut
  {NoStop}%
\bibitem [{\citenamefont {Aoki}\ \emph {et~al.}(2026)\citenamefont {Aoki} \emph
  {et~al.}}]{FlavourLatticeAveragingGroupFLAG:2024oxs}%
  \BibitemOpen
  \bibfield  {author} {\bibinfo {author} {\bibfnamefont {Y.}~\bibnamefont
  {Aoki}} \emph {et~al.} (\bibinfo {collaboration} {FLAG}),\ }\href {\doibase
  10.1103/nfzp-p5dn} {\bibfield  {journal} {\bibinfo  {journal} {Phys. Rev. D}\
  }\textbf {\bibinfo {volume} {113}},\ \bibinfo {pages} {014508} (\bibinfo
  {year} {2026})},\ \Eprint {http://arxiv.org/abs/2411.04268} {arXiv:2411.04268
  [hep-lat]} \BibitemShut {NoStop}%
\bibitem [{\citenamefont {Maltman}\ \emph {et~al.}(2008)\citenamefont
  {Maltman}, \citenamefont {Leinweber}, \citenamefont {Moran},\ and\
  \citenamefont {Sternbeck}}]{Maltman:2008bx}%
  \BibitemOpen
  \bibfield  {author} {\bibinfo {author} {\bibfnamefont {K.}~\bibnamefont
  {Maltman}}, \bibinfo {author} {\bibfnamefont {D.}~\bibnamefont {Leinweber}},
  \bibinfo {author} {\bibfnamefont {P.}~\bibnamefont {Moran}}, \ and\ \bibinfo
  {author} {\bibfnamefont {A.}~\bibnamefont {Sternbeck}},\ }\href {\doibase
  10.1103/PhysRevD.78.114504} {\bibfield  {journal} {\bibinfo  {journal} {Phys.
  Rev. D}\ }\textbf {\bibinfo {volume} {78}},\ \bibinfo {pages} {114504}
  (\bibinfo {year} {2008})},\ \Eprint {http://arxiv.org/abs/0807.2020}
  {arXiv:0807.2020 [hep-lat]} \BibitemShut {NoStop}%
\bibitem [{\citenamefont {Aoki}\ \emph {et~al.}(2009)\citenamefont {Aoki} \emph
  {et~al.}}]{PACS-CS:2009zxm}%
  \BibitemOpen
  \bibfield  {author} {\bibinfo {author} {\bibfnamefont {S.}~\bibnamefont
  {Aoki}} \emph {et~al.} (\bibinfo {collaboration} {PACS-CS}),\ }\href
  {\doibase 10.1088/1126-6708/2009/10/053} {\bibfield  {journal} {\bibinfo
  {journal} {JHEP}\ }\textbf {\bibinfo {volume} {10}},\ \bibinfo {pages} {053}
  (\bibinfo {year} {2009})},\ \Eprint {http://arxiv.org/abs/0906.3906}
  {arXiv:0906.3906 [hep-lat]} \BibitemShut {NoStop}%
\bibitem [{\citenamefont {McNeile}\ \emph {et~al.}(2010)\citenamefont
  {McNeile}, \citenamefont {Davies}, \citenamefont {Follana}, \citenamefont
  {Hornbostel},\ and\ \citenamefont {Lepage}}]{McNeile:2010ji}%
  \BibitemOpen
  \bibfield  {author} {\bibinfo {author} {\bibfnamefont {C.}~\bibnamefont
  {McNeile}}, \bibinfo {author} {\bibfnamefont {C.~T.~H.}\ \bibnamefont
  {Davies}}, \bibinfo {author} {\bibfnamefont {E.}~\bibnamefont {Follana}},
  \bibinfo {author} {\bibfnamefont {K.}~\bibnamefont {Hornbostel}}, \ and\
  \bibinfo {author} {\bibfnamefont {G.~P.}\ \bibnamefont {Lepage}},\ }\href
  {\doibase 10.1103/PhysRevD.82.034512} {\bibfield  {journal} {\bibinfo
  {journal} {Phys. Rev. D}\ }\textbf {\bibinfo {volume} {82}},\ \bibinfo
  {pages} {034512} (\bibinfo {year} {2010})},\ \Eprint
  {http://arxiv.org/abs/1004.4285} {arXiv:1004.4285 [hep-lat]} \BibitemShut
  {NoStop}%
\bibitem [{\citenamefont {Chakraborty}\ \emph {et~al.}(2015)\citenamefont
  {Chakraborty}, \citenamefont {Davies}, \citenamefont {Galloway},
  \citenamefont {Knecht}, \citenamefont {Koponen}, \citenamefont {Donald},
  \citenamefont {Dowdall}, \citenamefont {Lepage},\ and\ \citenamefont
  {McNeile}}]{Chakraborty:2014aca}%
  \BibitemOpen
  \bibfield  {author} {\bibinfo {author} {\bibfnamefont {B.}~\bibnamefont
  {Chakraborty}}, \bibinfo {author} {\bibfnamefont {C.~T.~H.}\ \bibnamefont
  {Davies}}, \bibinfo {author} {\bibfnamefont {B.}~\bibnamefont {Galloway}},
  \bibinfo {author} {\bibfnamefont {P.}~\bibnamefont {Knecht}}, \bibinfo
  {author} {\bibfnamefont {J.}~\bibnamefont {Koponen}}, \bibinfo {author}
  {\bibfnamefont {G.~C.}\ \bibnamefont {Donald}}, \bibinfo {author}
  {\bibfnamefont {R.~J.}\ \bibnamefont {Dowdall}}, \bibinfo {author}
  {\bibfnamefont {G.~P.}\ \bibnamefont {Lepage}}, \ and\ \bibinfo {author}
  {\bibfnamefont {C.}~\bibnamefont {McNeile}},\ }\href {\doibase
  10.1103/PhysRevD.91.054508} {\bibfield  {journal} {\bibinfo  {journal} {Phys.
  Rev. D}\ }\textbf {\bibinfo {volume} {91}},\ \bibinfo {pages} {054508}
  (\bibinfo {year} {2015})},\ \Eprint {http://arxiv.org/abs/1408.4169}
  {arXiv:1408.4169 [hep-lat]} \BibitemShut {NoStop}%
\bibitem [{\citenamefont {Bruno}\ \emph {et~al.}(2017)\citenamefont {Bruno},
  \citenamefont {Dalla~Brida}, \citenamefont {Fritzsch}, \citenamefont
  {Korzec}, \citenamefont {Ramos}, \citenamefont {Schaefer}, \citenamefont
  {Simma}, \citenamefont {Sint},\ and\ \citenamefont {Sommer}}]{Bruno:2017gxd}%
  \BibitemOpen
  \bibfield  {author} {\bibinfo {author} {\bibfnamefont {M.}~\bibnamefont
  {Bruno}}, \bibinfo {author} {\bibfnamefont {M.}~\bibnamefont {Dalla~Brida}},
  \bibinfo {author} {\bibfnamefont {P.}~\bibnamefont {Fritzsch}}, \bibinfo
  {author} {\bibfnamefont {T.}~\bibnamefont {Korzec}}, \bibinfo {author}
  {\bibfnamefont {A.}~\bibnamefont {Ramos}}, \bibinfo {author} {\bibfnamefont
  {S.}~\bibnamefont {Schaefer}}, \bibinfo {author} {\bibfnamefont
  {H.}~\bibnamefont {Simma}}, \bibinfo {author} {\bibfnamefont
  {S.}~\bibnamefont {Sint}}, \ and\ \bibinfo {author} {\bibfnamefont
  {R.}~\bibnamefont {Sommer}} (\bibinfo {collaboration} {ALPHA}),\ }\href
  {\doibase 10.1103/PhysRevLett.119.102001} {\bibfield  {journal} {\bibinfo
  {journal} {Phys. Rev. Lett.}\ }\textbf {\bibinfo {volume} {119}},\ \bibinfo
  {pages} {102001} (\bibinfo {year} {2017})},\ \Eprint
  {http://arxiv.org/abs/1706.03821} {arXiv:1706.03821 [hep-lat]} \BibitemShut
  {NoStop}%
\bibitem [{\citenamefont {Bazavov}\ \emph {et~al.}(2019)\citenamefont
  {Bazavov}, \citenamefont {Brambilla}, \citenamefont {Garcia~i Tormo},
  \citenamefont {Petreczky}, \citenamefont {Soto}, \citenamefont {Vairo},\ and\
  \citenamefont {Weber}}]{Bazavov:2019qoo}%
  \BibitemOpen
  \bibfield  {author} {\bibinfo {author} {\bibfnamefont {A.}~\bibnamefont
  {Bazavov}}, \bibinfo {author} {\bibfnamefont {N.}~\bibnamefont {Brambilla}},
  \bibinfo {author} {\bibfnamefont {X.}~\bibnamefont {Garcia~i Tormo}},
  \bibinfo {author} {\bibfnamefont {P.}~\bibnamefont {Petreczky}}, \bibinfo
  {author} {\bibfnamefont {J.}~\bibnamefont {Soto}}, \bibinfo {author}
  {\bibfnamefont {A.}~\bibnamefont {Vairo}}, \ and\ \bibinfo {author}
  {\bibfnamefont {J.~H.}\ \bibnamefont {Weber}} (\bibinfo {collaboration}
  {TUMQCD}),\ }\href {\doibase 10.1103/PhysRevD.100.114511} {\bibfield
  {journal} {\bibinfo  {journal} {Phys. Rev. D}\ }\textbf {\bibinfo {volume}
  {100}},\ \bibinfo {pages} {114511} (\bibinfo {year} {2019})},\ \Eprint
  {http://arxiv.org/abs/1907.11747} {arXiv:1907.11747 [hep-lat]} \BibitemShut
  {NoStop}%
\bibitem [{\citenamefont {Cali}\ \emph {et~al.}(2020)\citenamefont {Cali},
  \citenamefont {Cichy}, \citenamefont {Korcyl},\ and\ \citenamefont
  {Simeth}}]{Cali:2020hrj}%
  \BibitemOpen
  \bibfield  {author} {\bibinfo {author} {\bibfnamefont {S.}~\bibnamefont
  {Cali}}, \bibinfo {author} {\bibfnamefont {K.}~\bibnamefont {Cichy}},
  \bibinfo {author} {\bibfnamefont {P.}~\bibnamefont {Korcyl}}, \ and\ \bibinfo
  {author} {\bibfnamefont {J.}~\bibnamefont {Simeth}},\ }\href {\doibase
  10.1103/PhysRevLett.125.242002} {\bibfield  {journal} {\bibinfo  {journal}
  {Phys. Rev. Lett.}\ }\textbf {\bibinfo {volume} {125}},\ \bibinfo {pages}
  {242002} (\bibinfo {year} {2020})},\ \Eprint
  {http://arxiv.org/abs/2003.05781} {arXiv:2003.05781 [hep-lat]} \BibitemShut
  {NoStop}%
\bibitem [{\citenamefont {Ayala}\ \emph {et~al.}(2020)\citenamefont {Ayala},
  \citenamefont {Lobregat},\ and\ \citenamefont {Pineda}}]{Ayala:2020odx}%
  \BibitemOpen
  \bibfield  {author} {\bibinfo {author} {\bibfnamefont {C.}~\bibnamefont
  {Ayala}}, \bibinfo {author} {\bibfnamefont {X.}~\bibnamefont {Lobregat}}, \
  and\ \bibinfo {author} {\bibfnamefont {A.}~\bibnamefont {Pineda}},\ }\href
  {\doibase 10.1007/JHEP09(2020)016} {\bibfield  {journal} {\bibinfo  {journal}
  {JHEP}\ }\textbf {\bibinfo {volume} {09}},\ \bibinfo {pages} {016} (\bibinfo
  {year} {2020})},\ \Eprint {http://arxiv.org/abs/2005.12301} {arXiv:2005.12301
  [hep-ph]} \BibitemShut {NoStop}%
\bibitem [{\citenamefont {Petreczky}\ and\ \citenamefont
  {Weber}(2022)}]{Petreczky:2020tky}%
  \BibitemOpen
  \bibfield  {author} {\bibinfo {author} {\bibfnamefont {P.}~\bibnamefont
  {Petreczky}}\ and\ \bibinfo {author} {\bibfnamefont {J.~H.}\ \bibnamefont
  {Weber}},\ }\href {\doibase 10.1140/epjc/s10052-022-09998-0} {\bibfield
  {journal} {\bibinfo  {journal} {Eur. Phys. J. C}\ }\textbf {\bibinfo {volume}
  {82}},\ \bibinfo {pages} {64} (\bibinfo {year} {2022})},\ \Eprint
  {http://arxiv.org/abs/2012.06193} {arXiv:2012.06193 [hep-lat]} \BibitemShut
  {NoStop}%
\bibitem [{\citenamefont {Dalla~Brida}\ \emph {et~al.}(2022)\citenamefont
  {Dalla~Brida}, \citenamefont {H{\"o}llwieser}, \citenamefont {Knechtli},
  \citenamefont {Korzec}, \citenamefont {Nada}, \citenamefont {Ramos},
  \citenamefont {Sint},\ and\ \citenamefont {Sommer}}]{DallaBrida:2022eua}%
  \BibitemOpen
  \bibfield  {author} {\bibinfo {author} {\bibfnamefont {M.}~\bibnamefont
  {Dalla~Brida}}, \bibinfo {author} {\bibfnamefont {R.}~\bibnamefont
  {H{\"o}llwieser}}, \bibinfo {author} {\bibfnamefont {F.}~\bibnamefont
  {Knechtli}}, \bibinfo {author} {\bibfnamefont {T.}~\bibnamefont {Korzec}},
  \bibinfo {author} {\bibfnamefont {A.}~\bibnamefont {Nada}}, \bibinfo {author}
  {\bibfnamefont {A.}~\bibnamefont {Ramos}}, \bibinfo {author} {\bibfnamefont
  {S.}~\bibnamefont {Sint}}, \ and\ \bibinfo {author} {\bibfnamefont
  {R.}~\bibnamefont {Sommer}} (\bibinfo {collaboration} {ALPHA}),\ }\href
  {\doibase 10.1140/epjc/s10052-022-10998-3} {\bibfield  {journal} {\bibinfo
  {journal} {Eur. Phys. J. C}\ }\textbf {\bibinfo {volume} {82}},\ \bibinfo
  {pages} {1092} (\bibinfo {year} {2022})},\ \Eprint
  {http://arxiv.org/abs/2209.14204} {arXiv:2209.14204 [hep-lat]} \BibitemShut
  {NoStop}%
\bibitem [{\citenamefont {Parker}\ \emph {et~al.}(2018)\citenamefont {Parker},
  \citenamefont {Yu}, \citenamefont {Zhong}, \citenamefont {Estey},\ and\
  \citenamefont {M\"uller}}]{Parker:2018vye}%
  \BibitemOpen
  \bibfield  {author} {\bibinfo {author} {\bibfnamefont {R.~H.}\ \bibnamefont
  {Parker}}, \bibinfo {author} {\bibfnamefont {C.}~\bibnamefont {Yu}}, \bibinfo
  {author} {\bibfnamefont {W.}~\bibnamefont {Zhong}}, \bibinfo {author}
  {\bibfnamefont {B.}~\bibnamefont {Estey}}, \ and\ \bibinfo {author}
  {\bibfnamefont {H.}~\bibnamefont {M\"uller}},\ }\href {\doibase
  10.1126/science.aap7706} {\bibfield  {journal} {\bibinfo  {journal}
  {Science}\ }\textbf {\bibinfo {volume} {360}},\ \bibinfo {pages} {191}
  (\bibinfo {year} {2018})},\ \Eprint {http://arxiv.org/abs/1812.04130}
  {arXiv:1812.04130 [physics.atom-ph]} \BibitemShut {NoStop}%
\bibitem [{\citenamefont {Morel}\ \emph {et~al.}(2020)\citenamefont {Morel},
  \citenamefont {Yao}, \citenamefont {Clad\'e},\ and\ \citenamefont
  {Guellati-Kh\'elifa}}]{Morel:2020dww}%
  \BibitemOpen
  \bibfield  {author} {\bibinfo {author} {\bibfnamefont {L.}~\bibnamefont
  {Morel}}, \bibinfo {author} {\bibfnamefont {Z.}~\bibnamefont {Yao}}, \bibinfo
  {author} {\bibfnamefont {P.}~\bibnamefont {Clad\'e}}, \ and\ \bibinfo
  {author} {\bibfnamefont {S.}~\bibnamefont {Guellati-Kh\'elifa}},\ }\href
  {\doibase 10.1038/s41586-020-2964-7} {\bibfield  {journal} {\bibinfo
  {journal} {Nature}\ }\textbf {\bibinfo {volume} {588}},\ \bibinfo {pages}
  {61} (\bibinfo {year} {2020})}\BibitemShut {NoStop}%
\bibitem [{\citenamefont {Fan}\ \emph {et~al.}(2023)\citenamefont {Fan},
  \citenamefont {Myers}, \citenamefont {Sukra},\ and\ \citenamefont
  {Gabrielse}}]{Fan:2022eto}%
  \BibitemOpen
  \bibfield  {author} {\bibinfo {author} {\bibfnamefont {X.}~\bibnamefont
  {Fan}}, \bibinfo {author} {\bibfnamefont {T.~G.}\ \bibnamefont {Myers}},
  \bibinfo {author} {\bibfnamefont {B.~A.~D.}\ \bibnamefont {Sukra}}, \ and\
  \bibinfo {author} {\bibfnamefont {G.}~\bibnamefont {Gabrielse}},\ }\href
  {\doibase 10.1103/PhysRevLett.130.071801} {\bibfield  {journal} {\bibinfo
  {journal} {Phys. Rev. Lett.}\ }\textbf {\bibinfo {volume} {130}},\ \bibinfo
  {pages} {071801} (\bibinfo {year} {2023})},\ \Eprint
  {http://arxiv.org/abs/2209.13084} {arXiv:2209.13084 [physics.atom-ph]}
  \BibitemShut {NoStop}%
\bibitem [{\citenamefont {Sturm}(2013)}]{Sturm:2013uka}%
  \BibitemOpen
  \bibfield  {author} {\bibinfo {author} {\bibfnamefont {C.}~\bibnamefont
  {Sturm}},\ }\href {\doibase 10.1016/j.nuclphysb.2013.06.009} {\bibfield
  {journal} {\bibinfo  {journal} {Nucl. Phys. B}\ }\textbf {\bibinfo {volume}
  {874}},\ \bibinfo {pages} {698} (\bibinfo {year} {2013})},\ \Eprint
  {http://arxiv.org/abs/1305.0581} {arXiv:1305.0581 [hep-ph]} \BibitemShut
  {NoStop}%
\bibitem [{\citenamefont {Davier}\ \emph {et~al.}(2020)\citenamefont {Davier},
  \citenamefont {Hoecker}, \citenamefont {Malaescu},\ and\ \citenamefont
  {Zhang}}]{Davier:2019can}%
  \BibitemOpen
  \bibfield  {author} {\bibinfo {author} {\bibfnamefont {M.}~\bibnamefont
  {Davier}}, \bibinfo {author} {\bibfnamefont {A.}~\bibnamefont {Hoecker}},
  \bibinfo {author} {\bibfnamefont {B.}~\bibnamefont {Malaescu}}, \ and\
  \bibinfo {author} {\bibfnamefont {Z.}~\bibnamefont {Zhang}},\ }\href
  {\doibase 10.1140/epjc/s10052-020-7792-2} {\bibfield  {journal} {\bibinfo
  {journal} {Eur. Phys. J. C}\ }\textbf {\bibinfo {volume} {80}},\ \bibinfo
  {pages} {241} (\bibinfo {year} {2020})},\ \bibinfo {note} {[Erratum: Eur.
  Phys. J. C {\bf 80}, 410 (2020)]},\ \Eprint {http://arxiv.org/abs/1908.00921}
  {arXiv:1908.00921 [hep-ph]} \BibitemShut {NoStop}%
\bibitem [{\citenamefont {Keshavarzi}\ \emph {et~al.}(2020)\citenamefont
  {Keshavarzi}, \citenamefont {Nomura},\ and\ \citenamefont
  {Teubner}}]{Keshavarzi:2019abf}%
  \BibitemOpen
  \bibfield  {author} {\bibinfo {author} {\bibfnamefont {A.}~\bibnamefont
  {Keshavarzi}}, \bibinfo {author} {\bibfnamefont {D.}~\bibnamefont {Nomura}},
  \ and\ \bibinfo {author} {\bibfnamefont {T.}~\bibnamefont {Teubner}},\ }\href
  {\doibase 10.1103/PhysRevD.101.014029} {\bibfield  {journal} {\bibinfo
  {journal} {Phys. Rev. D}\ }\textbf {\bibinfo {volume} {101}},\ \bibinfo
  {pages} {014029} (\bibinfo {year} {2020})},\ \Eprint
  {http://arxiv.org/abs/1911.00367} {arXiv:1911.00367 [hep-ph]} \BibitemShut
  {NoStop}%
\bibitem [{\citenamefont {C\`e}\ \emph {et~al.}(2022)\citenamefont {C\`e} \emph
  {et~al.}}]{Ce:2022eix}%
  \BibitemOpen
  \bibfield  {author} {\bibinfo {author} {\bibfnamefont {M.}~\bibnamefont
  {C\`e}} \emph {et~al.},\ }\href {\doibase 10.1007/JHEP08(2022)220} {\bibfield
   {journal} {\bibinfo  {journal} {JHEP}\ }\textbf {\bibinfo {volume} {08}},\
  \bibinfo {pages} {220} (\bibinfo {year} {2022})},\ \Eprint
  {http://arxiv.org/abs/2203.08676} {arXiv:2203.08676 [hep-lat]} \BibitemShut
  {NoStop}%
\bibitem [{\citenamefont {Erler}\ and\ \citenamefont
  {Ferro-Hern{\'a}ndez}(2023)}]{Erler:2023hyi}%
  \BibitemOpen
  \bibfield  {author} {\bibinfo {author} {\bibfnamefont {J.}~\bibnamefont
  {Erler}}\ and\ \bibinfo {author} {\bibfnamefont {R.}~\bibnamefont
  {Ferro-Hern{\'a}ndez}},\ }\href {\doibase 10.1007/JHEP12(2023)131} {\bibfield
   {journal} {\bibinfo  {journal} {JHEP}\ }\textbf {\bibinfo {volume} {12}},\
  \bibinfo {pages} {131} (\bibinfo {year} {2023})},\ \Eprint
  {http://arxiv.org/abs/2308.05740} {arXiv:2308.05740 [hep-ph]} \BibitemShut
  {NoStop}%
\bibitem [{\citenamefont {Conigli}\ \emph {et~al.}(2025)\citenamefont
  {Conigli}, \citenamefont {Djukanovic}, \citenamefont {von Hippel},
  \citenamefont {Kuberski}, \citenamefont {Meyer}, \citenamefont {Miura},
  \citenamefont {Ottnad}, \citenamefont {Risch},\ and\ \citenamefont
  {Wittig}}]{Conigli:2025qvh}%
  \BibitemOpen
  \bibfield  {author} {\bibinfo {author} {\bibfnamefont {A.}~\bibnamefont
  {Conigli}}, \bibinfo {author} {\bibfnamefont {D.}~\bibnamefont {Djukanovic}},
  \bibinfo {author} {\bibfnamefont {G.}~\bibnamefont {von Hippel}}, \bibinfo
  {author} {\bibfnamefont {S.}~\bibnamefont {Kuberski}}, \bibinfo {author}
  {\bibfnamefont {H.~B.}\ \bibnamefont {Meyer}}, \bibinfo {author}
  {\bibfnamefont {K.}~\bibnamefont {Miura}}, \bibinfo {author} {\bibfnamefont
  {K.}~\bibnamefont {Ottnad}}, \bibinfo {author} {\bibfnamefont
  {A.}~\bibnamefont {Risch}}, \ and\ \bibinfo {author} {\bibfnamefont
  {H.}~\bibnamefont {Wittig}},\ }\href@noop {} {\  (\bibinfo {year} {2025})},\
  \Eprint {http://arxiv.org/abs/2511.01623} {arXiv:2511.01623 [hep-lat]}
  \BibitemShut {NoStop}%
\bibitem [{\citenamefont {Chetyrkin}\ \emph {et~al.}(1996)\citenamefont
  {Chetyrkin}, \citenamefont {K{\"u}hn},\ and\ \citenamefont
  {Steinhauser}}]{Chetyrkin:1996cf}%
  \BibitemOpen
  \bibfield  {author} {\bibinfo {author} {\bibfnamefont {K.~G.}\ \bibnamefont
  {Chetyrkin}}, \bibinfo {author} {\bibfnamefont {J.~H.}\ \bibnamefont
  {K{\"u}hn}}, \ and\ \bibinfo {author} {\bibfnamefont {M.}~\bibnamefont
  {Steinhauser}},\ }\href {\doibase 10.1016/S0550-3213(96)00534-2} {\bibfield
  {journal} {\bibinfo  {journal} {Nucl. Phys. B}\ }\textbf {\bibinfo {volume}
  {482}},\ \bibinfo {pages} {213} (\bibinfo {year} {1996})},\ \Eprint
  {http://arxiv.org/abs/hep-ph/9606230} {arXiv:hep-ph/9606230} \BibitemShut
  {NoStop}%
\bibitem [{\citenamefont {Fanchiotti}\ \emph {et~al.}(1993)\citenamefont
  {Fanchiotti}, \citenamefont {Kniehl},\ and\ \citenamefont
  {Sirlin}}]{Fanchiotti:1992tu}%
  \BibitemOpen
  \bibfield  {author} {\bibinfo {author} {\bibfnamefont {S.}~\bibnamefont
  {Fanchiotti}}, \bibinfo {author} {\bibfnamefont {B.~A.}\ \bibnamefont
  {Kniehl}}, \ and\ \bibinfo {author} {\bibfnamefont {A.}~\bibnamefont
  {Sirlin}},\ }\href {\doibase 10.1103/PhysRevD.48.307} {\bibfield  {journal}
  {\bibinfo  {journal} {Phys. Rev. D}\ }\textbf {\bibinfo {volume} {48}},\
  \bibinfo {pages} {307} (\bibinfo {year} {1993})},\ \Eprint
  {http://arxiv.org/abs/hep-ph/9212285} {arXiv:hep-ph/9212285} \BibitemShut
  {NoStop}%
\bibitem [{\citenamefont {Knecht}\ \emph {et~al.}(2000)\citenamefont {Knecht},
  \citenamefont {Neufeld}, \citenamefont {Rupertsberger},\ and\ \citenamefont
  {Talavera}}]{Knecht:1999ag}%
  \BibitemOpen
  \bibfield  {author} {\bibinfo {author} {\bibfnamefont {M.}~\bibnamefont
  {Knecht}}, \bibinfo {author} {\bibfnamefont {H.}~\bibnamefont {Neufeld}},
  \bibinfo {author} {\bibfnamefont {H.}~\bibnamefont {Rupertsberger}}, \ and\
  \bibinfo {author} {\bibfnamefont {P.}~\bibnamefont {Talavera}},\ }\href
  {\doibase 10.1007/s100529900265} {\bibfield  {journal} {\bibinfo  {journal}
  {Eur. Phys. J. C}\ }\textbf {\bibinfo {volume} {12}},\ \bibinfo {pages} {469}
  (\bibinfo {year} {2000})},\ \Eprint {http://arxiv.org/abs/hep-ph/9909284}
  {arXiv:hep-ph/9909284} \BibitemShut {NoStop}%
\bibitem [{\citenamefont {Urech}(1995)}]{Urech:1994hd}%
  \BibitemOpen
  \bibfield  {author} {\bibinfo {author} {\bibfnamefont {R.}~\bibnamefont
  {Urech}},\ }\href {\doibase 10.1016/0550-3213(95)90707-N} {\bibfield
  {journal} {\bibinfo  {journal} {Nucl. Phys. B}\ }\textbf {\bibinfo {volume}
  {433}},\ \bibinfo {pages} {234} (\bibinfo {year} {1995})},\ \Eprint
  {http://arxiv.org/abs/hep-ph/9405341} {arXiv:hep-ph/9405341} \BibitemShut
  {NoStop}%
\bibitem [{\citenamefont {Larin}\ and\ \citenamefont
  {Vermaseren}(1991)}]{Larin:1991tj}%
  \BibitemOpen
  \bibfield  {author} {\bibinfo {author} {\bibfnamefont {S.~A.}\ \bibnamefont
  {Larin}}\ and\ \bibinfo {author} {\bibfnamefont {J.~A.~M.}\ \bibnamefont
  {Vermaseren}},\ }\href {\doibase 10.1016/0370-2693(91)90839-I} {\bibfield
  {journal} {\bibinfo  {journal} {Phys. Lett. B}\ }\textbf {\bibinfo {volume}
  {259}},\ \bibinfo {pages} {345} (\bibinfo {year} {1991})}\BibitemShut
  {NoStop}%
\bibitem [{\citenamefont {Baikov}\ \emph {et~al.}(2010)\citenamefont {Baikov},
  \citenamefont {Chetyrkin},\ and\ \citenamefont {Kuhn}}]{Baikov:2010je}%
  \BibitemOpen
  \bibfield  {author} {\bibinfo {author} {\bibfnamefont {P.~A.}\ \bibnamefont
  {Baikov}}, \bibinfo {author} {\bibfnamefont {K.~G.}\ \bibnamefont
  {Chetyrkin}}, \ and\ \bibinfo {author} {\bibfnamefont {J.~H.}\ \bibnamefont
  {Kuhn}},\ }\href {\doibase 10.1103/PhysRevLett.104.132004} {\bibfield
  {journal} {\bibinfo  {journal} {Phys. Rev. Lett.}\ }\textbf {\bibinfo
  {volume} {104}},\ \bibinfo {pages} {132004} (\bibinfo {year} {2010})},\
  \Eprint {http://arxiv.org/abs/1001.3606} {arXiv:1001.3606 [hep-ph]}
  \BibitemShut {NoStop}%
\bibitem [{\citenamefont {Chetyrkin}\ \emph {et~al.}(2000)\citenamefont
  {Chetyrkin}, \citenamefont {K{\"u}hn},\ and\ \citenamefont
  {Steinhauser}}]{Chetyrkin:2000yt}%
  \BibitemOpen
  \bibfield  {author} {\bibinfo {author} {\bibfnamefont {K.~G.}\ \bibnamefont
  {Chetyrkin}}, \bibinfo {author} {\bibfnamefont {J.~H.}\ \bibnamefont
  {K{\"u}hn}}, \ and\ \bibinfo {author} {\bibfnamefont {M.}~\bibnamefont
  {Steinhauser}},\ }\href {\doibase 10.1016/S0010-4655(00)00155-7} {\bibfield
  {journal} {\bibinfo  {journal} {Comput. Phys. Commun.}\ }\textbf {\bibinfo
  {volume} {133}},\ \bibinfo {pages} {43} (\bibinfo {year} {2000})},\ \Eprint
  {http://arxiv.org/abs/hep-ph/0004189} {arXiv:hep-ph/0004189} \BibitemShut
  {NoStop}%
\bibitem [{\citenamefont {Herren}\ and\ \citenamefont
  {Steinhauser}(2018)}]{Herren:2017osy}%
  \BibitemOpen
  \bibfield  {author} {\bibinfo {author} {\bibfnamefont {F.}~\bibnamefont
  {Herren}}\ and\ \bibinfo {author} {\bibfnamefont {M.}~\bibnamefont
  {Steinhauser}},\ }\href {\doibase 10.1016/j.cpc.2017.11.014} {\bibfield
  {journal} {\bibinfo  {journal} {Comput. Phys. Commun.}\ }\textbf {\bibinfo
  {volume} {224}},\ \bibinfo {pages} {333} (\bibinfo {year} {2018})},\ \Eprint
  {http://arxiv.org/abs/1703.03751} {arXiv:1703.03751 [hep-ph]} \BibitemShut
  {NoStop}%
\bibitem [{\citenamefont {Takahashi}\ \emph {et~al.}(2026)\citenamefont
  {Takahashi} \emph {et~al.}}]{ParticleDataGroup:2026}%
  \BibitemOpen
  \bibfield  {author} {\bibinfo {author} {\bibfnamefont {F.}~\bibnamefont
  {Takahashi}} \emph {et~al.} (\bibinfo {collaboration} {Particle Data
  Group}),\ }\href {\doibase 10.1142/S0217751X26300115} {\bibfield  {journal}
  {\bibinfo  {journal} {Int. J. Mod. Phys. A}\ }\textbf {\bibinfo {volume}
  {41}},\ \bibinfo {pages} {2630011} (\bibinfo {year} {2026})}\BibitemShut
  {NoStop}%
\bibitem [{\citenamefont {Ferroglia}\ \emph {et~al.}(2013)\citenamefont
  {Ferroglia}, \citenamefont {Greub}, \citenamefont {Sirlin},\ and\
  \citenamefont {Zhang}}]{Ferroglia:2013dga}%
  \BibitemOpen
  \bibfield  {author} {\bibinfo {author} {\bibfnamefont {A.}~\bibnamefont
  {Ferroglia}}, \bibinfo {author} {\bibfnamefont {C.}~\bibnamefont {Greub}},
  \bibinfo {author} {\bibfnamefont {A.}~\bibnamefont {Sirlin}}, \ and\ \bibinfo
  {author} {\bibfnamefont {Z.}~\bibnamefont {Zhang}},\ }\href {\doibase
  10.1103/PhysRevD.88.033012} {\bibfield  {journal} {\bibinfo  {journal} {Phys.
  Rev. D}\ }\textbf {\bibinfo {volume} {88}},\ \bibinfo {pages} {033012}
  (\bibinfo {year} {2013})},\ \Eprint {http://arxiv.org/abs/1307.6900}
  {arXiv:1307.6900 [hep-ph]} \BibitemShut {NoStop}%
\bibitem [{\citenamefont {Fael}\ \emph {et~al.}(2013)\citenamefont {Fael},
  \citenamefont {Mercolli},\ and\ \citenamefont {Passera}}]{Fael:2013pja}%
  \BibitemOpen
  \bibfield  {author} {\bibinfo {author} {\bibfnamefont {M.}~\bibnamefont
  {Fael}}, \bibinfo {author} {\bibfnamefont {L.}~\bibnamefont {Mercolli}}, \
  and\ \bibinfo {author} {\bibfnamefont {M.}~\bibnamefont {Passera}},\ }\href
  {\doibase 10.1103/PhysRevD.88.093011} {\bibfield  {journal} {\bibinfo
  {journal} {Phys. Rev. D}\ }\textbf {\bibinfo {volume} {88}},\ \bibinfo
  {pages} {093011} (\bibinfo {year} {2013})},\ \Eprint
  {http://arxiv.org/abs/1310.1081} {arXiv:1310.1081 [hep-ph]} \BibitemShut
  {NoStop}%
\bibitem [{\citenamefont {Behrends}\ and\ \citenamefont
  {Sirlin}(1960)}]{Behrends:1960nf}%
  \BibitemOpen
  \bibfield  {author} {\bibinfo {author} {\bibfnamefont {R.~E.}\ \bibnamefont
  {Behrends}}\ and\ \bibinfo {author} {\bibfnamefont {A.}~\bibnamefont
  {Sirlin}},\ }\href {\doibase 10.1103/PhysRevLett.4.186} {\bibfield  {journal}
  {\bibinfo  {journal} {Phys. Rev. Lett.}\ }\textbf {\bibinfo {volume} {4}},\
  \bibinfo {pages} {186} (\bibinfo {year} {1960})}\BibitemShut {NoStop}%
\bibitem [{\citenamefont {Ademollo}\ and\ \citenamefont
  {Gatto}(1964)}]{Ademollo:1964sr}%
  \BibitemOpen
  \bibfield  {author} {\bibinfo {author} {\bibfnamefont {M.}~\bibnamefont
  {Ademollo}}\ and\ \bibinfo {author} {\bibfnamefont {R.}~\bibnamefont
  {Gatto}},\ }\href {\doibase 10.1103/PhysRevLett.13.264} {\bibfield  {journal}
  {\bibinfo  {journal} {Phys. Rev. Lett.}\ }\textbf {\bibinfo {volume} {13}},\
  \bibinfo {pages} {264} (\bibinfo {year} {1964})}\BibitemShut {NoStop}%
\bibitem [{\citenamefont {Gasser}\ and\ \citenamefont
  {Leutwyler}(1985{\natexlab{a}})}]{Gasser:1984ux}%
  \BibitemOpen
  \bibfield  {author} {\bibinfo {author} {\bibfnamefont {J.}~\bibnamefont
  {Gasser}}\ and\ \bibinfo {author} {\bibfnamefont {H.}~\bibnamefont
  {Leutwyler}},\ }\href {\doibase 10.1016/0550-3213(85)90493-6} {\bibfield
  {journal} {\bibinfo  {journal} {Nucl. Phys. B}\ }\textbf {\bibinfo {volume}
  {250}},\ \bibinfo {pages} {517} (\bibinfo {year}
  {1985}{\natexlab{a}})}\BibitemShut {NoStop}%
\bibitem [{\citenamefont {Crawford}\ \emph {et~al.}(1991)\citenamefont
  {Crawford}, \citenamefont {Daum}, \citenamefont {Frosch}, \citenamefont
  {Jost}, \citenamefont {Kettle}, \citenamefont {Marshall}, \citenamefont
  {Wright},\ and\ \citenamefont {Ziock}}]{Crawford:1990jc}%
  \BibitemOpen
  \bibfield  {author} {\bibinfo {author} {\bibfnamefont {J.~F.}\ \bibnamefont
  {Crawford}}, \bibinfo {author} {\bibfnamefont {M.}~\bibnamefont {Daum}},
  \bibinfo {author} {\bibfnamefont {R.}~\bibnamefont {Frosch}}, \bibinfo
  {author} {\bibfnamefont {B.}~\bibnamefont {Jost}}, \bibinfo {author}
  {\bibfnamefont {P.~R.}\ \bibnamefont {Kettle}}, \bibinfo {author}
  {\bibfnamefont {R.~M.}\ \bibnamefont {Marshall}}, \bibinfo {author}
  {\bibfnamefont {B.~K.}\ \bibnamefont {Wright}}, \ and\ \bibinfo {author}
  {\bibfnamefont {K.~O.~H.}\ \bibnamefont {Ziock}},\ }\href {\doibase
  10.1103/PhysRevD.43.46} {\bibfield  {journal} {\bibinfo  {journal} {Phys.
  Rev. D}\ }\textbf {\bibinfo {volume} {43}},\ \bibinfo {pages} {46} (\bibinfo
  {year} {1991})}\BibitemShut {NoStop}%
\bibitem [{\citenamefont {Gasser}\ and\ \citenamefont
  {Leutwyler}(1985{\natexlab{b}})}]{Gasser:1984gg}%
  \BibitemOpen
  \bibfield  {author} {\bibinfo {author} {\bibfnamefont {J.}~\bibnamefont
  {Gasser}}\ and\ \bibinfo {author} {\bibfnamefont {H.}~\bibnamefont
  {Leutwyler}},\ }\href {\doibase 10.1016/0550-3213(85)90492-4} {\bibfield
  {journal} {\bibinfo  {journal} {Nucl. Phys. B}\ }\textbf {\bibinfo {volume}
  {250}},\ \bibinfo {pages} {465} (\bibinfo {year}
  {1985}{\natexlab{b}})}\BibitemShut {NoStop}%
\bibitem [{\citenamefont {Bijnens}\ and\ \citenamefont
  {Talavera}(2002)}]{Bijnens:2002hp}%
  \BibitemOpen
  \bibfield  {author} {\bibinfo {author} {\bibfnamefont {J.}~\bibnamefont
  {Bijnens}}\ and\ \bibinfo {author} {\bibfnamefont {P.}~\bibnamefont
  {Talavera}},\ }\href {\doibase 10.1088/1126-6708/2002/03/046} {\bibfield
  {journal} {\bibinfo  {journal} {JHEP}\ }\textbf {\bibinfo {volume} {03}},\
  \bibinfo {pages} {046} (\bibinfo {year} {2002})},\ \Eprint
  {http://arxiv.org/abs/hep-ph/0203049} {arXiv:hep-ph/0203049} \BibitemShut
  {NoStop}%
\bibitem [{\citenamefont {Colangelo}\ \emph
  {et~al.}(2022{\natexlab{b}})\citenamefont {Colangelo}, \citenamefont
  {Hoferichter}, \citenamefont {Kubis}, \citenamefont {Niehus},\ and\
  \citenamefont {Ruiz~de Elvira}}]{Colangelo:2021moe}%
  \BibitemOpen
  \bibfield  {author} {\bibinfo {author} {\bibfnamefont {G.}~\bibnamefont
  {Colangelo}}, \bibinfo {author} {\bibfnamefont {M.}~\bibnamefont
  {Hoferichter}}, \bibinfo {author} {\bibfnamefont {B.}~\bibnamefont {Kubis}},
  \bibinfo {author} {\bibfnamefont {M.}~\bibnamefont {Niehus}}, \ and\ \bibinfo
  {author} {\bibfnamefont {J.}~\bibnamefont {Ruiz~de Elvira}},\ }\href
  {\doibase 10.1016/j.physletb.2021.136852} {\bibfield  {journal} {\bibinfo
  {journal} {Phys. Lett. B}\ }\textbf {\bibinfo {volume} {825}},\ \bibinfo
  {pages} {136852} (\bibinfo {year} {2022}{\natexlab{b}})},\ \Eprint
  {http://arxiv.org/abs/2110.05493} {arXiv:2110.05493 [hep-ph]} \BibitemShut
  {NoStop}%
\bibitem [{\citenamefont {Passera}\ \emph {et~al.}(2011)\citenamefont
  {Passera}, \citenamefont {Philippides},\ and\ \citenamefont
  {Sirlin}}]{Passera:2011ae}%
  \BibitemOpen
  \bibfield  {author} {\bibinfo {author} {\bibfnamefont {M.}~\bibnamefont
  {Passera}}, \bibinfo {author} {\bibfnamefont {K.}~\bibnamefont
  {Philippides}}, \ and\ \bibinfo {author} {\bibfnamefont {A.}~\bibnamefont
  {Sirlin}},\ }\href {\doibase 10.1103/PhysRevD.84.094030} {\bibfield
  {journal} {\bibinfo  {journal} {Phys. Rev. D}\ }\textbf {\bibinfo {volume}
  {84}},\ \bibinfo {pages} {094030} (\bibinfo {year} {2011})},\ \Eprint
  {http://arxiv.org/abs/1109.1069} {arXiv:1109.1069 [hep-ph]} \BibitemShut
  {NoStop}%
\bibitem [{\citenamefont {Kinoshita}(1962)}]{Kinoshita:1962ur}%
  \BibitemOpen
  \bibfield  {author} {\bibinfo {author} {\bibfnamefont {T.}~\bibnamefont
  {Kinoshita}},\ }\href {\doibase 10.1063/1.1724268} {\bibfield  {journal}
  {\bibinfo  {journal} {J. Math. Phys.}\ }\textbf {\bibinfo {volume} {3}},\
  \bibinfo {pages} {650} (\bibinfo {year} {1962})}\BibitemShut {NoStop}%
\bibitem [{\citenamefont {Lee}\ and\ \citenamefont
  {Nauenberg}(1964)}]{Lee:1964is}%
  \BibitemOpen
  \bibfield  {author} {\bibinfo {author} {\bibfnamefont {T.~D.}\ \bibnamefont
  {Lee}}\ and\ \bibinfo {author} {\bibfnamefont {M.}~\bibnamefont
  {Nauenberg}},\ }\href {\doibase 10.1103/PhysRev.133.B1549} {\bibfield
  {journal} {\bibinfo  {journal} {Phys. Rev.}\ }\textbf {\bibinfo {volume}
  {133}},\ \bibinfo {pages} {B1549} (\bibinfo {year} {1964})}\BibitemShut
  {NoStop}%
\bibitem [{\citenamefont {Aguilar-Arevalo}\ \emph {et~al.}(2015)\citenamefont
  {Aguilar-Arevalo} \emph {et~al.}}]{PiENu:2015seu}%
  \BibitemOpen
  \bibfield  {author} {\bibinfo {author} {\bibfnamefont {A.}~\bibnamefont
  {Aguilar-Arevalo}} \emph {et~al.} (\bibinfo {collaboration} {PiENu}),\ }\href
  {\doibase 10.1103/PhysRevLett.115.071801} {\bibfield  {journal} {\bibinfo
  {journal} {Phys. Rev. Lett.}\ }\textbf {\bibinfo {volume} {115}},\ \bibinfo
  {pages} {071801} (\bibinfo {year} {2015})},\ \Eprint
  {http://arxiv.org/abs/1506.05845} {arXiv:1506.05845 [hep-ex]} \BibitemShut
  {NoStop}%
\bibitem [{\citenamefont {Czapek}\ \emph {et~al.}(1993)\citenamefont {Czapek}
  \emph {et~al.}}]{Czapek:1993kc}%
  \BibitemOpen
  \bibfield  {author} {\bibinfo {author} {\bibfnamefont {G.}~\bibnamefont
  {Czapek}} \emph {et~al.},\ }\href {\doibase 10.1103/PhysRevLett.70.17}
  {\bibfield  {journal} {\bibinfo  {journal} {Phys. Rev. Lett.}\ }\textbf
  {\bibinfo {volume} {70}},\ \bibinfo {pages} {17} (\bibinfo {year}
  {1993})}\BibitemShut {NoStop}%
\bibitem [{\citenamefont {Britton}\ \emph {et~al.}(1992)\citenamefont {Britton}
  \emph {et~al.}}]{Britton:1992pg}%
  \BibitemOpen
  \bibfield  {author} {\bibinfo {author} {\bibfnamefont {D.~I.}\ \bibnamefont
  {Britton}} \emph {et~al.},\ }\href {\doibase 10.1103/PhysRevLett.68.3000}
  {\bibfield  {journal} {\bibinfo  {journal} {Phys. Rev. Lett.}\ }\textbf
  {\bibinfo {volume} {68}},\ \bibinfo {pages} {3000} (\bibinfo {year}
  {1992})}\BibitemShut {NoStop}%
\bibitem [{\citenamefont {Koptev}\ \emph {et~al.}(1995)\citenamefont {Koptev}
  \emph {et~al.}}]{Koptev:1995je}%
  \BibitemOpen
  \bibfield  {author} {\bibinfo {author} {\bibfnamefont {V.~P.}\ \bibnamefont
  {Koptev}} \emph {et~al.},\ }\href@noop {} {\bibfield  {journal} {\bibinfo
  {journal} {JETP Lett.}\ }\textbf {\bibinfo {volume} {61}},\ \bibinfo {pages}
  {877} (\bibinfo {year} {1995})}\BibitemShut {NoStop}%
\bibitem [{\citenamefont {Numao}\ \emph {et~al.}(1995)\citenamefont {Numao},
  \citenamefont {Macdonald}, \citenamefont {Marshall}, \citenamefont {Olin},\
  and\ \citenamefont {Fujiwara}}]{Numao:1995qf}%
  \BibitemOpen
  \bibfield  {author} {\bibinfo {author} {\bibfnamefont {T.}~\bibnamefont
  {Numao}}, \bibinfo {author} {\bibfnamefont {J.~A.}\ \bibnamefont
  {Macdonald}}, \bibinfo {author} {\bibfnamefont {G.~M.}\ \bibnamefont
  {Marshall}}, \bibinfo {author} {\bibfnamefont {A.}~\bibnamefont {Olin}}, \
  and\ \bibinfo {author} {\bibfnamefont {M.~C.}\ \bibnamefont {Fujiwara}},\
  }\href {\doibase 10.1103/PhysRevD.52.4855} {\bibfield  {journal} {\bibinfo
  {journal} {Phys. Rev. D}\ }\textbf {\bibinfo {volume} {52}},\ \bibinfo
  {pages} {4855} (\bibinfo {year} {1995})}\BibitemShut {NoStop}%
\bibitem [{\citenamefont {Marciano}\ and\ \citenamefont
  {Sirlin}(1986)}]{Marciano:1985pd}%
  \BibitemOpen
  \bibfield  {author} {\bibinfo {author} {\bibfnamefont {W.~J.}\ \bibnamefont
  {Marciano}}\ and\ \bibinfo {author} {\bibfnamefont {A.}~\bibnamefont
  {Sirlin}},\ }\href {\doibase 10.1103/PhysRevLett.56.22} {\bibfield  {journal}
  {\bibinfo  {journal} {Phys. Rev. Lett.}\ }\textbf {\bibinfo {volume} {56}},\
  \bibinfo {pages} {22} (\bibinfo {year} {1986})}\BibitemShut {NoStop}%
\bibitem [{\citenamefont {Marciano}\ and\ \citenamefont
  {Sirlin}(1988)}]{Marciano:1988vm}%
  \BibitemOpen
  \bibfield  {author} {\bibinfo {author} {\bibfnamefont {W.~J.}\ \bibnamefont
  {Marciano}}\ and\ \bibinfo {author} {\bibfnamefont {A.}~\bibnamefont
  {Sirlin}},\ }\href {\doibase 10.1103/PhysRevLett.61.1815} {\bibfield
  {journal} {\bibinfo  {journal} {Phys. Rev. Lett.}\ }\textbf {\bibinfo
  {volume} {61}},\ \bibinfo {pages} {1815} (\bibinfo {year}
  {1988})}\BibitemShut {NoStop}%
\bibitem [{\citenamefont {Marciano}\ and\ \citenamefont
  {Sirlin}(1993)}]{Marciano:1993sh}%
  \BibitemOpen
  \bibfield  {author} {\bibinfo {author} {\bibfnamefont {W.~J.}\ \bibnamefont
  {Marciano}}\ and\ \bibinfo {author} {\bibfnamefont {A.}~\bibnamefont
  {Sirlin}},\ }\href {\doibase 10.1103/PhysRevLett.71.3629} {\bibfield
  {journal} {\bibinfo  {journal} {Phys. Rev. Lett.}\ }\textbf {\bibinfo
  {volume} {71}},\ \bibinfo {pages} {3629} (\bibinfo {year}
  {1993})}\BibitemShut {NoStop}%
\bibitem [{\citenamefont {Braaten}\ and\ \citenamefont
  {Li}(1990)}]{Braaten:1990ef}%
  \BibitemOpen
  \bibfield  {author} {\bibinfo {author} {\bibfnamefont {E.}~\bibnamefont
  {Braaten}}\ and\ \bibinfo {author} {\bibfnamefont {C.-S.}\ \bibnamefont
  {Li}},\ }\href {\doibase 10.1103/PhysRevD.42.3888} {\bibfield  {journal}
  {\bibinfo  {journal} {Phys. Rev. D}\ }\textbf {\bibinfo {volume} {42}},\
  \bibinfo {pages} {3888} (\bibinfo {year} {1990})}\BibitemShut {NoStop}%
\bibitem [{\citenamefont {Erler}(2004)}]{Erler:2002mv}%
  \BibitemOpen
  \bibfield  {author} {\bibinfo {author} {\bibfnamefont {J.}~\bibnamefont
  {Erler}},\ }\href@noop {} {\bibfield  {journal} {\bibinfo  {journal} {Rev.
  Mex. Fis.}\ }\textbf {\bibinfo {volume} {50}},\ \bibinfo {pages} {200}
  (\bibinfo {year} {2004})},\ \Eprint {http://arxiv.org/abs/hep-ph/0211345}
  {arXiv:hep-ph/0211345} \BibitemShut {NoStop}%
\bibitem [{\citenamefont {Davier}\ \emph {et~al.}(2003)\citenamefont {Davier},
  \citenamefont {Eidelman}, \citenamefont {Hocker},\ and\ \citenamefont
  {Zhang}}]{Davier:2002dy}%
  \BibitemOpen
  \bibfield  {author} {\bibinfo {author} {\bibfnamefont {M.}~\bibnamefont
  {Davier}}, \bibinfo {author} {\bibfnamefont {S.}~\bibnamefont {Eidelman}},
  \bibinfo {author} {\bibfnamefont {A.}~\bibnamefont {Hocker}}, \ and\ \bibinfo
  {author} {\bibfnamefont {Z.}~\bibnamefont {Zhang}},\ }\href {\doibase
  10.1140/epjc/s2003-01136-2} {\bibfield  {journal} {\bibinfo  {journal} {Eur.
  Phys. J. C}\ }\textbf {\bibinfo {volume} {27}},\ \bibinfo {pages} {497}
  (\bibinfo {year} {2003})},\ \Eprint {http://arxiv.org/abs/hep-ph/0208177}
  {arXiv:hep-ph/0208177} \BibitemShut {NoStop}%
\bibitem [{\citenamefont {Cirigliano}\ \emph
  {et~al.}(2022{\natexlab{b}})\citenamefont {Cirigliano}, \citenamefont
  {de~Vries}, \citenamefont {Hayen}, \citenamefont {Mereghetti},\ and\
  \citenamefont {Walker-Loud}}]{Cirigliano:2022hob}%
  \BibitemOpen
  \bibfield  {author} {\bibinfo {author} {\bibfnamefont {V.}~\bibnamefont
  {Cirigliano}}, \bibinfo {author} {\bibfnamefont {J.}~\bibnamefont
  {de~Vries}}, \bibinfo {author} {\bibfnamefont {L.}~\bibnamefont {Hayen}},
  \bibinfo {author} {\bibfnamefont {E.}~\bibnamefont {Mereghetti}}, \ and\
  \bibinfo {author} {\bibfnamefont {A.}~\bibnamefont {Walker-Loud}},\ }\href
  {\doibase 10.1103/PhysRevLett.129.121801} {\bibfield  {journal} {\bibinfo
  {journal} {Phys. Rev. Lett.}\ }\textbf {\bibinfo {volume} {129}},\ \bibinfo
  {pages} {121801} (\bibinfo {year} {2022}{\natexlab{b}})},\ \Eprint
  {http://arxiv.org/abs/2202.10439} {arXiv:2202.10439 [nucl-th]} \BibitemShut
  {NoStop}%
\bibitem [{\citenamefont {Ji}\ and\ \citenamefont {Musolf}(1991)}]{Ji:1991pr}%
  \BibitemOpen
  \bibfield  {author} {\bibinfo {author} {\bibfnamefont {X.-D.}\ \bibnamefont
  {Ji}}\ and\ \bibinfo {author} {\bibfnamefont {M.~J.}\ \bibnamefont
  {Musolf}},\ }\href {\doibase 10.1016/0370-2693(91)91916-J} {\bibfield
  {journal} {\bibinfo  {journal} {Phys. Lett. B}\ }\textbf {\bibinfo {volume}
  {257}},\ \bibinfo {pages} {409} (\bibinfo {year} {1991})}\BibitemShut
  {NoStop}%
\end{thebibliography}%

\end{document}